\documentclass[11pt]{article}
\pdfoutput=1
\usepackage[centertags]{amsmath}
\usepackage[square, comma, sort&compress,numbers]{natbib}
\usepackage{array,multirow}
\numberwithin{equation}{section}
\usepackage{amssymb,amsfonts}
\usepackage{graphicx}
\usepackage{color}
\usepackage{mathtools,bm}
\usepackage{accents}
\usepackage[dvipsnames]{xcolor}
\usepackage{oplotsymbl}

\usepackage{epsfig}
\usepackage{bbold}

\usepackage{wrapfig}
\usepackage{float}
\usepackage{soul}
\usepackage{tkz-euclide}
\usepackage{braket}

\usepackage{tikz,pgf}
\usetikzlibrary{shapes}
\usetikzlibrary{calc}
\usetikzlibrary{decorations.pathmorphing}
\usetikzlibrary{decorations.pathreplacing,shapes.misc}
\usetikzlibrary{positioning}
\usetikzlibrary{arrows}
\usetikzlibrary{decorations.markings}
\usetikzlibrary{shadings}

\usetikzlibrary{intersections}

\def\be{\begin{equation}}
\def\ee{\end{equation}}
\def\ba{\begin{array}}
\def\ea{\end{array}}

\def\bst{\begin{split}}
\def\est{\end{split}}

\def\dps{\displaystyle}

\def\Li2{\operatorname{Li_2}}

\newdimen\tableauside\tableauside=1.0ex
\newdimen\tableaurule\tableaurule=0.4pt
\newdimen\tableaustep
\def\phantomhrule#1{\hbox{\vbox to0pt{\hrule height\tableaurule
width#1\vss}}}
\def\phantomvrule#1{\vbox{\hbox to0pt{\vrule width\tableaurule
height#1\hss}}}
\def\sqr{\vbox{%
\phantomhrule\tableaustep

\hbox{\phantomvrule\tableaustep\kern\tableaustep\phantomvrule\tableaustep}%
\hbox{\vbox{\phantomhrule\tableauside}\kern-\tableaurule}}}
\def\squares#1{\hbox{\count0=#1\noindent\loop\sqr
\advance\count0 by-1 \ifnum\count0>0\repeat}}
\def\tableau#1{\vcenter{\offinterlineskip
\tableaustep=\tableauside\advance\tableaustep by-\tableaurule
\kern\normallineskip\hbox
{\kern\normallineskip\vbox
{\gettableau#1 0 }%
\kern\normallineskip\kern\tableaurule}%
\kern\normallineskip\kern\tableaurule}}
\def\gettableau#1 {\ifnum#1=0\let\next=\null\else
\squares{#1}\let\next=\gettableau\fi\next}

\newcommand{\bref}[1]{\textbf{\ref{#1}}}

\newcommand{\RR}{\mathbb{R}}

\newcommand{\ZZ}{\mathbb{Z}}

\def\cA{\mathcal{A}}
\def\cB{\mathcal{B}}

\def\cN{\mathcal{N}}

\def\cS{\mathcal{S}}

\def\rA{{\rm A}}

\def\rR{{\rm R}}
\def\rS{{\rm S}}
\def\rT{{\rm T}}
\def\rU{{\rm U}}
\def\rV{{\rm V}}
\def\rW{{\rm W}}

\def\rY{{\rm Y}}

\numberwithin{equation}{section} \makeatletter
\@addtoreset{equation}{section}

\def\be{\begin{equation}}
\def\ee{\end{equation}}
\def\ba{\begin{array}}
\def\ea{\end{array}}

\def\dps{\displaystyle}

\def\hd{\frac{D}{2}}

\newcommand{\dl}{\Delta}

\def\C2{\text{C}_2}

\def\HG{{\rm H}}
\def\BF{\Phi}

\newcommand{\triple}[1]{\langle #1 \rangle} 
\newcommand{\cycle}{\mathrm{C}}
\newcommand{\cham}{{\mathbb{C}}}

\usepackage{jheppub}
\makeatletter
\def\@fpheader{\vspace{-.1cm}}
\makeatother

\title{\centering{Multipoint conformal integrals in $D$ dimensions. Part II\\ {\Large Polygons   and basis functions}}}

\author{Konstantin\ Alkalaev and}
\author{Semyon\ Mandrygin}

\affiliation{I.E. Tamm Department of Theoretical Physics, \\P.N. Lebedev Physical
Institute, 119991 Moscow, Russia}

\emailAdd{alkalaev@lpi.ru}
\emailAdd{semyon.mandrygin@gmail.com}

\abstract{We explicitly construct a  class of  multivariate generalized hypergeometric series  which is conjectured in our previous paper \cite{Alkalaev:2025fgn} to  calculate   multipoint one-loop parametric conformal integrals in $D$  dimensions. Our approach is based on a simple  diagrammatic algorithm which systematically builds both arguments and series coefficients in terms of a convex polygon which is part of the  Baxter lattice. The examples of the box, pentagon, and hexagon integrals are considered in detail.
}

\begin{document}

\maketitle
\flushbottom

\section{Introduction}

Feynman integrals calculate transition amplitudes of elementary particles that makes their study pivotal for our understanding of  quantum field theory and high-energy physics.  Despite many remarkable  advances in recent  decades (for review see e.g. \cite{Smirnov:2004ym,Weinzierl:2022eaz}), which  revealed a number of underlying algebraic and geometric structures,  the explicit evaluation of general Feynman  integrals and the study of their analytical  properties remain a challenging task. 

In favourable cases, Feynman integrals can be evaluated in terms of known special functions and their various generalizations \cite{Kalmykov:2008gq, Bourjaily:2022bwx}. Frequently, the presence of a particular (in)finite-dimensional symmetry may simplify the calculation and organize the final expression  in terms of relevant variables.   For example,  among other things, one can explicitly calculate the three-point (in any dimensions) and four-point (in four dimensions) conformally invariant Feynman integrals \cite{Usyukina:1992jd}\footnote{The  conformal integrals \cite{Symanzik:1972wj}  appear in various field theories including general CFTs on both planar \cite{Ferrara:1972uq,Ferrara:1972kab, Dolan:2000uw, Dolan:2000ut,Dolan:2011dv,Fateev:2011qa,SimmonsDuffin:2012uy,Rosenhaus:2018zqn} and thermal \cite{Petkou:2021zhg, Karydas:2023ufs, Alkalaev:2023evp, Alkalaev:2024jxh, Belavin:2024mzd, Belavin:2024nnw} backgrounds, $\cN=4$ SYM theory  \cite{Drummond:2006rz,Drummond:2008vq}, the fishnet CFT models  \cite{Zamolodchikov:1980mb, Gurdogan:2015csr, Loebbert:2022nfu}.} as the rational function and the Bloch-Wigner function, respectively.

In this paper, a class of special functions relevant to the calculation of multipoint conformal integrals is proposed. Namely, we introduce $n$-parametric multivariate generalized hypergeometric series in $n(n-3)/2$ conformal cross-ratios associated with arbitrary $n$ points in $\RR^{D}$.\footnote{\label{f1} A number of independent cross-ratios equals  $n(n-3)/2$ at $n \leq D+2$ or $n D - (D+1)(D+2)/2$ at $n > D+2$, see e.g. \cite{Osborn}. These inequalities can be easily understood using the ambient space approach, where $\RR^{D}$ space can be realized as a null cone's section in $\RR^{D+1,1}$ space. Then, more then $D+2$ points are linearly  dependent that reduces a number of independent cross-ratios. A convenient tool here is the Gram determinant, its zero surfaces determine all functional dependencies of cross-ratios \cite{Buric:2021kgy}.  
In the  present paper, we will assume that a complete set of cross-ratios for  $n$ points in $\RR^{D}$ for any $n$ and $D$ consists of $n(n-3)/2$ elements. The issue of dependent cross-ratios and the corresponding reduced hypergeometric series  will be considered elsewhere.} Our construction uses a diagrammatic algorithm which consists of a number of geometric operations  applied to a convex  $n$-gon on the plane with specified angles. Such a polygon provides a minimal set of geometric data from which one can build particular sets of independent cross-ratios and define series coefficients as rational functions of $n$ parameters.

We conjecture that the $n$-point one-loop parametric conformal integrals in particular coordinate domains are evaluated in terms of such basis functions. Moreover, a set of functions which calculates a given conformal integral is invariant under the action of the cyclic group $\ZZ_n$. It follows that  the set of functions can be split into $\ZZ_n$-orbits of which representatives we call master functions. It is clear that knowing master functions one can build the whole conformal integral simply  by acting with cyclic transformations. This phenomenon was observed in explicit calculations of some lower-point conformal integrals in our previous paper \cite{Alkalaev:2025fgn}, where it was  called the reconstruction conjecture.\footnote{A similar permutation generated representation is known in the literature on Feynman integrals: e.g. see the non-parametric hexagon integral with three massive corners \cite{DelDuca:2011wh}; the non-parametric pentagon integral \cite{Nandan:2013ip}; the Yangian bootstrap for the lower point conformal integrals in \cite{Loebbert:2019vcj}.}

The paper is organized as follows. In section \bref{sec:conf_int} we briefly review the conformal integrals and formulate the reconstruction conjecture. Here, we introduce our main geometric tools which are convex polygons and associated plane geometry constructions. In section \bref{sec:basis} we systematically build a new class of multivariate hypergeometric series using a  diagrammatic algorithm. Finally, in section \bref{sec:examples} we illustrate our approach in all detail by considering the box, pentagon, and hexagon conformal integrals. Here, the final expressions for the pentagon and hexagon parametric conformal integrals are given by eqs. \eqref{pentagon_reconstruction} and \eqref{hexagon_reconstruction}, respectively. In the concluding section \bref{sec:concl} we summarize our results and discuss possible future directions. In Appendix \bref{app:orbit} we study the action of the symmetric groups on cross-ratios. The generalized  hypergeometric  functions which represent the pentagon and hexagon conformal integrals are collected in Appendix \bref{app:expl}.

\section{Conformal integrals}
\label{sec:conf_int}

\begin{figure}
\centering
\includegraphics[scale=0.8]{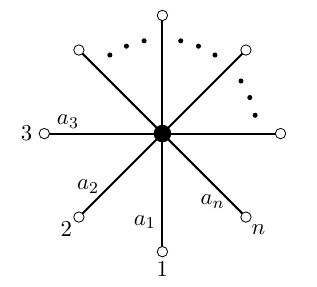}
\caption{The conformal integral  $I_n^{{\bm a}}({\bm x})$ represented as  the labelled  $n$-valent vertex. The  $i$-th leg depicts the propagator $X_{0i}^{-a_i}$ which is characterized by the position $x_i$ and the propagator power $a_i$; the vertex denotes integration over $x_0$.} 
\label{fig:vertex}
\end{figure}

We study a parametric $n$-point one-loop conformal integral in $\RR^{D}$  (see fig. \bref{fig:vertex}): 
\be
\label{indef}
I_n^{{\bm a}}({\bm x}) = \int_{\mathbb{R}^D} \frac{{\rm d}^D x_0}{\pi^{D/2}} \,\prod_{i=1}^n  X_{0i}^{-a_i} 
 \,, \qquad X_{ij} \equiv X_{i,j} = (x_i - x_j)^2\,.
\ee
Here, ${\bm x} = \{x_i \in \RR^{D},\, i=1,..., n\}$, the propagator powers   ${\bm a} = \{a_i \in \RR,\, i=1,..., n\}$ obey the conformality  constraint
\be
\label{confcond}
\sum_{i=1}^n a_i = D \,,
\ee
which ensures the covariant transformation law of the conformal integral  under the conformal group $O(D+1,1)$:
\be
\label{in_transform}
I_n^{\bm a}(\bm x') = \prod_{i=1}^{n} \Omega^{-a_i}(x_i) \, I_n^{\bm a}(\bm x)\,,
\ee
where $\Omega(x)$ is a local scale factor: ${\rm d} x'^2 = \Omega(x)^2 {\rm d} x^2$. This transformation law partially fixes the conformal integral up to an unknown function depending only on conformally invariant cross-ratios. Another constraint comes from the invariance of the conformal integral under the symmetric group:\footnote{{The symmetric group of degree $n$ is denoted as $\cS_n$; arbitrary permutations and transpositions $(ij)$ are denoted as $\pi$ and $\sigma_{ij}$; an identity permutation is $e$, a longest cycle is $C_n = (12...n)$, its representation will be denoted $\rR(C_n)\equiv \cycle_n$.}} 
\be
\label{s_invariant}
\forall \pi \in \cS_n : \qquad \rR(\pi) \circ  I_n^{\bm a}(\bm x) =  I_n^{\bm a}(\bm x)\,,
\ee
where $\rR(\pi)$ represents a permutation $\pi$ acting as $x_i \to x_{\pi(i)}$, $a_i \to a_{\pi(i)}$.

Up to date, only a handful of conformal integrals are known explicitly since the evaluation of \eqref{indef} quickly becomes difficult as  $n$ increases. Actually, the only well-understood example is the  $4$-point (box) parametric conformal integral (see e.g. \cite{Dolan:2000uw} and references therein). Also, the 5-point (pentagon) non-parametric conformal integral was calculated  in \cite{Nandan:2013ip} by using a geometric approach which treats Feynman integrals as volumes of simplices \cite{Davydychev:1997wa,Mason:2010pg,Schnetz:2010pd,Nandan:2013ip, Bourjaily:2019exo, Ren:2023tuj}; the parametric version was calculated in \cite{Alkalaev:2025fgn} by using the bipartite Mellin-Barnes representation, where it was also checked that choosing unit propagator powers one reproduces the non-parametric result. The $6$-point (hexagon) parametric conformal integral was  calculated in \cite{Ananthanarayan:2020ncn} by using the method of multifold  Mellin-Barnes integrals \cite{Ananthanarayan:2020fhl}.\footnote{A refined method was also proposed which resulted  in a simpler expression for the hexagon integral \cite{Banik:2023rrz} (see also \cite{Banik:2024ann} for the discussion of relevant issues).} Note that parametric conformal integrals were  recently  considered  using the Gelfand-Kapranov-Zelevinsky (GKZ) hypergeometric systems   \cite{Pal:2021llg, Pal:2023kgu, Levkovich-Maslyuk:2024zdy}. This   latter approach can be viewed as a natural development of the Yangian bootstrap \cite{Loebbert:2019vcj}, where the Yangian invariance  of  conformal integrals  \cite{Chicherin:2017cns, Chicherin:2017frs} is supplemented by the permutation invariance  condition \eqref{s_invariant}.

\subsection{Reconstruction}
\label{sec:rec}

In \cite{Alkalaev:2025fgn}  we  conjectured  that the conformal integral can be decomposed over a system of {\it basis} functions $\BF_n^{\triple{ijk}}({\bm a}|\bm x)$ as follows  
\be
\label{rep1}
I_n^{{\bm a}}({\bm x}) = \sum_{\triple{ijk} \in {\rm R}_n} \BF_n^{\triple{ijk}}({\bm a}|\bm x)\,,
\ee
where the basis functions are labelled  by ordered index triples
\be
\label{Rn}
{\rm R}_n = \big\{\triple{ijk}, 1\leqslant i<j<k \leqslant n\big\}\,.
\ee 
Thus, a number of  basis functions in \eqref{rep1} equals 
\be
\label{num_basis}
|{\rm R}_n| = \frac{n(n-1)(n-2)}{6}\,.
\ee 
Each basis function is the product of three factors,  
\be
\label{rep2}
\BF_n^{\langle ijk\rangle}({\bm a}|\bm x) 
= {\rm S}_{n}^{\triple{ijk}}({\bm a})\,
  {\rm V}_n^{\triple{ijk}}({\bm a}|\bm x)\,
  \HG_n^{\triple{ijk}}\big( {\bm a}\big|\bm \rY_n^{\triple{ijk}} \big)\,,
\ee  
where 
\begin{itemize}

\item $\rS_{n}^{\triple{ijk}}({\bm a})$ is a {\it triangle-factor} which  is a particular product of $\Gamma$-functions depending  on parameters $\bm a$;

\item $\rV_n^{\triple{ijk}}({\bm a}|\bm x)$ is a {\it leg-factor} which ensures correct transformation law of the conformal integral under conformal transformations  \eqref{in_transform}; 

Both factors are given in eqs. \eqref{star_n}  and \eqref{V_n} below.  

\item  $\HG_n^{\triple{ijk}}\big( {\bm a}\big|\bm \rY_n^{\triple{ijk}} \big)$ is {a} {\it polygonal } function which is a multivariate hypergeometric series\footnote{One may expect that in the case $a_l=1$,  $l=1,...,n$, a polygonal function can be expressed in terms of (multiple) polylogarithms. At least, this holds true in the case of  the lower-point $(n=4,5)$ conformal integrals, see e.g. \cite{Alkalaev:2025fgn}.} with  arguments being cross-ratios collectively denoted as   
\be
\label{cross-set}
\bm \rY_n^{\triple{ijk}}  = \Big\{\rY_l^{\triple{ijk}}, \, l=1,..., \frac{n(n-3)}{2} \Big\}\,.
\ee

\end{itemize}

The choice of a particular set \eqref{cross-set} and, hence, the choice of a particular polygonal  function partially breaks the invariance condition \eqref{s_invariant}. The residual permutation symmetry is governed  by the cyclic subgroup $\ZZ_n \subset \cS_n$ which  splits  ${\rm R}_n$ into $\ZZ_n$-orbits,
\be
\label{2.9}
\big\{\ZZ_n \circ \triple{ijk}\big\} \subset {\rm R}_n \,, \qquad \mathbb{Z}_{n} = \big\{e, C_n, (C_n)^2, ..., (C_n)^{n-1}\big\}\,.
\ee
From each orbit we choose a representative $\triple{i j k}$ determined by the minimum value of the sum $i+j+k$. The resulting set ${\rm T}_n \subset {\rm R}_n$ contains  all ordered index triples which are not related to each other by permutations from the cyclic group $\ZZ_n$. By construction, all triples $\triple{i j k}$  from ${\rm T}_n$ have the property that $i,j,k \neq n$ which is important for subsequent analysis. The number of such triples is given by
\be
\label{cord}
\left| {\rm T}_n \right| = \begin{cases}
\dps \frac{(n-2)(n-1)}{6}\,, \quad \text{if} \quad \frac{n}{3} \notin \mathbb{N}\,, 
\vspace{2mm}
\\
\dps \frac{n(n-3)}{6}+1 \,, \qquad \text{if} \quad \frac{n}{3} \in \mathbb{N}\,.
\end{cases}
\ee 

Thus, there are $\left| {\rm T}_n \right|$ orbits, but, in general, they can have different lengths. If $n/3 \notin \mathbb{N}$, then each orbit has length $n$. If $n/3 \in \mathbb{N}$, then all orbits but one  have  length $n$, while one orbit is shorter and has length $n/3$. As a consequence, the set of basis functions is also split into $\ZZ_n$-orbits of which representatives are called  {\it master functions} $\BF_n^{\langle ijk\rangle}({\bm a}|\bm x)$, $\triple{ijk} \in \rT_n$. In this way, we can formulate the following

\paragraph{Reconstruction conjecture.}
{\it
The $n$-point conformal integral \eqref{indef} is the $\mathbb{Z}_n$-invariant sum of basis  functions
\be
\label{conj_fin}
I_n^{{\bm a}}({\bm x})\, \overset{\mathrm{nr}}{=}\; \sum_{m=0}^{n-1} (\cycle_n)^m  \circ \sum_{\triple{ijk} \in \rT_n} \BF_n^{\triple{ijk}}({\bm a}|\bm x) \,.
\ee
Here, the symbol $\overset{\mathrm{nr}}{=}$ takes into account the orbit shortening for $n/3 \in \mathbb{N}$ and implies that among all basis functions produced by acting with $\mathbb{Z}_n$ on the master functions one keeps only non-repeating ones which number equals \eqref{num_basis}. 
}

The main idea underlying the reconstruction is that instead of finding the full set of $|\rR_n|$ basis functions it is sufficient to have  $|\rT_n|$ master functions. Note that cardinalities of the two sets grow according to different power laws: $|\rR_n| \sim n^3$, while $|\rT_n| \sim n^2$. On the other hand, the conformal integral solves the partial differential equations which follow from the Yangian invariance condition \cite{Chicherin:2017frs, Chicherin:2017cns, Loebbert:2019vcj}. In this respect, our reconstruction approach suggests a method for  classifying  solutions to the Yangian equations according to  their properties under  the action of the cyclic group.  Finally, note that any element of the symmetric group $\cS_n$ can be represented as a composition of a transposition and a longest cycle, see e.g. \cite{Isaev_Rubakov_1}. Then, the action of transpositions on the  manifestly $\ZZ_n$-invariant expression \eqref{conj_fin} should provide a reasonable way to find various analytic continuation formulas for the conformal integral.

\paragraph{Motivation and consistency checks.} 
Using the standard techniques the $n$-point conformal integral can be expressed through a $n(n-3)/2$-fold integral of the Mellin-Barnes type \cite{Symanzik:1972wj}. Notably, the integral is balanced \cite{Dubovyk:2022obc} which implies that there are  multiple choices to close the integration contours. As a result, the Cauchy's integral formula applied to different sets of poles leads to different  multivariate power series each of which converges  on its own  coordinate domain. If the corresponding  analytic continuation formulas are known then these power series can be related to each other. On the other hand, the choice of a convergence domain where one wants to obtain an asymptotic expansion of the conformal integral may help to select a specific closure of the integration contours. A general issue with this approach is that there is no obvious criterion for choosing the way to close contours which would lead to the most streamlined and systemized expression for the integral in a given coordinate domain. Such a multidimensional problem for the general  balanced  multifold Mellin–Barnes integral is poorly tractable, although  it has recently been probed by several methods \cite{Ananthanarayan:2020fhl, Banik:2023rrz}.  The only case understood in great details is the two-fold Mellin-Barnes integral \cite{Zhdanov1998, Friot:2011ic}, which in our context calculates the 4-point conformal integral \cite{Loebbert:2019vcj}. 

The problem can be partially avoided by introducing  the bipartite Mellin–Barnes representation \cite{Alkalaev:2025fgn}, which splits a given  $n$-point conformal integral into two additive parts, each given by a $(n-2)(n-3)/2$-fold Mellin-Barnes integral.\footnote{Compared to the standard $n(n-3)/2$-fold Mellin-Barnes  representation, the bipartite representation allows one to evaluate $n-3$ integrals explicitly in terms of the Lauricella functions. In  particular, it follows that the novel methods \cite{Ananthanarayan:2020fhl, Banik:2023rrz} of handling multiple Mellin-Barnes integrals are not directly applicable to the bipartite representation due to the presence of a multivariate hypergeometric function in the integrand.} The clear pole structure of one part and known analytic continuation formulas make it computable, thereby  providing us with a subset of basis functions. In turn, the uncomputable part is suggested to be restored using the permutation invariance property \eqref{s_invariant}. This calculation scheme works well as long as the basis functions coming from the computable part contain all necessary master functions. In other words, a complete  set of basis functions can be generated from them by acting with the cyclic subgroup $\ZZ_n$. 

To examine the resulting expressions one imposes another constraint which naturally follows from the very definition of the conformal integral.  Namely, setting in \eqref{indef} one of propagator powers to zero, the $n$-point conformal integral reduces  to the $(n-1)$-point conformal integral. E.g. for $a_n = 0$:
\be
\label{in_reduction}
I_n^{a_1, ... ,a_{n-1},a_n}(x_1,...,x_{n-1},x_n)\Big|_{a_n=0} = 
I_{n-1}^{a_1, ... ,a_{n-1}}(x_1,...,x_{n-1})\,, 
\qquad \;
\sum_{i=1}^{n-1} a_i = D\,.
\ee 
It should be stressed out that this condition works for  any subset of  propagator powers $\{a_{i_1}, ..., a_{i_k}\} \subset \{a_1, ..., a_n\}$ which are set to zero. In other words, the reduction comprises a parametric tower of conditions which reduce a given $n$-point integral as a function of ${\bm x}$ and ${\bm a}$ down to the 3-point conformal integral explicitly calculated by the star-triangle relation (see \eqref{star-triangle} below). 

One can explicitly show that the lower-point conformal integrals ($n=4,5$) calculated by means of  the bipartite Mellin-Barnes representation do satisfy the reduction condition \eqref{in_reduction}, see  \cite{Alkalaev:2025fgn} for more details. In  the non-parametric limit ($\forall a_i\to 1$) these two integrals  are reduced to  polylogarithmic expressions obtained by different methods in the  earlier  literature.\footnote{It is important to note that the non-parametric pentagon conformal integral was calculated in \cite{Nandan:2013ip} in manifestly  $\ZZ_5$-invariant form. It is straightforward to compare this result with that one obtained by the reconstruction method (see Appendix {\bf B} in \cite{Alkalaev:2025fgn}). In the case of the box conformal integral the presence of $\ZZ_4$ has not been explicitly noted in the literature though the existing expression in terms of the Bloch-Wigner function in the appropriate analyticity domain makes this symmetry manifest too.} Thus, consideration of the non-parametric limit, at which the conformal integral expressions are simplified, may provide yet another non-trivial consistency check.      
 
Considering  $n=6$ we observe an obstacle for using the above calculation scheme directly.  The reason is that the computable part of the bipartite Mellin-Barnes representation does not contain a complete set of master functions. This can be seen by using available master functions in order to generate basis functions by cyclic permutations and check that in this case the reduction condition \eqref{in_reduction} is violated. A simple counting  based on the proposed enumeration  of master functions  shows that one master function is missing \cite{Alkalaev:2025fgn}. One can satisfy  the reduction condition by adding this function to those  master functions which came from the computable part of the conformal integral. Below \eqref{cord} such a counting is  formulated  in terms of $\ZZ_n$-orbits and their lengths. 

Thus, there is a series of exact non-trivial results which are reproduced by the reconstruction method. Any analytic expression for the $n$-point parametric conformal integrals for any $n$ can be verified  in at least two possible ways: (1) the reduction condition; (2) the non-parametric limit. In section \bref{sec:red} we show that the $n$-point reconstruction formula \eqref{conj_fin} conforms the reduction condition. Up to date, the non-parametric limit check is not available  beyond $n=4,5$.

In this paper, relying on  explicit calculations and observations  from \cite{Alkalaev:2025fgn}, we propose the alternative procedure of finding both the master and  basis functions which is conveniently described by  a diagrammatic representation closely related to the Baxter lattice \cite{Baxter:1978xr, Zamolodchikov:1980mb}. It is based on examining exact forms of  $n=4,5,6$ basis functions obtained by direct  calculation and highlighting hidden patterns which control series coefficients and arguments. The suggested parameterization of basis functions in terms of index triples \eqref{Rn} plays here a key role.

\subsection{Conformal polygon and basis triangles}
\label{sec:polygon}

\begin{figure}
\centering
\includegraphics[scale=0.25]{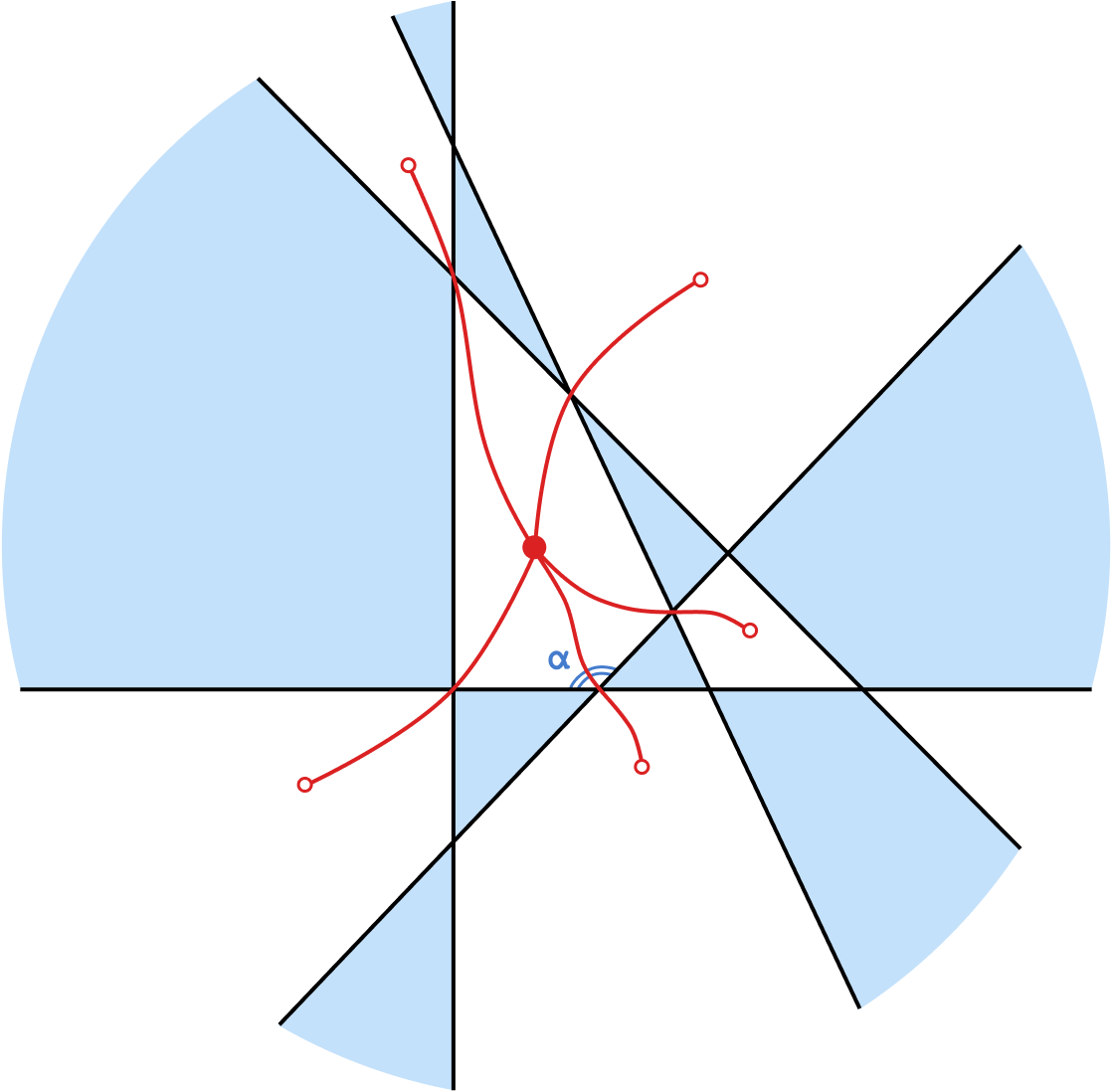}
\caption{The Baxter lattice and the conformal $n$-valent vertex drawn in the central polygon. The angle $\alpha$ is related to the propagator power \eqref{angle}. Note that compared to the conformal graph in fig. \bref{fig:vertex} the emphasis here is shifted to the polygon as such, and, in particular, to its angles which encode the conformality condition.} 
\label{fig:baxter}
\end{figure}

The conformal integral depicted as the  labelled  graph in fig. \bref{fig:vertex} has an alternative pictorial representation which for our purposes of constructing  basis functions proves to be more effective. Such a representation is directly related to the fishnet diagrams \cite{Zamolodchikov:1980mb}  and the loom construction  \cite{Kazakov:2022dbd}, where the key  element is the Baxter lattice \cite{Baxter:1978xr, Zamolodchikov:1980mb}.\footnote{For the further development of the loom construction see, e.g. \cite{Kazakov:2023nyu, Alfimov:2023vev}.} The lattice is  formed  by straight lines on the plane which can intersect at arbitrary slopes and  split the plane into domains with the "checkerboard" colouring, see fig. \bref{fig:baxter}. Then, the conformal integral is depicted as the $n$-valent tree graph centered  in one of polygons generated by the Baxter lattice: (1) a vertex lies inside a polygon, its edges go through the polygon vertices; (2) the angles of the polygon are related to the propagator powers as follows 
\be
\label{angle}
\alpha_i = \frac{2\pi}{D}\left(\frac{D}{2} -  a_i \right) \equiv \frac{2\pi}{D}\,a'_i \;,
\qquad
i \in \mathbb{N}_n = \{ 1,2,...,n\} \,;
\qquad
\sum_{i=1}^n \alpha_i =  (n-2)\pi\,,
\ee
where the total angle condition is just  the conformality constraint \eqref{confcond}. Note that for $a_i = 1$, $\forall i \in \mathbb{N}_n$, the polygon is regular because the conformality constraint \eqref{confcond} claims $n=D$ and all the angles become equal, $\alpha_i = \pi(n-2)/n$. The corresponding conformal integral is called non-parametric.

\begin{figure}
\centering
\includegraphics[scale=0.15]{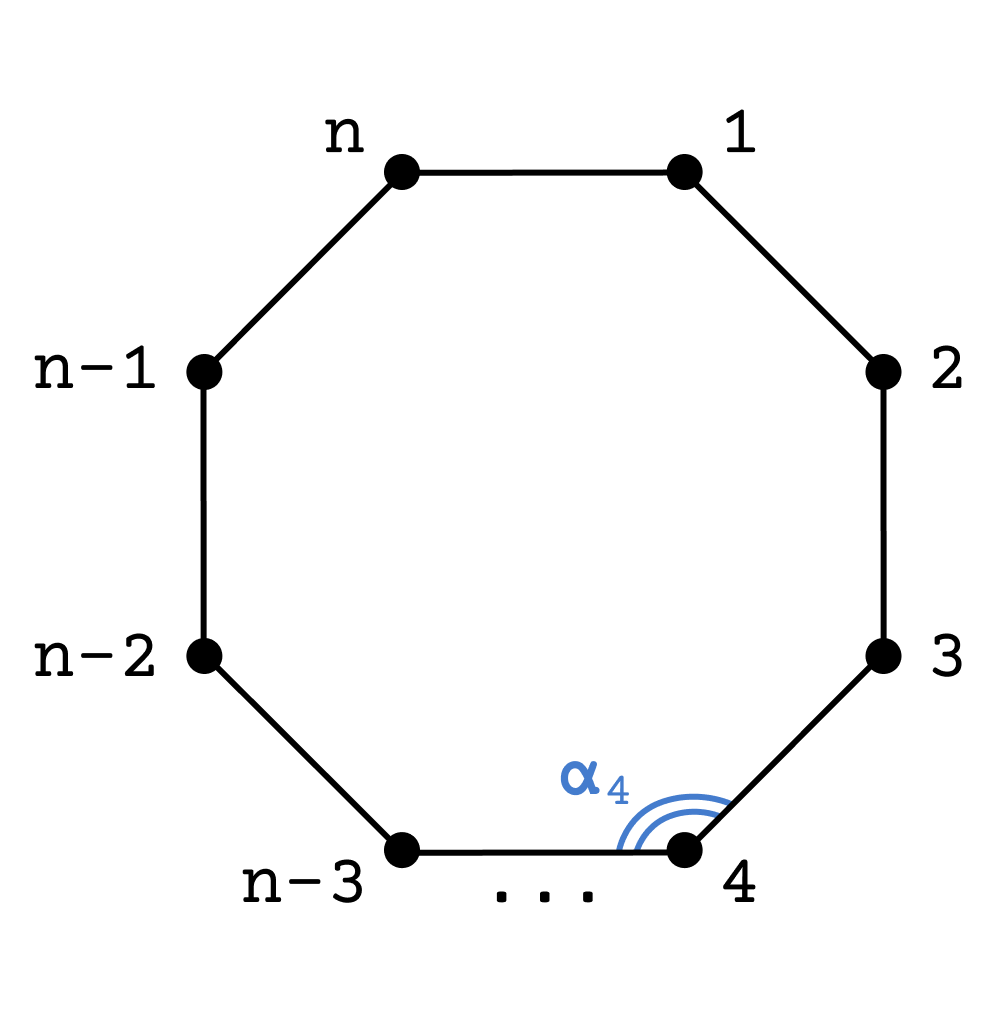} \hspace{20mm}
\includegraphics[scale=0.15]{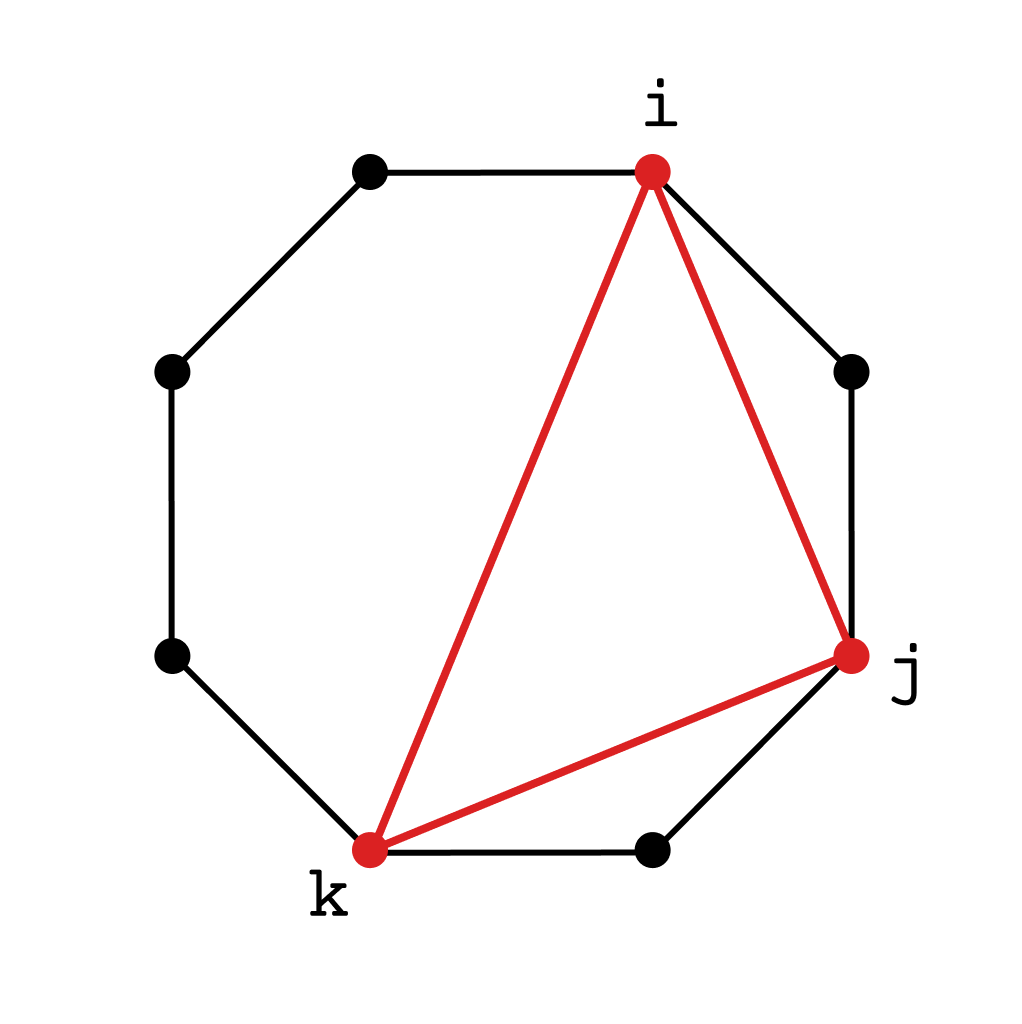}
\caption{{\it Left:} Irregular conformal polygon from the Baxter lattice in fig. \bref{fig:baxter} shown here for simplicity  as the regular $n$-gon. The sum of angles $(n-2)\pi$ reproduces the conformality constraint. {\it Right:} The triangle $\triangle_{n}^{\triple{ijk}}\subset P_n$ (shown in red) represents the ordered index triple $\triple{ijk} \in {\rm R}_n$.} 
\label{fig:polygon}
\end{figure}

The Baxter lattice brings to light the main geometric object of our study which we call a {\it conformal polygon} $P_n$. By definition, this is a polygon on the Baxter lattice which contains a given conformal $n$-valent vertex, fig. \bref{fig:baxter}.  It is convenient to draw this polygon separately. Moreover, without loss of generality, we can simplify our drawings and  depict  $P_n$ as a regular $n$-gon by assuming that all angles are arbitrary and less than $\pi$ but subject to the total angle constraint \eqref{angle}; also, we label vertices from 1 to $n$, see fig. \bref{fig:polygon}. Note that the  lengths of edges are inessential here,  so   the angles are the only geometric parameters which define the conformal polygon. In fact, one considers equivalence classes of homothetic convex polygons. It turns out that the conformal polygon encodes all geometric structures needed for constructing basis functions which calculate the conformal integral. Here, we introduce  some of them and elaborate a complete  geometric description latter in section \bref{sec:basis}. 

\begin{figure}[h]
\centering
\includegraphics[scale=0.2]{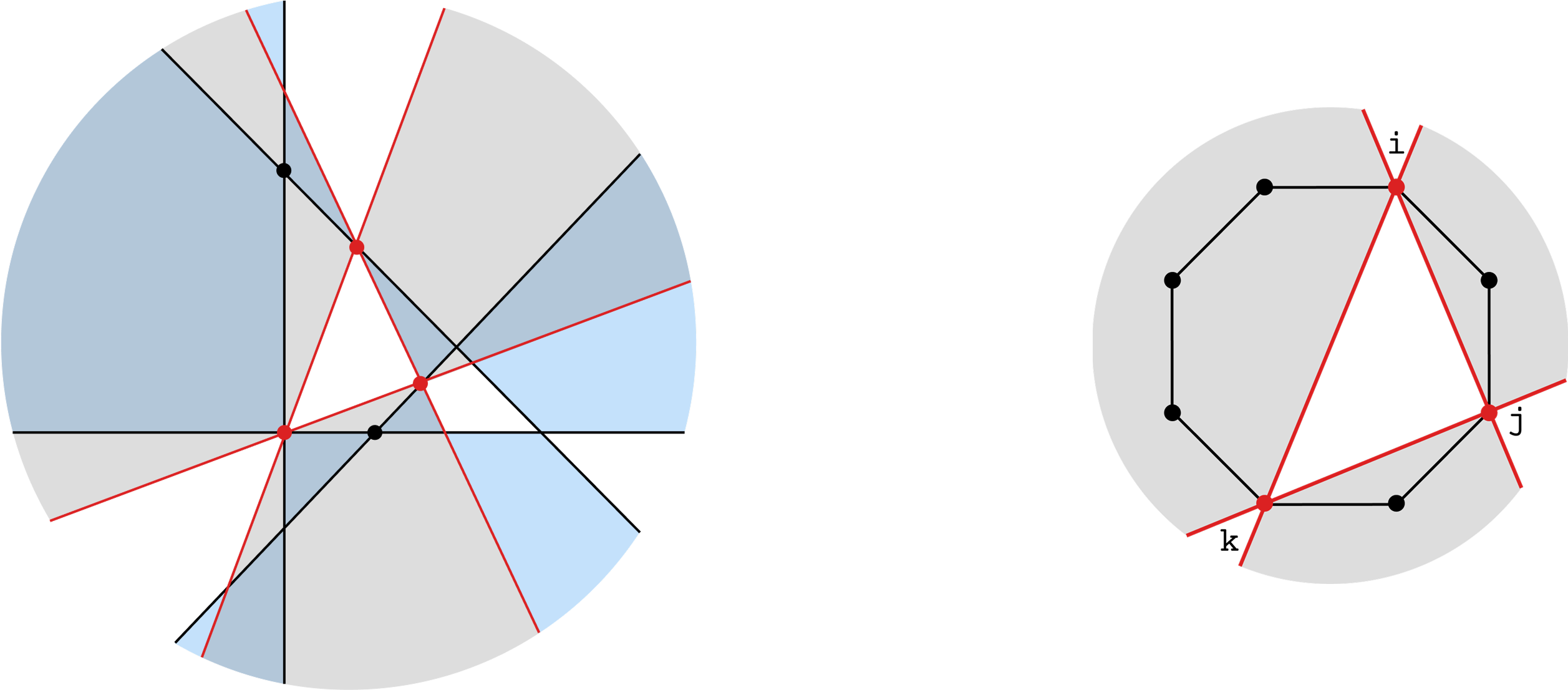} 
\caption{{\it Left:} The superposition of two Baxter lattices leads to the notion  of chambers (shown in grey). The first Baxter lattice is the same as in fig. \bref{fig:baxter}, the second Baxter lattice is given by three red lines. {\it Right:} Open chambers are shown in grey. Adding red boundaries yields closed chambers.} 
\label{fig:baxter_aux}
\end{figure}

An ordered index  triple $\triple{ijk} \in {\rm R}_n$ can be conveniently depicted as a {\it basis triangle} $\triangle_{n}^{\triple{ijk}}$ inscribed in the conformal polygon $P_n$ as shown in  fig. \bref{fig:polygon}. In turn, the triangle $\triangle_{n}^{\triple{ijk}} \subset P_n$ can also be viewed as a domain of another  Baxter lattice,  see fig. \bref{fig:baxter_aux}.
We will refer to open regions of this new Baxter lattice that contain vertices of the conformal polygon $P_n$ as {\it open chambers}, see fig. \bref{fig:baxter_aux}. A  chamber lying opposite to the $i$-th vertex of $\triangle_{n}^{\triple{ijk}}$ will be  denoted as $\cham_n^{( i )}$. It contains $L_n^{( i )}$ vertices of the conformal polygon  which are located   between $j$-th and $k$-th vertices. A number of the conformal  polygon vertices located in a particular chamber is constrained  by the balance relation: 
\be
\label{balance}
L_n^{( i )}+L_n^{( j )}+L_n^{( k )} + 3  = n\,,
\ee
with  $L_n^{( j )}$ and $L_n^{( k )}$ being numbers of vertices located  in $\cham_n^{( j )}$ and $\cham_n^{( k )}$, respectively.

Adding boundaries (red lines in fig. \bref{fig:baxter_aux}) to an open chamber, e.g. $\cham_n^{( i )}$, one obtains a {\it closed chamber} $\bar\cham_{n}^{(i)}$, which contains two additional vertices $k$ and $j$ lying on the basis triangle $\triangle_{n}^{\triple{ijk}}$.  Then, any two closed chambers intersect at just one point, which is one of the vertices of the basis triangle $\triangle_{n}^{\triple{ijk}}$, e.g.   $\bar\cham_n^{(i)} \cap \bar\cham_n^{(j)}=k$. Using the notion of chambers, the  basis triangle and the  conformal polygon are related  by the set equation 
$\triangle_{n}^{\triple{ijk}} = P_n  \big\backslash \Big( \cham_n^{( i )} \cup \cham_n^{( j )} \cup \cham_n^{( k )}\Big)$. In particular, the balance condition \eqref{balance} is a direct consequence.

Finally, it is worth noting that there is a theorem by Caratheodory \cite{Caratheodory:1919zz} which in one of its forms states  that a (convex) polygon on the two-plane is the union of triangles inscribed in this polygon in all possible ways. In the present context, the theorem is formulated as the following decomposition    
\be
P_n  = \dps\bigcup_{\triple{ijk} \in {\rm R}_n} \triangle_{n}^{\triple{ijk}}\,.
\ee
It is tempting to speculate that the Caratheodory theorem may underlie  a geometric realization of the conformal integral expanded over  basis functions \eqref{rep1}.

\section{Diagrammatic algorithm} 
\label{sec:basis}

In this section, we develop a diagrammatic algorithm, i.e. a series  of simple operations  that successively construct all three factors in \eqref{rep2}, which triple product results in a basis function $\BF_n^{\langle ijk\rangle}({\bm a}|\bm x)$. In this respect, the diagrammatic algorithm is naturally divided into three parts. Firstly, we define the triangle-factor and the leg-factor. Then, we propose an algorithm  which allows one to explicitly build a set of cross-ratios \eqref{cross-set}, which are realized as particular quadrilaterals and pentagons inscribed in the conformal polygon. Finally, we introduce a number of  numerical-geometric  characteristics of the constructed set of variables aimed to define a polygonal function. 

\subsection{Triangle-factors and leg-factors}
\label{sec:prefactor}

The triangle-factor is given  by 
\be
\label{star_n}
{\rm S}_{n}^{\triple{ijk}}({\bm a}) =  
\frac{\Gamma(-|\bm a_{i,j}|') \Gamma(-|\bm a_{j,k}|') \Gamma(-|\bm a_{k,i}|') }
{\Gamma(a_i) \Gamma(a_j) \Gamma(a_k)}
\equiv \Gamma
\Bigg[
\begin{array}{l l}
-|\bm a_{i,j}|' \,, -|\bm a_{j,k}|'\,, -|\bm a_{k,i}|' \\
\qquad a_i\,, \qquad a_{j}\,, \qquad a_k
\end{array}
\Bigg] 
\,,
\ee
and the leg-factor is conveniently given by using the open chambers as  
\be 
\label{V_n}
{\rm V}_n^{\triple{ijk}}({\bm a}|\bm x) = 
X_{ij}^{|\bm a_{i,j}|'} 
X_{jk}^{|\bm a_{j,k}|'}
X_{ki}^{|\bm a_{k,i}|'}\,  
\prod_{p \in \cham_{n}^{(i)} } X_{ip}^{-a_p} 
\prod_{s \in \cham_{n}^{(j)} } X_{js}^{-a_s}
\prod_{l \in \cham_{n}^{(k)} } X_{kl}^{-a_l} \,,
\ee
where we introduced the notation 
\be
\label{notations_params}
|\bm a_{i,j}| = \begin{cases}
				\dps
				\sum_{l=i}^j a_l \,, \quad i<j\,, \vspace{1mm }\\
				\dps
				\sum_{l=i}^n a_l + \sum_{l=1}^j a_l\,, \quad i>j\,,
				\end{cases} 
\quad
|\bm a_{i,j}|' = \hd - |\bm a_{i,j}|\,. 
\ee
One can show that under conformal transformations the leg-factor \eqref{V_n} transforms as
\be
\label{V_transform}
{\rm V}_n^{\triple{ijk}}({\bm a}|\bm x') = \prod_{l=1}^{n} \Omega^{-a_l}(x_l) \,{\rm V}_n^{\triple{ijk}}({\bm a}|\bm x)\,,
\qquad
\text{cf. \eqref{in_transform}}.
\ee

Having in mind the geometrical interpretation  of the propagator powers $a_i$ in terms of  the conformal polygon angles $\alpha_i$  \eqref{angle} one can rewrite \eqref{notations_params} as combinations  of angles  
\be
|\bm a_{i,j}| =\frac{D}{2\pi}\Big(\pi + \pi\, L_n^{( k )}- \cA_n^{( k )} \Big) \,,
\qquad
|{\bm a}_{i,j}|' = \frac{D}{2\pi}\Big(-\pi\, L_n^{( k )}+ \cA_n^{( k )} \Big)\,, 
\ee
where $L_n^{( k )}$ is the number of vertices in the open chamber $\cham_{n}^{(k)}$ and $\cA_n^{( k )}$ is the total angle at vertices in the closed chamber $\bar \cham_{n}^{(k)}$. Similar to the balance relation \eqref{balance} the total angles satisfy their own constraint: $\cA_n^{( i )}+\cA_n^{( j )} +\cA_n^{( k )}  = (n-2)\pi - \left(\alpha_i + \alpha_j +\alpha_k\right)$.

At $n=3$, there exists only one basis triangle $\triangle_3^{\triple{123}}$ which at the same time coincides with the conformal polygon $P_3$. Since there are no cross-ratios for three points and the polygonal function in this case equals 1, the product of \eqref{star_n} and \eqref{V_n} reproduces the star-triangle relation \cite{Symanzik:1972wj}:\footnote{The star-triangle relation, also known as the uniqueness relation, arises in various contexts, see e.g. \cite{Vasiliev:1981dg, Kazakov:1983dyk, Gorishnii:1984te, Isaev:2003tk, Chicherin:2012yn, Derkachov:2022ytx, Derkachov:2023xqq}.}
\be
\label{star-triangle}
I_3^{\bm a}(\bm x) = 
\Gamma
\left[
\begin{array}{l l}
a'_1, a'_2, a'_3 \\
a_1, a_2, a_3
\end{array}
\right] \, X_{12}^{-a'_3} X_{13}^{-a'_2} X_{23}^{-a'_1}\,,
\ee 
where we used the conformality constraint \eqref{confcond}.

\subsection{Conformal  diagrammatics}
\label{sec:UW}

Let us introduce two types of functions invariant against conformal transformations of $x_i \in \RR^D$, $i\in \mathbb{N}_n \equiv  \{1,2,...,n\}$, $n\geqslant4$, 
\be
\label{U_W}
U[i_1, i_2, i_3, i_4] := \frac{X_{i_1,i_2} X_{i_3,i_4}}{X_{i_1,i_3} X_{i_2, i_4}}\,, 
\qquad
W[j_1, j_2, j_3, j_4, j_5] := \frac{X_{j_1,j_2} X_{j_2,j_3} X_{j_4, j_5} }{ X_{j_1,j_3} X_{j_2, j_4} X_{j_2, j_5} }\,,
\ee
which we call quadratic and cubic cross-ratios, respectively. Here,  sets of indices  $\{i_1, i_2, i_3, i_4\}\subset \mathbb{N}_n$ and  $\{j_1, j_2, j_3, j_4, j_5\} \subset \mathbb{N}_n$. These functions are related to each other by means of two types of relations. First, the cubic cross-ratios reduce to quadratic ones if some particular pairs of indices coincide, namely, 
\be
\label{U_W_reduction}
\ba{c}
W[i_1, i_2, \underline{i_3}, \underline{i_3}, i_5] = U[i_1, i_2, \underline{i_3}, i_5] \,,
\qquad
W[\underline{i_1}, i_2, i_3, \underline{i_1}, i_5] = U[i_5, \underline{i_1},  i_2, i_3] \,,
\\
W[\underline{i_1}, i_2, i_3, i_4, \underline{i_1}] = U[i_4, \underline{i_1}, i_2, i_3] \,,
\qquad
W[i_1, i_2, \underline{i_3}, i_4, \underline{i_3}] = U[i_1, i_2, \underline{i_3}, i_4] \,.
\\
\ea
\ee
Second, any cubic cross-ratio can be represented as a product of two quadratic cross-ratios:
\be
\label{WUU}
W[j_1, j_2, j_3, j_4, j_5] = U[j_4,j_5, j_1, j_2] \, U[j_2, j_3, j_4, j_1] 
= U[j_3, j_5, j_1, j_2] \, U[j_2, j_3, j_4, j_5]\,.
\ee
It is worth noting that although a set of independent cross-ratios can consist of quadratic cross-ratios  only (see e.g. \cite{Osborn}) it may happen that introducing cubic and even higher-order cross-ratios can be a convenient {\it ad hoc} choice. This is the case within  the diagrammatic algorithm discussed in the next section.\footnote{Also, higher-order cross-ratios turn out to be convenient when analysing the OPE regimes of higher-point correlation functions  in CFT$_D$ \cite{Rosenhaus:2018zqn, Buric:2021kgy}. } 

\begin{figure}
\centering
\includegraphics[scale=0.15]{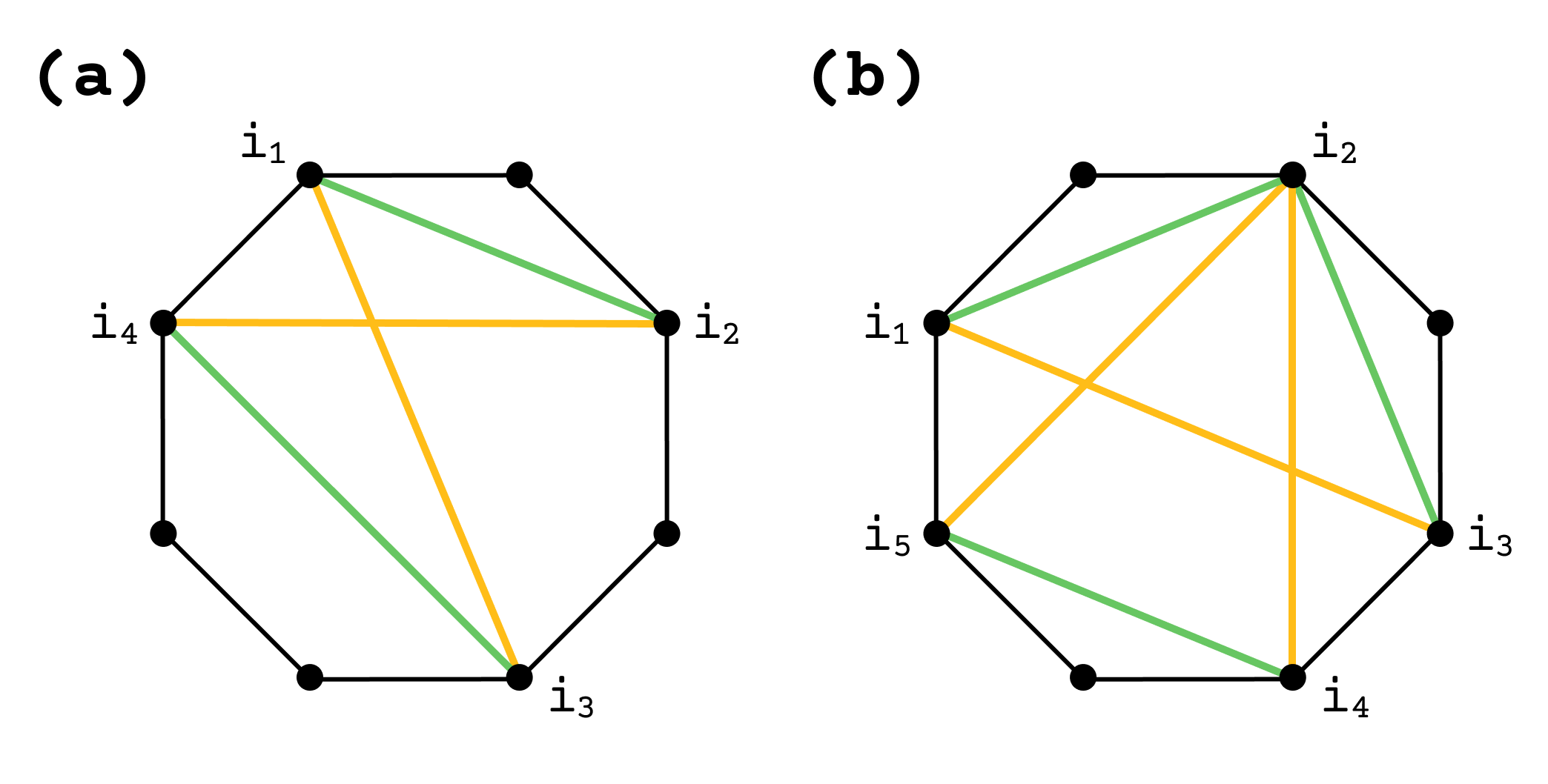}
\caption{The quadratic {\bf (a)} and cubic {\bf (b)} cross-ratio diagrams. The numerators and denominators are colored with green and  orange, respectively.} 
\label{fig:cross}
\end{figure}

The symmetric group $\cS_n$ acts on functions $U$ and $W$ \eqref{U_W} by permuting indices:  $\pi \circ \{ i_1, i_2, ..., i_n\} = \{\pi(i_1),\pi(i_2),..., \pi(i_n)\}$. In what follows, we will be interested in the cyclic subgroup $\ZZ_n \subset \cS_n$ since its action on  cross-ratios is part of the reconstruction formula \eqref{conj_fin}.  Also, there are other two  symmetric groups $\cS_4$ and $\cS_5$  which act on $U$ and $W$  by permuting  labels of indices:  $\tilde \pi \circ \{ i_1, i_2, ... \} = \{i_{\tilde\pi(1)},i_{\tilde\pi(2)},...\}$. Equivalently, they act on a given subset of either four or five elements  leaving other elements  in  $\mathbb{N}_n$ intact. This allows one to describe all possible cross-ratios which can be built from four or five points in $\RR^D$, see Appendix  \bref{app:orbit}.

The cross-ratios \eqref{U_W} can be represented as quadrilaterals and pentagons inscribed in the conformal polygon $P_n$  between marked vertices  $\{i_1,..., i_4\}$ and $\{j_1,..., j_5\}$ as shown in  fig. \bref{fig:cross}.\footnote{Similar diagrams for cross-ratios can be found in \cite{Buric:2021kgy}.} Then, the respective numerators and denominators are colored chords connecting vertices. The action of $\cS_4$ and $\cS_5$ on given sets of indices $\{i_1,..., i_4\}$ and $\{j_1,..., j_5\}$ changes the ways in which they are connected by (colored) chords. For the future convenience, we denote these {\it cross-ratio diagrams} as
\be
\label{box_penta}
\Box_{n|\alpha}^{i_1i_2i_3i_4} 
\qquad\text{and}\qquad
\pentago_{n| \beta}^{i_1i_2i_3i_4i_5} \;,
\ee
where $\alpha$ and $\beta$ specify a particular way of colouring.


\subsubsection{Cross-ratio sets}

One chooses an index triple $\triple{ijk} \in {\rm R}_n$ and considers  the associated basis triangle inscribed in the conformal polygon, $\triangle_n^{\triple{ijk}} \subset P_n$, along with the corresponding chambers $\cham_n^{( i )},\cham_n^{( j )}, \cham_n^{( k )}$, see fig. \bref{fig:baxter_aux}. The  cross-ratio set $\bm \rY_n^{\triple{ijk}}$ is split into two parts
\be
\bm \rY_n^{\triple{ijk}} = \bm \rU_n^{\triple{ijk}} \cup  \bm \rW_n^{\triple{ijk}}\,,
\ee
of quadratic and cubic cross-ratios \eqref{U_W}. Below we build the two subsets separately.

\begin{figure}
\centering
\hspace{-5mm}\includegraphics[scale=0.15]{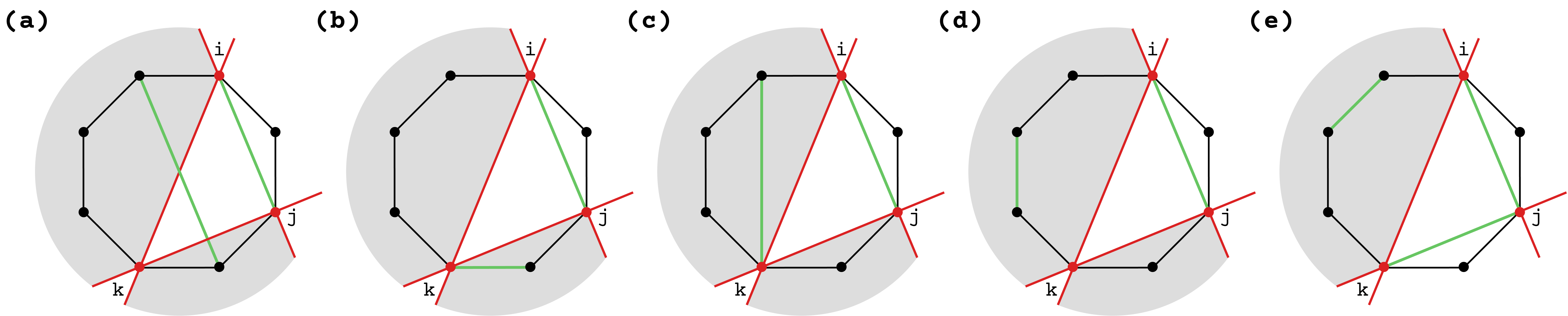}
\caption{The diagrammatic algorithm for  cross-ratios (numerators). From left to right: {\bf (a)-(c)} admissible numerators  in quadratic cross-ratios; {\bf (d)} forbidden numerators in quadratic  cross-ratios (two vertices in one open chamber); {\bf (e)} admissible numerators in cubic cross-ratios.  } 
\label{fig:quadra_rules1}
\end{figure}

\begin{figure}
\centering
\includegraphics[scale=0.15]{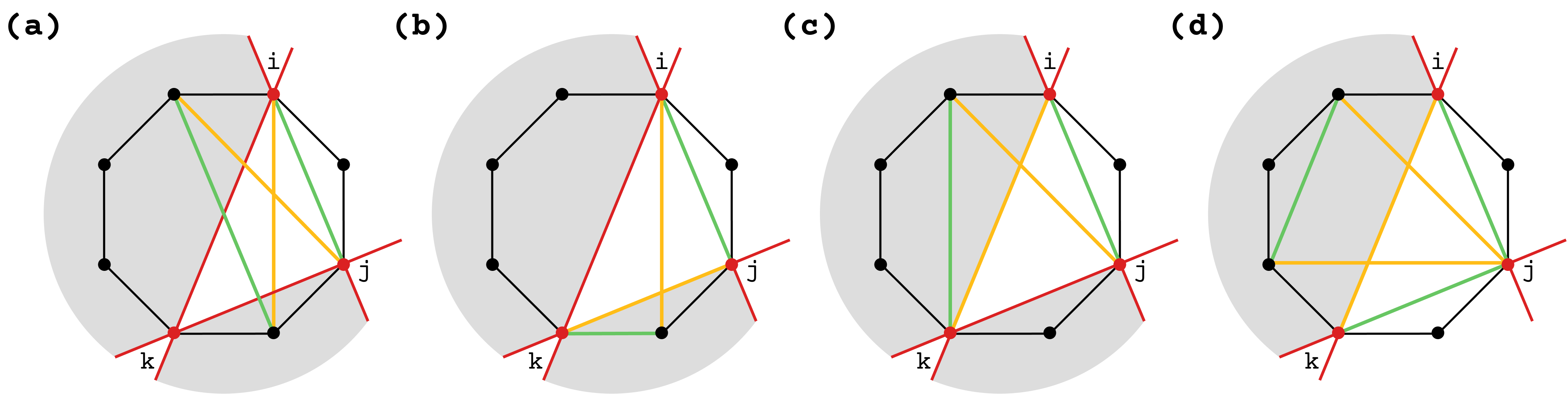}
\caption{The diagrammatic algorithm for  cross-ratios  (denominators). From left to right: {\bf (a)-(c)} admissible quadratic cross-ratios; {\bf (d)} admissible cubic cross-ratios. Note that all the resulting cross-ratios are of the types shown in fig. \bref{fig:cross}. } 
\label{fig:quadra_rules2}
\end{figure}

\paragraph{Quadratic cross-ratios.} Choose any two vertices of the basis triangle $\triangle_n^{\triple{ijk}}$, e.g. $i$ and $j$, and connect them with a green chord which means that $X_{ij}$ will be in the  numerator, see \eqref{U_W}.  Then, consider two closed chambers $\bar \cham_n^{(i)} \cup\, \bar \cham_n^{(j)}$ and connect  any two vertices $p \neq i,j$ and $s\neq i,j$ from adjacent chambers with a green chord. It is not allowed to connect vertices from one open chamber $\cham_n^{(i)}$ or $\cham_n^{(j)}$. The resulting $X_{ps}$ is also in the numerator, see fig. \bref{fig:quadra_rules1} {\bf (a)}-{\bf (d)}. The denominator is build by connecting the vertices by orange chords diagonally,  see fig. \bref{fig:quadra_rules2} {\bf (a)}-{\bf(c)}. In this way we obtain a cross-ratio $\rY_l^{\triple{ijk}}$ for some $l$ in the set \eqref{cross-set}. Now, repeat these steps  for other pairs of vertices, $j$ and $k$, $k$ and $i$. This will give a complete set of quadratic cross-ratios $\bm \rU_n^{\triple{ijk}}$. According to the just described algorithm this set is naturally split into two parts as 
\be
\bm \rU_n^{\triple{ijk}} = {}^{_2}\hspace{-0.5mm}\bm \rU_n^{\triple{ijk}}\, \cup \, {}^{_1}\hspace{-0.5mm}\bm \rU_n^{\triple{ijk}}\,,
\ee
where indices 2 and 1 refer to cross-ratios defined by a green chord connecting two vertices either in two open chambers (index 2) or otherwise (index 1). Examples of the cross-ratios from ${}^{_2}\hspace{-0.5mm}\bm \rU_n^{\triple{ijk}}$ and ${}^{_1}\hspace{-0.5mm}\bm \rU_n^{\triple{ijk}}$ are shown in fig. \bref{fig:quadra_rules2} {\bf (a)} and  {\bf (b)},{\bf(c)}, respectively.   

The total number of quadratic cross-ratios equals 
\be
\label{U_cardinality}
\ba{c}
|\bm \rU_n^{\triple{ijk}}| = L_n^{(j)} L_n^{(i)} + L_n^{(j)}+L_n^{(i)}
\\
\hspace{15mm} + L_n^{(k)} L_n^{(j)} + L_n^{(k)}+L_n^{(j)}\,
\\
\hspace{17mm} + L_n^{(i)} L_n^{(k)} + L_n^{(i)}+L_n^{(k)}\,,
\ea
\ee  
where $L_n^{(j)}$ is the  number of vertices located  in $\cham_n^{( j )}$, etc. 


\paragraph{Cubic cross-ratios.} Connect vertices $i,j$ and $j,k$ of the basis triangle $\triangle_n^{\triple{ijk}}$ by green chords. Then, choose any two vertices $p$ and $s$ in the open chamber $\cham_n^{(j)}$ and connect them by a green chord.\footnote{Obviously,  the chamber should contain  at least two vertices. Open chambers with less than two vertices do not allow for constructing cubic-ratios, see the example in section \bref{sec:pentagon}. }  In this way one obtains a numerator of the cubic cross-ratio, i.e. $X_{ij} X_{jk} X_{ps}$, see fig. \bref{fig:quadra_rules1} {\bf (e)}.  To build a denominator one connects the five vertices $i,j,k,p,s$ as shown in fig. \bref{fig:quadra_rules2} {\bf (d)}. Repeat these steps for other pairs of vertices: $j,k$ and $k,i$;  $k,i$ and $i,j$. This will give a complete set of cubic cross-ratios $\bm \rW_n^{\triple{ijk}}$.

The total number of cubic cross-ratios equals 
\be
\label{W_cardinality}
|\bm \rW_n^{\triple{ijk}}| = \binom{ L_n^{(i)}}{2}+ \binom{L_n^{(j)}}{2} +\binom{L_n^{(k)}}{2} \,.
\ee
Summing up  \eqref{U_cardinality} and \eqref{W_cardinality} and taking into account the balance relation \eqref{balance} we conclude that the total number of cross-ratios in $\bm \rY_n^{\triple{ijk}}$  is given by   
\be
|\bm \rY_n^{\triple{ijk}}| = |\bm \rU_n^{\triple{ijk}}| + |\bm \rW_n^{\triple{ijk}}| = \frac{n(n-3)}{2}\,,
\qquad \forall \triple{ijk} \in  {\rm R}_n\,.
\ee  
It is important that this combinatorial counting is independent of choosing a particular index triple.

\subsubsection{Properties}

Here, we  discuss a series of characteristic  properties of  the cross-ratio set $\bm \rY_n^{\triple{ijk}}$.

\begin{itemize}

\item Any cross-ratio from $\bm \rY_n^{\triple{ijk}}$ \eqref{cross-set} can be depicted by one of the cross-ratio diagrams  \eqref{box_penta}: either  $\Box_{n|\alpha}^{s_1...s_4}\subset P_n$ or  $\pentago_{n|\beta}^{p_1...p_5}\subset P_n$, where possible combinations of vertices $\{s_1,...,s_4\} \subset \mathbb{N}_n$, $\{p_1,..., p_5\} \subset \mathbb{N}_n$ along with  the way of colouring chords $\alpha$, $\beta$ follow from the diagrammatic algorithm. 

\item Extending a given conformal polygon $P_n$ by adding one more vertex, $P_n \to P_{n+1}$, results in increasing the size of one of its chambers, e.g. $L_n^{(j)} \to L_{n+1}^{(j)} = L_n^{(j)} + 1$. Recalling the cardinalities  \eqref{U_cardinality} and  \eqref{W_cardinality} one can directly verify that
\be
|\bm \rY_{n+1}^{\triple{ijk}}| = |\bm \rY_n^{\triple{ijk}}| + n-1 = \frac{(n+1)(n-2)}{2}\,, \qquad n = 4,5\,...\,,
\ee
whence we conclude that the sets of variables form the sequence of embeddings:
\be
\label{Y_subset}
\bm \rY_4^{\triple{ijk}} \subset \ldots \subset \bm \rY_{n-1}^{\triple{ijk}} \subset \bm \rY_n^{\triple{ijk}} \subset \ldots  \,.
\ee
The cross-ratios from  $\bm \rY_4^{\triple{ijk}}$ become the first two elements of the set $\bm \rY_5^{\triple{ijk}}$, which in its turn constitutes the first five elements of $\bm \rY_6^{\triple{ijk}}$, etc (see examples in section \bref{sec:examples}).

\item Among all {\it quadratic} cross-ratios only the elements from the subset ${}^{_1}\hspace{-0.5mm}\bm \rU_n^{\triple{ijk}} \subset \bm \rU_n^{\triple{ijk}}$  can be expressed as {\it cubic} cross-ratios with coinciding points. Graphically, the presence of two matching points in  \eqref{U_W_reduction}  means that the corresponding  edges, which always have the opposite color, cancel each other.  

\begin{figure}
\centering
\includegraphics[scale=0.15]{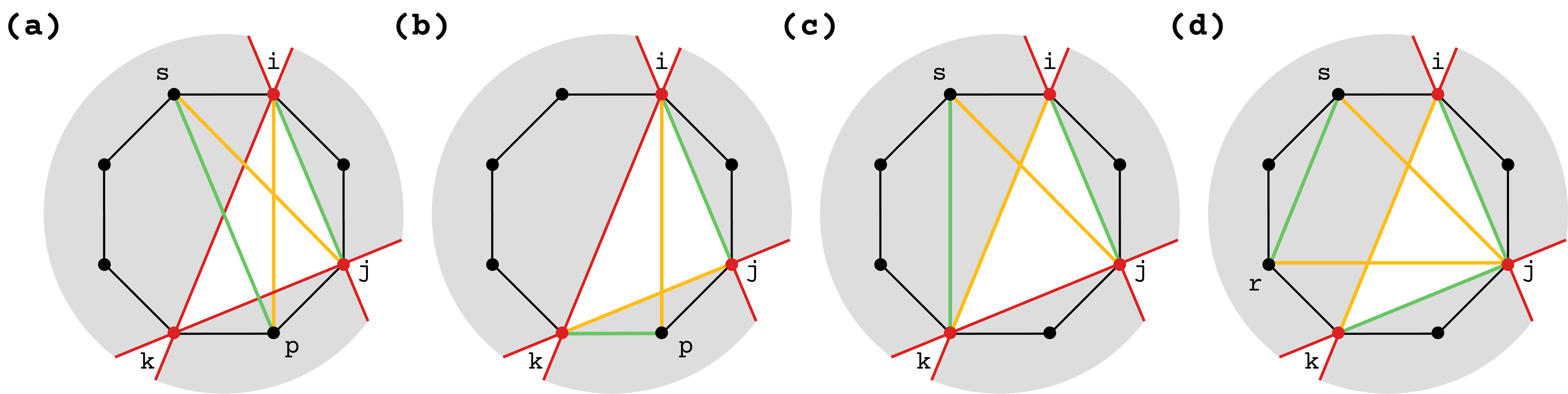}
\caption{Three types of cross-ratios from $\bm \rY_n^{\triple{ijk}}$ which belong to ${}^{_2}\hspace{-0.5mm}\bm \rU_{n}^{\triple{ijk}}$ {\bf (a)}, ${}^{_1}\hspace{-0.5mm}\bm \rU_{n}^{\triple{ijk}}$ {\bf (b)}, {\bf(c)}, and $\bm \rW_n^{\triple{ijk}}$ {\bf (d)}.  } 
\label{fig:funct}
\end{figure}

\item Let us validate that for a given triple $\triple{ijk} \in {\rm R}_n$ the diagrammatic algorithm yields a set of $\frac{n(n-3)}{2}$ functionally independent cross-ratios constituting the set   $\bm \rY_n^{\triple{ijk}}$. If this were not the case, then  there would be an element expressed through other elements of the same set,
\be
\label{funct_depend}
Y_q^{\triple{ijk}} = \prod_{r=1}^{\frac{n(n-3)}{2}} \left(Y_{l_r}^{\triple{ijk}}\right)^{\alpha_r}  
\quad 
\text{for some $\alpha_r \in \ZZ$}\,.
\ee  
Consider the diagrams on fig. \bref{fig:funct} as particular representatives of $\bm  \rU_{n}^{\triple{ijk}}$ and $\bm  \rW_n^{\triple{ijk}}$. To begin with, consider the diagram {\bf (a)} $\in {}^{_2}\hspace{-0.5mm}\bm  \rU_{n}^{\triple{ijk}}$ which depicts the following cross-ratio
\be
\label{U_depend}
U[i,j, p,s]  = \frac{X_{ij}X_{ps}}{X_{ip}X_{js}}\;.
\ee
If $U[i,j, p,s]$ is represented as \eqref{funct_depend} then the combination of cross-ratios on the right-hand side of \eqref{funct_depend} is identically reduced to that of \eqref{U_depend}. In particular, the distance $X_{ps}$ can arise in two possible ways (regardless  of particular $\alpha_r$)
\be
U[i,j, p,s]  = \frac{X_{ps}}{...} \times (...)  \equiv \frac{...}{\dps\frac{1}{X_{ps}}}\times(...)\;.
\ee        
Both $X_{ps}$ in the numerator and $1/X_{ps}$ in the denominator correspond to cross-ratios from $\bm \rY_n^{\triple{ijk}}$, where $X_{ps}$ is realized either by green or orange chords. If $X_{ps}$ is in the numerator (i.e. it is a green chord) then by construction the only element of $\bm \rY_n^{\triple{ijk}}$ with this property is $U[i,j, p,s]$ itself that contradicts our assumption that this cross-ratio is expressed in terms of others. The second possibility when $X_{ps}$ is in the denominator of some cross-ratio (i.e. it is an orange chord) also cannot be realized  because by construction any orange chord  always has at least one of its ends in one of vertices of the basis triangle $\triangle_n^{\triple{ijk}}$. Thus, a cross-ratio from the subset ${}^{_2}\hspace{-0.5mm}\bm \rU_{n}^{\triple{ijk}}$ cannot be expressed in terms of other elements of the set $\bm \rY_n^{\triple{ijk}}$.  

This analysis can be directly extended to  elements from ${}^{_1}\hspace{-0.5mm}\bm \rU_{n}^{\triple{ijk}}$  (diagrams of the type {\bf (b)} and {\bf (c)} on fig. \bref{fig:funct}) and $\bm \rW_n^{\triple{ijk}}$ (diagrams of the type {\bf (d)} on fig. \bref{fig:funct}).\footnote{It is worth noting here that the relation \eqref{WUU}  represents a cubic cross-ratio from $\bm  \rW_n^{\triple{ijk}}$ as a product of two quadratic cross-ratios one of which does not belong to $\bm \rY_n^{\triple{ijk}}$.}  In summary, $\bm \rY_n^{\triple{ijk}}$ is indeed a set of independent cross-ratios, both quadratic and cubic, as guaranteed by construction.

\item Since for given $n$ points $\bm x$ one can build exactly $\frac{n(n-3)}{2}$ independent cross-ratios (see footnote \bref{f1}) it follows that the cross-ratios from two sets $\bm \rY_n^{\triple{ijk}}$ and $\bm \rY_n^{\triple{i'j'k'}}$ are generally expressed in terms of each other as rational functions.  

\end{itemize}

\subsection{Transianic indices}
\label{sec:indexes}

A few geometric shapes can be associated with the  Baxter lattice: conformal polygons, basis triangles, and cross-ratio diagrams. It turns out that the polygonal functions in \eqref{rep2} can be  completely defined by   relative positions of these planar figures which can be conveniently characterized  by a set of numbers.  

To this end, one describes all these figures uniformly as particular planar graphs, $G = (V, E)$, where $V$ is  a set of vertices,  $E$ is a set of edges $(r,s)$ connecting two vertices $r,s \in V$. Then, there are three types of graphs related by the diagrammatic algorithm for some $\triple{ijk} \in {\rm R}_n$:\footnote{To simplify notation, we omit labels $n$, $\triple{ijk}$, and $m = 1,...,n(n-3)/2$ in the respective graphs.}   
\be
\label{graph_polygon}
\ba{l}
\dps
G_{{\rm polygon}} = \big(V_{{\rm polygon}}, E_{{\rm polygon}}\big) = \big(\{1,2,...,n \}\,, \{(1,2),(2,3),...,(n-1,n),(n,1) \}\big) \,,
\vspace{2mm}
\\
\dps
 G_{{\rm triangle}} =  \big(V_{{\rm triangle}}, E_{{\rm triangle}}\big) =  \big( \{i,j,k\}\,, \{ (i,j),(j,k),(i,k) \} \big)\,,
\vspace{2mm}
\\
\dps
G_{{\rm cross-ratio}}  = \big( V_{{\rm cross-ratio}}\,, \{E_{{\rm cross-ratio}}^{^{({\rm \textcolor{green}{green}})}}\,, E_{{\rm cross-ratio}}^{^{(\textcolor{orange}{\rm orange})}} \} \big)\,.
\ea
\ee
The vertices and edges in $G_{{\rm cross-ratio}}$ are read off from the cross-ratio diagrams  \eqref{box_penta},   
\be
\label{vertices_cross}
V_{{\rm cross-ratio}} \;\;=\;\; 
\begin{cases}
\{q_1,q_2,q_3,q_4 \} \quad \text{for} \quad \Box_{n|\alpha}^{q_1q_2q_3q_4} \,, \\
\{p_1,p_2,p_3,p_4, p_5 \} \quad \text{for} \quad \pentago_{n|\beta}^{p_1p_2p_3p_4p_5} \,,
\end{cases}
\ee
and
\be
\label{edges_cross}
\begin{split}
E_{{\rm cross-ratio}}^{^{({\rm \textcolor{green}{green}})}} &\;\;=\;\; 
\begin{cases}
\{(q_1,q_2),(q_3,q_4) \} \quad \text{for} \quad \Box_{n|\alpha}^{q_1q_2q_3q_4} \,, \\
\{(p_1,p_2),(p_2,p_3),(p_4,p_5) \} \quad \text{for} \quad \pentago_{n|\beta}^{p_1p_2p_3p_4p_5} \,,
\end{cases} \\
E_{{\rm cross-ratio}}^{^{({\rm \textcolor{orange}{orange}})}} & \;\;=\;\; 
\begin{cases}
\{(q_1,q_3),(q_2,q_4) \} \quad \text{for} \quad \Box_{n|\alpha}^{q_1q_2q_3q_4} \,, \\
\{(p_1,p_3),(p_2,p_4),(p_2,p_5) \} \quad \text{for} \quad \pentago_{n|\beta}^{p_1p_2p_3p_4p_5} \,,
\end{cases}
\end{split}
\ee
where the edge colouring  distinguishes between numerators (green chords) and denominators (orange chords) of the respective cross-ratio function \eqref{U_W}.

There is always a number of set relations which describe relative positions of $G_{{\rm polygon}}$, $G_{{\rm triangle}}$, $G_{{\rm cross-ratio}}$. E.g. one  always  has  $V_{{\rm triangle}}\,, V_{{\rm cross-ratio}} \subset V_{{\rm polygon}}$ and $V_{{\rm triangle}} \cap V_{{\rm cross-ratio}} \neq \varnothing$, while $E_{{\rm triangle}} \cap E_{{\rm cross-ratio}}^{^{(\textcolor{orange}{\rm orange})}} \neq \varnothing$ depends on particular  $G_{{\rm cross-ratio}}$, see fig. \bref{fig:funct}, where all three graphs are represented.  However, there is a more precise  characterization that can help formalize their relative positions: this is a number of (non-)matching vertices or edges. We call any such number {\it a transianic index}.


\paragraph{Definition.} {\it For every $l \in V_{\rm polygon}$, $l\neq i,j,k$ and for every $(r,s) \in  E_{\rm triangle}$ there are two transianic indices:\footnote{To the best of our knowledge, such indices related to the edge colouring were not previously discussed in the literature. Nevertheless, the second transianic index $b_{rs}$ seems to be a generalization of the adjacency matrix in graph theory.}
\be
\label{transianic}
\ba{l}
b_l = 
\begin{cases}
1, \quad \text{if} \quad l \in V_{{\rm cross-ratio}}\,, \\
0, \quad \text{otherwise} \,,
\end{cases}
\vspace{2mm}
\\
\dps
b_{rs} = 
\begin{cases}
1, \quad \text{if} \quad (r,s) \in E_{{\rm cross-ratio}}^{^{({\rm \textcolor{green}{green}})}}\,, \\
-1, \quad \text{if} \quad (r,s) \in E_{{\rm cross-ratio}}^{^{({\rm \textcolor{orange}{orange}})}}\,, \\
0, \quad \text{otherwise} \,.
\end{cases}
\ea
\ee}

Thus, a sequence of $n$ numbers equal to $0, \pm 1$ is assigned to any element of $\bm \rY_n^{\triple{ijk}}$. These sequences can be organised into {\it a transianic  matrix} $\cB\big(\bm \rY_n^{\triple{ijk}}\big)$, which  has $n(n-3)/2$ columns corresponding to elements of $\bm \rY_n^{\triple{ijk}}$ and $n$ rows corresponding  to their transianic indices, e.g. 

\be
\label{tr_mat}
\cB\big(\bm \rY_n^{\triple{ijk}}\big)\;  =\quad  
\begin{tabular}{|c||c|c|c|c|c|}
\hline
 & $\rY^{\triple{ijk} }_1$ & $\rY^{\triple{ijk} }_2$ & $\cdots\cdots$ & $\cdots\cdots$ & $\rY^{\triple{ijk} }_{n(n-3)/2}$  \\
\hline
\hline
$\vdots$ & $1$ & $1$ & $0$ & $0$ & $1$ \\
\hline
\hline
$b_l$ & $0$ & $0$ & $1$ & $1$ & $1$ \\
\hline
\hline
$\vdots$ & $0$ & $1$ & $0$ & $1$ & $1$ \\
\hline
\hline
$b_{ij}$ & $-1$ & $-1$ & $1$ & $0$ & $0$ \\
\hline
\hline
$b_{jk}$ & $1$ & $-1$ & $-1$ & $0$ & $-1$ \\
\hline
\hline
$b_{ki}$ & $-1$ & $-1$ & $-1$ & $0$ & $0$ \\
\hline
\end{tabular}
\ee
\hspace{2mm} 

\begin{figure}
\centering
\includegraphics[scale=0.15]{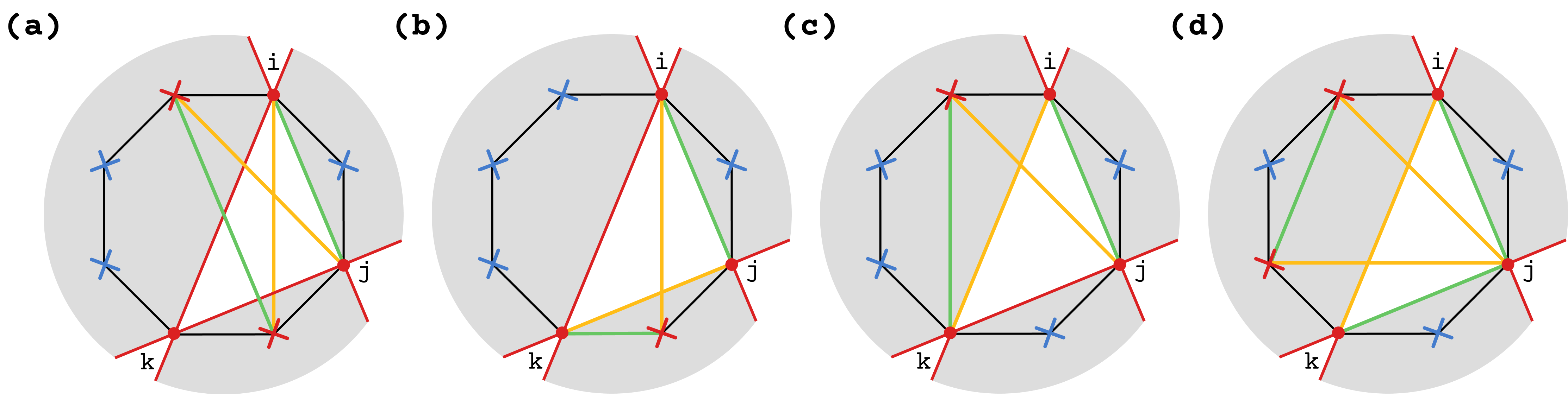}
\caption{Examples of calculating the first transianic  index. Here, the red crosses give $b_l=1$, the blue crosses give $b_l=0$.} 
\label{fig:index2}
\end{figure}

Let us  showcase how the transianic indices are calculated. Consider all open chambers associated to a given triangle $\triangle_n^{\triple{ijk}}$: $\cham_n^{(i)} \cup \cham_n^{(j)} \cup  \cham_n^{(k)}$,  see fig. \bref{fig:baxter_aux}. In these chambers one sequently go through all vertices $l\in V_{\rm polygon}$ by checking whether they belong to a given cross-ratio diagram or not. Such a division of the conformal polygon vertices by the cross-ratio diagram into two subsets is described by the first transianic index $b_l$. This is illustrated  in fig. \bref{fig:index2}. 
The second transianic index describes how chords of a given cross-ratio diagram are located with respect to the basis triangle $\triangle_n^{\triple{ijk}} \subset P_n$. E.g., the cross-ratio diagrams shown in fig. \bref{fig:index2}  have: $b_{ij}=1, b_{jk} = b_{ki}=0$ {\bf (a)}; $b_{ij}=1, b_{jk} = -1,  b_{ki}=0$ {\bf (b)}; $b_{ij}=1, b_{jk} =0,  b_{ki}=-1$ {\bf (c)}; $b_{ij}=b_{jk} =1,  b_{ki}=-1$ {\bf (d)}.

Recalling the embedding structure of the cross-ratios sets \eqref{Y_subset} one concludes that the transianic matrix $\cB\big(\bm \rY_n^{\triple{ijk}}\big)$ contains $\cB\big(\bm \rY_{n-1}^{\triple{ijk}}\big)$ as a submatrix  which is obtained by deleting some of its rows and columns. For instance, consider a triple $\triple{ijk} \in \rT_n$, for which, by definition, we have  $i,j,k \neq n$, see section \bref{sec:rec}. Then, from the embedding \eqref{Y_subset} it follows that a cross-ratio $\rY_l^{\triple{ijk}}$ which contains the vertex  $n$ appears in $\bm \rY_n^{\triple{ijk}}$ only when $l > (n-1)(n-4)/2$. For the first transianic index we then have
\be
\label{transianic_embed}
\begin{tabular}{|c||c|c|c|c|c|c|}
\hline
 & $\rY^{\triple{ijk} }_1$ & $\cdots\cdots$ & $\rY^{\triple{ijk} }_{(n-1)(n-4)/2}$ & $\rY^{\triple{ijk} }_{(n-1)(n-4)/2+1}$ & $\cdots\cdots$ & $\rY^{\triple{ijk} }_{n(n-3)/2}$  \\
\hline
\hline
$b_n$ & $0$ & $0$ & $0$ & $1$ & $1$ & $1$ \\
\hline
\end{tabular}
\ee
Thus, deleting from $\cB\big(\bm \rY_{n}^{\triple{ijk}}\big)$ the row corresponding to $b_n$ and columns corresponding to $\rY^{\triple{ijk} }_{(n-1)(n-4)/2+1},...,\rY^{\triple{ijk} }_{n(n-3)/2}$ one gets  the submatrix $\cB\big(\bm \rY_{n-1}^{\triple{ijk}}\big)$. See examples in section \bref{sec:examples}.

\subsection{Polygonal functions}

Let  $\triple{ijk} \in \rR_n$ and  $\bm \rY_n^{\triple{ijk}}$ be the cross-ratio set built by the diagrammatic algorithm.  Then, one computes  transianic indices \eqref{transianic} for all elements of $\bm \rY_n^{\triple{ijk}}$ and composes  the corresponding transianic matrix \eqref{tr_mat}. Given all this data, a $n$-point polygonal  function can be defined as the following hypergeometric series
\be
\label{bare_n_1}
\HG_n^{\triple{ijk}}\big( {\bm a}\big|\bm \rY_n^{\triple{ijk}} \big) = 
\sum_{{\bm m}=0}^\infty 
{\rm A}^{\triple{ijk}} ({\bm a}|\bm m) 
\prod_{s=1}^{\frac{n(n-3)}{2}} \frac{\left( \rY_s^{\triple{ijk}} \right)^{m_s}}{m_s!} \,,
\ee
where $\bm m = \{ m_1, ... , m_{n(n-3)/2} \in \ZZ^{0+} \}$, and 
\be
\label{bare_n_2} 
{\rm A}^{\triple{ijk}} ({\bm a}|\bm m) = 
\frac{\dps \prod_{l=1,\, l\neq i,j,k }^{n} \, (-)^{M_l} \, (a_l)_{M_l} }
{
(1+|\bm a_{i,j}|')_{M_{ij}} 
(1+|\bm a_{j,k}|')_{M_{jk}} 
(1+|\bm a_{k,i}|')_{M_{ki}} 
} \,.
\ee
Here: $(a)_M = \Gamma(a+M)/\Gamma(a)$ is the Pochhammer symbol; $|\bm a_{q,r}|'$ is defined in \eqref{notations_params}; $M_l$  and $M_{qr}$ are  linear functions in  $\bm m$:
\be
\label{bare_n_3}
M_l = \sum_{p=1}^{\frac{n(n-3)}{2}} b_l^{(p)}\, m_p\,, 
\qquad 
M_{qr} = \sum_{p=1}^{\frac{n(n-3)}{2}} b_{qr}^{(p)}\, m_p\,,
\ee
with coefficients being the  transianic indices  $b_l^{(p)}$ and $b^{(p)}_{qr}$   \eqref{transianic}; the labels $s,p$ enumerate cross-ratios from the set $\bm \rY_n^{\triple{ijk}}$.

\subsubsection{Convergence domains}
\label{sec:domain} 

The convergence of the multivariate hypergeometric series \eqref{bare_n_1} can be analyzed using the methods of \cite{Horn1889,exton1976multiple}. To this end, one introduces  the ratios of series coefficients
\be
\label{coeff_ratios_1}
f_q(\bm m) = \frac{1}{m_q+1} \frac{\rA^{\triple{ijk}}(\bm a| \bm m + \bm e_q) }{\rA^{\triple{ijk}}(\bm a| \bm m )}\,,
\qquad
q = 1,...\,, \frac{n(n-3)}{2}\,,
\ee
where $\bm e_q = \{0,...,1,...,0 \}$ is a unit vector with all components but $q$-th equal to zero. The power series \eqref{bare_n_1} absolutely converges  for $|\rY_q^{\triple{ijk}}| < r_q$, where the radii of convergence $r_q$ can be expressed in terms of   \eqref{coeff_ratios_1} as follows 
\be
\label{coeff_ratios_2}
r_q = |F_q(\bm m)|^{-1}\,, 
\quad \text{where} \quad
F_q (\bm m) = \lim_{t\to \infty} f_q( t \, \bm m)\,.
\ee
By construction, the  rational functions \eqref{coeff_ratios_1} are not independent, they satisfy a certain system of functional identities. It follows that the radii of convergence  obey   algebraic  equations which  describe a surface in the space of variables. A particular form of this surface which defines the domain of convergence is quite challenging to identify.
 
Leaving aside the question of finding the domain of convergence explicitly, one can introduce a weaker characteristic, the convergence index. Using the relation \cite{Bateman:100233}
\be
\label{gammas_ratio}
\frac{\Gamma(\tau + a)}{\Gamma(\tau + b)} = \tau^{a-b}\big(1 + O(\tau^{-1})\big)\,,
\qquad |\tau| \to \infty\,,
\ee
one calculates the limit in \eqref{coeff_ratios_2}
\be
\label{coeff_ratios_3}
F_q (\bm m) = \frac{\dps \prod_{l=1,\, l\neq i,j,k }^{n} \, (-)^{b_l^{(q)}} \, (M_l)^{b_l^{(q)}} }{m_q \, (M_{ij})^{b_{ij}^{(q)}} (M_{jk})^{b_{jk}^{(q)}} (M_{ki})^{b_{ki}^{(q)}}}\,
\lim_{t \to \infty} t^{-\dl_q^{\triple{ijk}}}\,,
\ee
where the convergence index is given by 
\be
\label{coeff_ratios_4}
\dl_q^{\triple{ijk}} = 1 + b_{ij}^{(q)}+b_{jk}^{(q)}+b_{ki}^{(q)} - \dps \sum_{l=1,\, l\neq i,j,k }^{n} b_l^{(q)}\,.
\ee
In fact, it is  defined by the weighted sum of elements from the $q$-th column of the transianic matrix \eqref{tr_mat}.   We see that the convergence indices $\dl_q^{\triple{ijk}}$ are completely  defined in terms of the transianic indices and determine whether the polygonal series $\HG_n^{\triple{ijk}}\big( {\bm a}\big|\bm \rY_n^{\triple{ijk}} \big)$ converges or not. If at least for one variable $\dl_q^{\triple{ijk}} < 0$, then the power series \eqref{bare_n_1} diverges, while  $\dl_q^{\triple{ijk}} > 0$ for all variables means that it converges everywhere. When $\dl_q^{\triple{ijk}} = 0$  the radius of convergence \eqref{coeff_ratios_2} is defined by
\be
\label{convegence_radius}
F_q (\bm m) = \frac{\dps \prod_{l=1,\, l\neq i,j,k }^{n} \, (-)^{b_l^{(q)}} \, (M_l)^{b_l^{(q)}} }{m_q \, (M_{ij})^{b_{ij}^{(q)}} (M_{jk})^{b_{jk}^{(q)}} (M_{ki})^{b_{ki}^{(q)}}} \,.
\ee

\subsubsection{Reduction properties}
\label{sec:red}

We have explicitly constructed all the factors needed to compose the basis functions \eqref{rep2}:
\be
\label{rep22}
\BF_n^{\langle ijk\rangle}({\bm a}|\bm x) 
= {\rm S}_{n}^{\triple{ijk}}({\bm a})\,
  {\rm V}_n^{\triple{ijk}}({\bm a}|\bm x)\,
  \HG_n^{\triple{ijk}}\big( {\bm a}\big|\bm \rY_n^{\triple{ijk}} \big)\,.
\ee  
Below we show that in terms of  basis functions the reduction formula  \eqref{in_reduction} takes the form  
\be
\label{Phi_reduction}
\BF_{n}^{\triple{ijk}}(a_1,...,a_{n-1},a_n | x_1,...,x_{n-1},x_n)\Big|_{a_n=0} \hspace{-2mm} = 
\BF_{n-1}^{\triple{ijk}}(a_1,...,a_{n-1}| x_{1},...,x_{n-1})\,,
\quad
\sum_{i=1}^{n-1} a_i = D\,.
\ee 
Since $k$ on the right-hand side cannot be equal to $n$, it follows that the $k=n$ basis function on the left-hand side should necessarily vanish at $a_n=0$.  In fact, the formula \eqref{Phi_reduction} says that the reduction of the conformal integral in the reconstruction representation goes termwise, when a part of $n$-point basis functions is mapped one-to-one onto all $(n-1)$-point basis functions, while the remaining part goes to zero. Moreover, due to the flag structure \eqref{Y_subset} of the cross-ratios there is no need to apply analytic continuation formulas to map the reduced basis functions back to the original convergence domain. It is also worth noting that taking the limit $a_n\to 0 $ does not require any regularization.\footnote{In general, the behavior of  basis functions near particular points or surfaces in the space of propagator powers ${\bm a}$ may be singular. E.g. the box conformal integral  in the non-parametric limit $a_i\to1$ diverges that requires introducing a dimensional regularization, see  \cite{Dolan:2000uw,Davydychev:1992eww,Alkalaev:2025fgn}.}

Consider first the case $k \neq n$. Each factor in \eqref{rep22} can be reduced separately. 

\begin{itemize}

\item Recalling  the notation \eqref{notations_params} one directly finds that the triangle-factor \eqref{star_n} reduces as   
\be
\label{reduction_prefactors1}
\rS_{n}^{\triple{ijk}}(a_1,...,a_{n-1},a_n)\Big|_{a_n=0} = \rS_{n-1}^{\triple{ijk}}(a_1,...,a_{n-1})\,. 
\ee
\item  Considering the leg-factor one notes that the size of the chamber $\cham_{n}^{(j)}$ is decreased, i.e. $L_n^{(j)} \to L_{n-1}^{(j)} = L_{n}^{(j)}-1$. Then,   
\be
\label{reduction_prefactors2}
\rV_{n}^{\triple{ijk}}(a_1,...,a_{n-1},a_n | x_1,...,x_{n-1},x_n)\Big|_{a_n=0} = \rV_{n-1}^{\triple{ijk}}(a_1,...,a_{n-1}| x_{1},...,x_{n-1})\,.
\ee
\item  Since $k \neq n$, then the polygonal function \eqref{bare_n_1} contains $(a_n)_{M_n}$ in the numerator. Combining \eqref{Y_subset} with the submatrix structure of the transianic matrix \eqref{transianic_embed} one obtains  
\be
\label{reduction}
\HG_n^{\triple{ijk}}\big(a_1,...,a_n\big|\bm \rY_n^{\triple{ijk}} \big) \Big|_{a_n=0}
\; =\; 
\HG_{n-1}^{\triple{ijk}}\big( a_1,...,a_{n-1}\big|\bm \rY_{n-1}^{\triple{ijk}} \big)\,.
\ee

\end{itemize}

When $k=n$ the triangle-factor $\rS_{n}^{\triple{ijn}}$ vanishes due to $\Gamma(a_n \to 0) \to \infty$ in the denominator of \eqref{star_n}. This leads to that the corresponding basis function $\BF_{n}^{\triple{ijn}}(a_n \to 0)$ vanishes as well despite the leg-factor along with the polygonal function can stay  non-vanishing.

\section{Examples}
\label{sec:examples}

The diagrammatic algorithm described in the previous sections  defines a triple product of the triangle-factor \eqref{star_n}, the leg-factor \eqref{V_n}, and the polygonal function  \eqref{bare_n_1} that gives the basis function $\BF_n^{\langle ijk\rangle}({\bm a}|\bm x)$ \eqref{rep2}. Going over all index triples $\triple{ijk} \in \rR_n$ \eqref{Rn} one builds a complete set of  basis functions that results in  finding  the $n$-point conformal integral in the form \eqref{rep1}. On the other hand, following the reconstruction formula \eqref{conj_fin} it is sufficient to build the master functions which are basis functions $\BF_n^{\triple{ijk}}({\bm a}|\bm x)$ for $\triple{ijk} \in \rT_n$. In this section we illustrate the diagrammatic algorithm and the reconstruction procedure  with examples of the  lower-point  conformal integrals: $n=4 \text{ (box)}, 5 \text{ (pentagon)}, 6\text{ (hexagon)}$.

\subsection{Box conformal integral}
\label{sec:box}

\begin{figure}[h]
\centering
\includegraphics[scale=0.15]{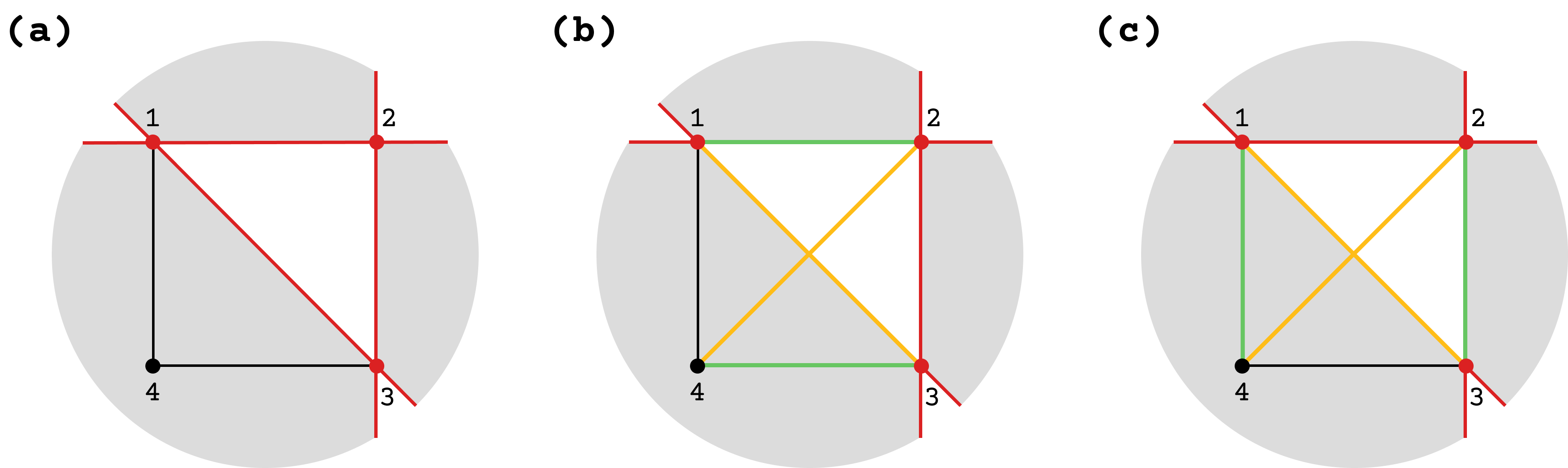}
\caption{{\bf (a)}: Basis triangle $\triangle_4^{\triple{123}}$ inscribed in the conformal polygon $P_4$ with two of three open chambers being empty. {\bf (b)}, {\bf (c)}: The two cross-ratios for the box master function.} 
\label{fig:box_kinematic}
\end{figure}

In the four-point case, the set of all ordered index triples $\rR_4$ consists of four elements:
\be
\rR_4 = \Big\{ \triple{123}, \triple{234}, \triple{134}, \triple{124} \Big \}\,.
\ee 
The whole set is one $\ZZ_4$-orbit of which representative we choose  $\rT_4 = \{ \triple{123} \}$ since the sum of indices is minimal (see our convention below \eqref{2.9}). 
Thus, there are four basis functions and  one master function  given by  
\be
\label{master_4_123}
\BF_4^{\langle 123\rangle}({\bm a}|\bm x) =  
\rS_4^{\triple{123}}(\bm a)\, 
\rV_4^{\triple{123}}(\bm a| \bm x)\,
\HG_4^{\triple{123}}\left( {\bm a}\big|\rY^{\langle 123 \rangle }_1\,, \rY^{\langle 123 \rangle }_2 \right)\,.
\ee
Here, the triangle factor \eqref{star_n} is 
\be
\label{S_4_123}
\rS_4^{\triple{123}}(\bm a) = \Gamma
\left[
\begin{array}{l l}
-|\bm a_{1,2}|', \,-|\bm a_{2,3}|',\, -|\bm a_{3,1}|' \\
\qquad a_1,\qquad  a_2, \qquad a_3
\end{array}
\right]. 
\ee
The conformal polygon $P_4$, the basis triangle $\triangle_4^{\triple{123}} \subset P_4$, and the corresponding chambers are shown in fig. \bref{fig:box_kinematic} {\bf (a)}. Among the three  chambers only $\cham_4^{(2)}$ contains the vertex  4, the other two $\cham_4^{(1)}$ and $\cham_4^{(3)}$ are empty, i.e. they do not contain any vertices: $L_4^{(2)} = 1$,  $L_4^{(1)} = L_4^{(3)} = 0$.

\noindent Then, the general formula \eqref{V_n} gives the following expression for the four-point leg-factor:
\be
\label{V_4_123}
\rV_4^{\triple{123}}(\bm a| \bm x) = 
X_{12}^{|\bm a_{1,2}|'} 
X_{23}^{|\bm a_{2,3}|'}
X_{31}^{|\bm a_{3,1}|'} \, 
X_{24}^{-a_4}\,.
\ee
To write down the polygonal function in \eqref{master_4_123} one introduces the set $\bm \rY_4^{\triple{123}}$ which consists of two cross-ratios\footnote{The cross-ratios \eqref{cros_4} coincide with the  cross-ratios $u$ and $v$  introduced in \cite{Dolan:2000ut}: $\rY^{\langle 123 \rangle }_1 = u$ and $\rY^{\langle 123 \rangle }_2 = v$.}
\be
\label{cros_4}
\rY^{\langle 123 \rangle }_1 = U[1,2,3,4] = \frac{X_{12} X_{34} }{X_{13} X_{24}} \,, 
\qquad 
\rY^{\langle 123 \rangle }_2 = U[3,2,1,4] = \frac{X_{14} X_{23} }{X_{13} X_{24}}\,,
\ee
where function $U$ is defined in \eqref{U_W}, the corresponding diagrams are shown in fig. \bref{fig:box_kinematic} {\bf (b)}, {\bf (c)}. The cubic cross-ratios are obviously absent in the four-point case. By calculating the transianic indices \eqref{transianic} one  composes  the transianic matrix
\be
\label{B4_123}
\cB\big(\bm \rY_4^{\triple{123}}\big)\;  =\quad  
\begin{tabular}{|c||c|c|}
\hline
 	  & $\rY^{\triple{123} }_1$ & $\rY^{\triple{123} }_2$   \\
\hline
\hline
$b_4$ & $1$ & $1$  \\
\hline
\hline
$b_{12}$ & $1$ & $0$  \\
\hline
\hline
$b_{23}$ & $0$ & $1$ \\
\hline
\hline
$b_{31}$ & $-1$ & $-1$  \\
\hline
\end{tabular}
\ee

The polygonal   function in \eqref{master_4_123} can now be built  using the general formulas \eqref{bare_n_1}--\eqref{bare_n_3} and the resulting hypergeometric series is given by
\begin{multline}
\label{H4_123}
\HG_4^{\triple{123}}\left( {\bm a}\big|\rY^{\langle 123 \rangle }_1\,, \rY^{\langle 123 \rangle }_2 \right)  \\
= \sum_{m_1, m_2=0}^\infty \frac{(-)^{m_1 + m_2}\,(a_4)_{m_1+m_2}  }{(1+|\bm a_{1,2}|')_{m_1} (1+|\bm a_{2,3}|')_{m_2} (1+|\bm a_{3,1}|')_{-m_1-m_2} } 
\frac{\left( \rY^{\langle 123 \rangle }_1 \right)^{m_1}}{m_1!}
\frac{\left( \rY^{\langle 123 \rangle }_2 \right)^{m_2}}{m_2!}\,.
\end{multline}

Using the transianic matrix \eqref{B4_123} one verifies that the convergence indices \eqref{coeff_ratios_4} are equal to zero for both variables. Then, the radii of convergence \eqref{coeff_ratios_2}, \eqref{convegence_radius} are given by 
\be
r_1 = \frac{m_1^2}{(m_1+m_2)^2}\,,
\qquad
r_2 = \frac{m_2^2}{(m_1+m_2)^2}\,,
\ee
that implies that they lie on a curve $\sqrt{r_1}+\sqrt{r_2} = 1$. Thus, the convergence domain is given by
\be
\label{convergence_F4}
\sqrt{|\rY_1^{\triple{123}}|} + \sqrt{|\rY_2^{\triple{123}}|} < 1\,.
\ee

There are three more basis functions which can be obtained from the master function \eqref{master_4_123}  by acting with the cyclic group $\ZZ_4$ elements: 
\be
\label{basis_4_1}
\begin{split}
\BF_4^{\langle 234\rangle}({\bm a}|\bm x) 
&= 
(\cycle_4)^1 \circ \BF_4^{\langle 123\rangle}({\bm a}|\bm x) 
= 
\BF_4^{\langle 123\rangle}((\cycle_4)^1 \circ {\bm a}|(\cycle_4)^1 \circ \bm x)\,, \\
\BF_4^{\langle 134\rangle}({\bm a}|\bm x) 
&= 
(\cycle_4)^2 \circ \BF_4^{\langle 123\rangle}({\bm a}|\bm x) 
= 
\BF_4^{\langle 123\rangle}((\cycle_4)^2 \circ {\bm a}|(\cycle_4)^2 \circ \bm x)\,, \\
\BF_4^{\langle 124\rangle}({\bm a}|\bm x) 
&= 
(\cycle_4)^3 \circ \BF_4^{\langle 123\rangle}({\bm a}|\bm x) 
= 
\BF_4^{\langle 123\rangle}((\cycle_4)^3 \circ {\bm a}|(\cycle_4)^3 \circ \bm x)\,.
\end{split}
\ee
This completes the construction of  basis functions. Then, the box conformal integral can be evaluated  by means of the reconstruction formula \eqref{conj_fin}:
\be
\label{box_reconstruction}
\begin{split}
I_4^{\bm a} (\bm x) &=  \sum_{l=0}^{3} (\cycle_4)^l \circ \BF_4^{\langle 123\rangle}({\bm a}|\bm x)
\\
&=\BF_4^{\langle 123\rangle}({\bm a}|\bm x)  + 
\BF_4^{\langle 234\rangle}({\bm a}|\bm x) + 
\BF_4^{\langle 134\rangle}({\bm a}|\bm x) + 
\BF_4^{\langle 124\rangle}({\bm a}|\bm x) \,.
\end{split}
\ee

Note that the double hypergeometric series \eqref{H4_123} can be reduced to the known special function by making identical transformations. To this end, one applies the Pochhammer symbol relation $(1-a)_{-M} = {(-)^M}/{(a)_M}$ to the factor $(1+|\bm a_{3,1}|')_{-m_1-m_2}$ in \eqref{H4_123} to obtain:
\be
\label{H4_123_F4}
\HG_4^{\triple{123}}\left( {\bm a}\big|\rY^{\langle 123 \rangle }_1\,, \rY^{\langle 123 \rangle }_2 \right) 
=
F_4 
\left[
\begin{array}{l l}
\qquad a_4\,, \quad  -|\bm a_{3,1}|' \\
1+|\bm a_{1,2}|',\, 1+|\bm a_{2,3}|'
\end{array}\Bigg|\, \rY_1^{\triple{123}}\,, \rY_2^{\triple{123}}
\right],
\ee
where the right-hand side is the fourth Appell function defined by the hypergeometric series
\be
\label{appell_F4}
F_4 \Bigg[
\begin{array}{l l}
a_1,a_2  \\
c_1,c_2
\end{array}\bigg| 
\xi_1, \xi_2
\Bigg] 
= 
\sum_{m_1,m_2 = 0}^{\infty} \frac{(a_1)_{m_1+m_2} (a_2)_{m_2+m_1}  }{(c_1)_{m_1} (c_2)_{m_2} } \frac{\xi_1^{m_1} }{m_1!} \frac{\xi_2^{m_2} }{m_2!} \,,
\ee
converging in the domain $\sqrt{|\xi_1|}+\sqrt{|\xi_2|} <1$, which coincides with   \eqref{convergence_F4}. Applying the identical transformation \eqref{H4_123_F4} to other polygonal functions  one derives that
\be
\begin{split}
\HG_4^{\triple{234}}\left( {\bm a}\big|\rY^{\langle 234 \rangle }_1\,, \rY^{\langle 234 \rangle }_2 \right) 
&= 
F_4 
\left[
\begin{array}{l l}
\qquad a_1\,, \quad-|\bm a_{4,2}|' \\
1+|\bm a_{2,3}|', 1+|\bm a_{3,4}|'
\end{array}\Bigg|\, \rY_2^{\triple{123}}\,, \rY_1^{\triple{123}}
\right]\,, \\
\HG_4^{\triple{134}}\left( {\bm a}\big|\rY^{\langle 134 \rangle }_1\,, \rY^{\langle 134 \rangle }_2 \right) 
&= 
F_4 
\left[
\begin{array}{l l}
\qquad a_2\,,\quad  -|\bm a_{1,3}|' \\
1+|\bm a_{3,4}|',\, 1+|\bm a_{4,1}|'
\end{array}\Bigg|\, \rY_1^{\triple{123}}\,, \rY_2^{\triple{123}}
\right]\,, \\
\HG_4^{\triple{124}}\left( {\bm a}\big|\rY^{\langle 124 \rangle }_1\,, \rY^{\langle 124 \rangle }_2 \right) 
&= 
F_4 
\left[
\begin{array}{l l}
\qquad a_3\,, \quad  -|\bm a_{2,4}|' \\
1+|\bm a_{4,1}|', \, 1+|\bm a_{1,2}|'
\end{array}\Bigg|\, \rY_2^{\triple{123}}\,, \rY_1^{\triple{123}}
\right] \,,
\end{split}
\ee
where we used that the longest cycle $C_4 \in \ZZ_4$ permutes the two cross-ratios \eqref{cros_4}:
\be
\label{cross_4_cycled}
\begin{split}
\rY^{\langle 234 \rangle }_1 &= (\cycle_4)^1 \circ \rY^{\langle 123 \rangle }_1 = \rY^{\langle 123 \rangle }_2  \,,  
\qquad 
\rY^{\langle 234 \rangle }_2 = (\cycle_4)^1 \circ \rY^{\langle 123 \rangle }_2 = \rY^{\langle 123 \rangle }_1  \,, \\
\rY^{\langle 134 \rangle }_1 &= (\cycle_4)^2 \circ \rY^{\langle 123 \rangle }_1 = \rY^{\langle 123 \rangle }_1 \,, 
\qquad 
\rY^{\langle 134 \rangle }_2 = (\cycle_4)^2 \circ \rY^{\langle 123 \rangle }_2 = \rY^{\langle 123 \rangle }_2\,, \\
\rY^{\langle 124 \rangle }_1 &= (\cycle_4)^3 \circ \rY^{\langle 123 \rangle }_1 = \rY^{\langle 123 \rangle }_2 \,, 
\qquad 
\rY^{\langle 124 \rangle }_2 =(\cycle_4)^3 \circ \rY^{\langle 123 \rangle }_2 = \rY^{\langle 123 \rangle }_1  \,. 
\end{split}
\ee

Thus, using the Appell functions we see that the reconstruction formula \eqref{box_reconstruction} now reproduces the well-known expression for the box conformal integral \cite{Dolan:2000uw}. Since all basis functions are power series in the same variables $\rY_1^{\triple{123}}, \rY_2^{\triple{123}}$, this expression is valid within the convergence domain \eqref{convergence_F4}.

The resulting expression \eqref{box_reconstruction} can be examined against the consistency conditions discussed in section \bref{sec:rec}. Setting e.g. $a_4=0$ and combining \eqref{reduction_prefactors1}--\eqref{reduction} one concludes that $\BF_4^{\langle 123\rangle}$ reproduces the star-triangle relation \eqref{star-triangle}, while  the other three  basis functions vanish. Setting any other parameter to zero  means that some other basis function is non-zero which again boils down to the corresponding star-triangle relation. On the other hand, in the non-parametric regime ($\forall a_i=1$) one verifies that the reconstruction formula \eqref{box_reconstruction} gives  the Bloch-Wigner function \cite{Davydychev:1992eww, Dolan:2000uw} (all relevant  technical details can be found e.g. in \cite{Alkalaev:2025fgn}).

\subsection{Pentagon conformal integral}
\label{sec:pentagon}

In the five-point case, the set of all ordered index triples $\rR_5$ consists of  ten elements:
\be
\label{R_5}
\begin{split}
\rR_5 = \Big\{ 
&\triple{123}, \triple{234}, \triple{345}, \triple{145}, \triple{125}, \\
&\triple{124}, \triple{235},\triple{134}, \triple{245}, \triple{135} \Big\}\,.
\end{split}
\ee
The two lines here  are two $\ZZ_5$-orbits. By choosing representatives from each orbit we compose the set  
\be
\rT_5 = \Big\{\triple{123}, \triple{124} \Big\}\,.
\ee 
Thus, there are two master functions, 
\be
\label{master_5}
\begin{split}
\BF_5^{\langle 123\rangle}({\bm a}|\bm x)
&=  
\rS_5^{\triple{123}}(\bm a)\, 
\rV_5^{\triple{123}}(\bm a| \bm x)\,
\HG_5^{\triple{123}}\left( {\bm a}\big|\rY^{\langle 123 \rangle }_1\,, ...\,, \rY^{\langle 123 \rangle }_5 \right)\,, \\
\BF_5^{\langle 124\rangle}({\bm a}|\bm x) 
&=  
\rS_5^{\triple{124}}(\bm a)\, 
\rV_5^{\triple{124}}(\bm a| \bm x)\,
\HG_5^{\triple{124}}\left( {\bm a}\big|\rY^{\langle 124 \rangle }_1\,, ...\,, \rY^{\langle 124 \rangle }_5 \right)\,.
\end{split}
\ee
The triangle-factors read off from the general formula \eqref{star_n} are given by 
\be
\label{S_5}
\rS_5^{\triple{123}} = \Gamma
\left[
\begin{array}{l l}
-|\bm a_{1,2}|', -|\bm a_{2,3}|', -|\bm a_{3,1}|' \\
\qquad a_1, \qquad a_2, \qquad a_3
\end{array}
\right],
\quad
\rS_5^{\triple{124}} = \Gamma
\left[
\begin{array}{l l}
-|\bm a_{1,2}|', -|\bm a_{2,4}|', -|\bm a_{4,1}|' \\
\qquad a_1, \qquad a_2, \qquad a_4
\end{array}
\right].
\ee
Note that despite the same arguments, the five-point triangle-factor in \eqref{S_5} is different from the four-point triangle-factors \eqref{S_4_123}, because the sums $|\bm a_{p,q}|'$ generally depend on additional parameter $a_5$, cf. \eqref{notations_params}. 
\begin{figure}
\centering
\includegraphics[scale=0.15]{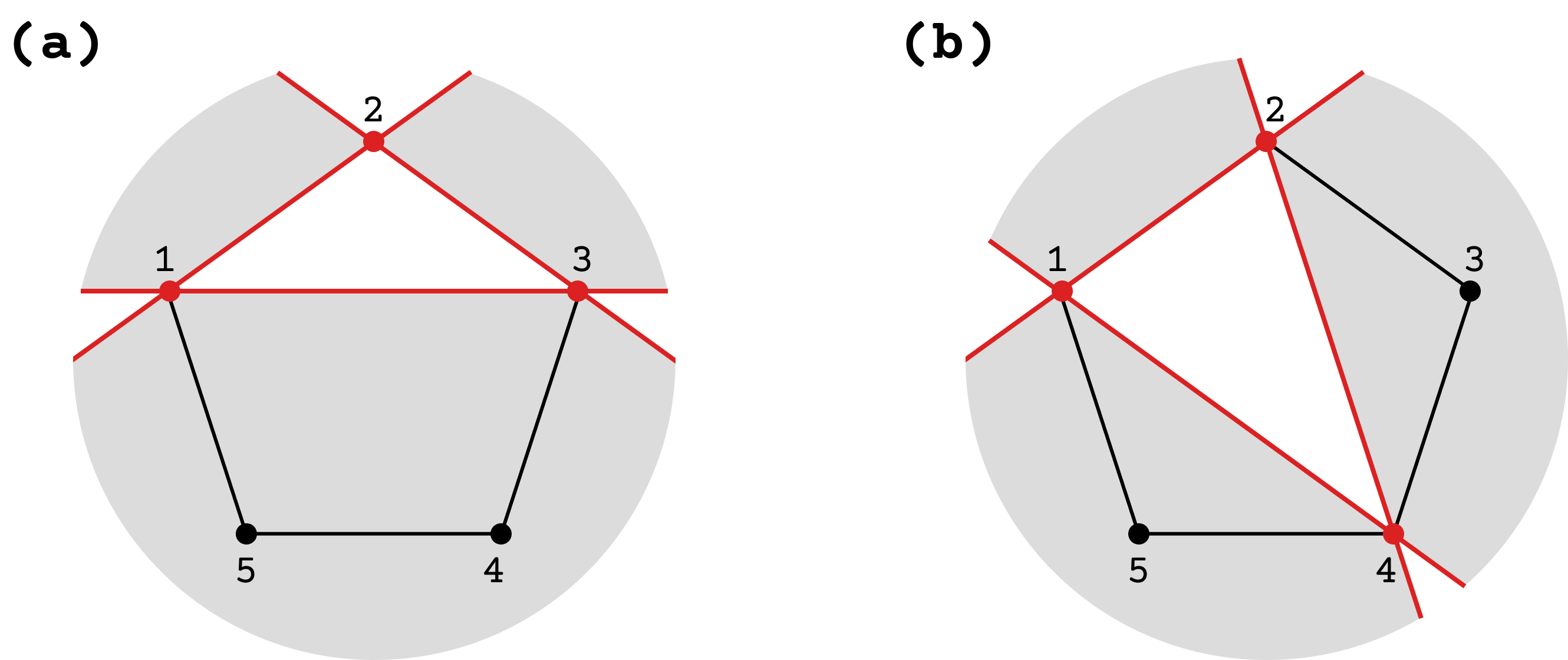}
\caption{Two basis triangles $\triangle_5^{\triple{123}}$ and $\triangle_5^{\triple{124}}$ inscribed in the conformal polygon $P_5$. {\bf (a)}: There are two empty open chambers,  $L_5^{(1)} = L_5^{(3)} = 0$, while the presence of two vertices in the third open chamber, $L_5^{(2)} = 2$,  allows one to construct a cubic cross-ratio. {\bf (b)}: One open chamber is empty, while the other two are non-empty, $L_5^{(1)} = L_5^{(2)} = 1$. No cubic cross-ratios are possible in this case.  }  
\label{fig:pentagon_kinematic}
\end{figure}

In order to build the leg-factors in \eqref{master_5} one has to count a number of vertices in open chambers associated to each of two basis triangles.  The first basis triangle $\triangle_5^{\triple{123}}$: there is only one non-empty open chamber $\cham_5^{(2)}$ which  contains vertices  $4$ and $5$. The second basis triangle  $\triangle_5^{\triple{124}}$:  the open  chamber $\cham_5^{(4)}$ is empty and the open chambers $\cham_5^{(1)}$ and $\cham_5^{(2)}$ contain vertices  $3$ and $5$, respectively. This is illustrated in fig. \bref{fig:pentagon_kinematic}. Therefore, from \eqref{V_n} we find that
\be
\label{V_5}
\begin{split}
\rV_5^{\triple{123}}(\bm a|\bm x) 
&=
X_{12}^{|\bm a_{1,2}|'} X_{23}^{|\bm a_{2,3}|'}  X_{31}^{|\bm a_{3,1}|'}\, X_{24}^{-a_4} X_{25}^{-a_5}\,,\\
\rV_5^{\triple{124}}(\bm a|\bm x) 
&=
X_{12}^{|\bm a_{1,2}|'} X_{24}^{|\bm a_{2,4}|'} X_{41}^{|\bm a_{4,1}|'}\, X_{31}^{-a_3} X_{25}^{-a_5} \,.
\end{split}
\ee 

Now, one constructs arguments of the polygonal functions  in \eqref{master_5}. 
\begin{itemize}
\item
The basis triangle $\triangle_5^{\triple{123}}$, see fig. \bref{fig:pentagon_kinematic} {\bf (a)}. The corresponding set $\bm \rY_5^{\triple{123}}$ consists of the following five elements
\be
\label{cross_5_123}
\begin{split}
\rY^{\langle 123 \rangle }_1 &= U[1,2,3,4] = \frac{X_{12} X_{34} }{X_{13} X_{24}}\,, \quad \rY^{\langle 123 \rangle }_2 = U[3,2,1,4] = \frac{X_{23} X_{14} }{X_{13} X_{24}}\,, \\
\rY^{\langle 123 \rangle }_3 &= U[1,2,3,5] = \frac{X_{12} X_{35} }{X_{13} X_{25}}\,, \quad
\rY^{\langle 123 \rangle }_4 = U[3,2,1,5] = \frac{X_{23} X_{15} }{X_{13} X_{25}}\,,  \\
\rY^{\langle 123 \rangle }_5 &= W[1,2,3,4,5] = \frac{X_{12} X_{23} X_{45} }{X_{13} X_{24} X_{25}}\,,
\end{split}
\ee
where functions $U$ and $W$ are defined in \eqref{U_W}. The respective cross-ratio diagrams  are shown in fig. \bref{fig:cross_pentagon}. Note that $\rY^{\langle 123 \rangle }_{1,2} \in \bm \rY_5^{\triple{123}}$ are the same cross-ratios which were previously constructed in  the four-point case \eqref{cros_4}, $\rY^{\langle 123 \rangle }_{1,2} \in  \bm \rY_4^{\triple{123}}$, i.e.  there is the embedding $\bm \rY_4^{\triple{123}} \subset \bm \rY_5^{\triple{123}}$, cf. \eqref{Y_subset}.

\begin{figure}
\centering
\includegraphics[scale=0.15]{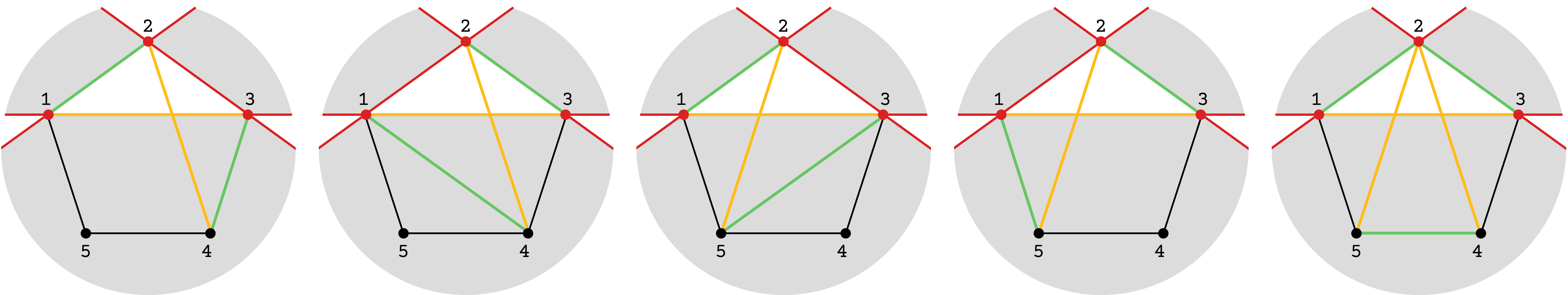}
\caption{Cross-ratios for the  polygonal  function labelled by $\triple{123} \in \rT_5$.} 
\label{fig:cross_pentagon}
\end{figure}

The  transianic matrix for the set $\bm \rY_5^{\triple{123}}$ calculated by means of  \eqref{transianic} is given by 
\be
\label{B5_123}
\cB\big(\bm \rY_5^{\triple{123}}\big)\;  =\quad  
\begin{tabular}{|c||c|c|c|c|c|}
\hline
 & $\rY^{\triple{123} }_1$ & $\rY^{\triple{123} }_2$ & $\rY^{\triple{123} }_3$ & $\rY^{\triple{123} }_4$ & $\rY^{\triple{123} }_5$  \\
\hline
\hline
$b_4$ & $1$ & $1$ & $0$ & $0$ & $1$ \\
\hline
\hline
$b_5$ & $0$ & $0$ & $1$ & $1$ & $1$ \\
\hline
\hline
$b_{12}$ & $1$ & $0$ & $1$ & $0$ & $1$ \\
\hline
\hline
$b_{23}$ & $0$ & $1$ & $0$ & $1$ & $1$ \\
\hline
\hline
$b_{31}$ & $-1$ & $-1$ & $-1$ & $-1$ & $-1$ \\
\hline
\end{tabular}
\ee
Deleting  the columns labelled by $\rY^{\triple{123} }_3$, $\rY^{\triple{123} }_4$, $\rY^{\triple{123}}_5$ and the row labelled by $b_5$ one restores the transianic matrix for the set $\bm \rY_4^{\triple{123}}$ \eqref{B4_123}.

\item The basis triangle $\triangle_5^{\triple{124}}$, see fig. \bref{fig:pentagon_kinematic} {\bf (b)}. In this case, the set $\bm \rY_5^{\triple{124}}$ contains quadratic cross-ratios only:
\be
\label{cross_5_124}
\begin{split}
\rY^{\langle 124 \rangle }_1 &= U[4,1,2,3] = \frac{X_{23} X_{14} }{X_{13} X_{24}} \,, \quad
\rY^{\langle 124 \rangle }_2 = U[2,1,4,3]  = \frac{X_{12} X_{34} }{X_{13} X_{24}} \,,  \\
\rY^{\langle 124 \rangle }_3 &= U[4,2,1,5] = \frac{X_{24} X_{15} }{X_{14} X_{25}} \,, \quad
\rY^{\langle 124 \rangle }_4  = U[1,2,4,5] = \frac{X_{12} X_{45} }{X_{14} X_{25}} \,,   \\
\rY^{\langle 124 \rangle }_5 &= U[1,2,3,5] = \frac{X_{12} X_{35} }{X_{13} X_{25}} \,,
\end{split}
\ee
For the sake of brevity, we do not present  the corresponding cross-ratio diagrams. Note again that the cross-ratios from the first line in \eqref{cross_5_124}  coincide with those in the last line in \eqref{cross_4_cycled}, i.e. we have $\bm \rY_4^{\triple{124}} \subset \bm \rY_5^{\triple{124}}$, cf. \eqref{Y_subset}.

The transianic matrix  for the set $\bm \rY_5^{\triple{124}}$ is given by 
\be
\cB\big(\bm \rY_5^{\triple{124}}\big)\;  =\quad  
\begin{tabular}{|c||c|c|c|c|c|}
\hline
 & $\rY^{\triple{124} }_1$ & $\rY^{\triple{124} }_2$ & $\rY^{\triple{124} }_3$ & $\rY^{\triple{124} }_4$ & $\rY^{\triple{124} }_5$  \\
\hline
\hline
$b_3$ & $1$ & $1$ & $0$ & $0$ & $1$ \\
\hline
\hline
$b_5$ & $0$ & $0$ & $1$ & $1$ & $1$ \\
\hline
\hline
$b_{12}$ & $0$ & $1$ & $0$ & $1$ & $1$ \\
\hline
\hline
$b_{24}$ & $-1$ & $-1$ & $1$ & $0$ & $0$ \\
\hline
\hline
$b_{41}$ & $1$ & $0$ & $-1$ & $-1$ & $0$ \\
\hline
\end{tabular}
\ee

\end{itemize}

\noindent Using the general formulas \eqref{bare_n_1}--\eqref{bare_n_3} we build  two polygonal functions:
\begin{multline}
\label{H5_123}
\HG_5^{\triple{123}}\left( {\bm a}\big|\rY^{\triple{123} }_1\,,...\,, \rY^{\triple{123}}_5 \right)  
= \sum_{m_1,...,m_5=0}^{\infty} \;\;
\prod_{l=1}^{5} \frac{\left(\rY^{\triple{123}}_l \right)^{m_l}}{m_l!}\\
\times \frac{(-)^{m_1+m_2+m_3+m_4} (a_4)_{m_1+m_2+m_5} (a_5)_{m_3+m_4+m_5} }{(1+|\bm a_{12}|')_{m_1+m_3+m_5} (1+|\bm a_{23}|')_{m_2+m_4+m_5} (1+|\bm a_{31}|')_{-m_1-m_2-m_3-m_4-m_5} } \,,
\end{multline}
\begin{multline}
\label{H5_122}
\HG_5^{\triple{124}}\left( {\bm a}\big|\rY^{\triple{124} }_1\,,...\,, \rY^{\triple{124}}_5 \right)  
= \sum_{m_1,..., m_5=0}^\infty \;\; \prod_{l=1}^5 \frac{\left( \rY_l^{\langle 124 \rangle }\right)^{m_l}}{m_l!}
\\
\times
\frac{(-)^{m_1+m_2+m_3+m_4} (a_3)_{m_1+m_2+m_5} (a_5)_{m_3+m_4+m_5}  }{(1+|\bm a_{1,2}|')_{m_2+m_4+m_5} (1+|\bm a_{2,4}|')_{m_3-m_1-m_2} (1+|\bm a_{4,1}|')_{m_1-m_3-m_4} } \,.
\end{multline}

This step completes the construction of master functions \eqref{master_5}. The reconstruction formula \eqref{rep2} now represents the pentagon conformal integral as 
\be
\label{pentagon_reconstruction}
\begin{split}
I_5^{\bm a}(\bm x) 
&= 
\sum_{l=0}^{4} (\cycle_5)^l \circ \left( \BF_5^{\langle 123\rangle}({\bm a}|\bm x)  + \BF_5^{\langle 124\rangle}({\bm a}|\bm x) \right) \\
&= \BF_5^{\triple{123}}({\bm a}|\bm x) + \BF_5^{\triple{234}}({\bm a}|\bm x) + \BF_5^{\triple{345}}({\bm a}|\bm x) + \BF_5^{\triple{145}}({\bm a}|\bm x) + \BF_5^{\triple{125}}({\bm a}|\bm x) \\
&+ \BF_5^{\triple{124}}({\bm a}|\bm x) + \BF_5^{\triple{235}}({\bm a}|\bm x) + \BF_5^{\triple{134}}({\bm a}|\bm x) + \BF_5^{\triple{245}}({\bm a}|\bm x) + \BF_5^{\triple{135}}({\bm a}|\bm x)
\,.
\end{split}
\ee
Here, one derives four more basis functions from each master function by acting with $\ZZ_5$. Each of two $\ZZ_5$-orbits (the two lines in \eqref{pentagon_reconstruction}) has 5 elements, which are listed below.  

\begin{itemize}
\item Basis functions labelled by elements of the orbit $\{\ZZ_5 \circ \triple{123} \}$:
\be
\label{basis_5_orbit_123}
\begin{split}
\BF_5^{\langle 123\rangle}({\bm a}|\bm x) 
&= \BF_5^{\langle 123\rangle}((\cycle_5)^0 \circ {\bm a}|(\cycle_5)^0 \circ \bm x)\,, \\
\BF_5^{\langle 234\rangle}({\bm a}|\bm x) 
&= \BF_5^{\langle 123\rangle}((\cycle_5)^1 \circ {\bm a}|(\cycle_5)^1 \circ \bm x)\,, \\
\BF_5^{\langle 345\rangle}({\bm a}|\bm x) 
&= \BF_5^{\langle 123\rangle}((\cycle_5)^2 \circ {\bm a}|(\cycle_5)^2 \circ \bm x)\,, \\
\BF_5^{\langle 145\rangle}({\bm a}|\bm x) 
&= \BF_5^{\langle 123\rangle}((\cycle_5)^3 \circ {\bm a}|(\cycle_5)^3 \circ \bm x)\,, \\
\BF_5^{\langle 125\rangle}({\bm a}|\bm x) 
&= \BF_5^{\langle 123\rangle}((\cycle_5)^4 \circ {\bm a}|(\cycle_5)^4 \circ \bm x)\,.
\end{split}
\ee

\item Basis functions labelled by elements of the orbit $\{\ZZ_5 \circ \triple{124} \}$:
\be
\label{basis_5_orbit_124}
\begin{split}
\BF_5^{\langle 124\rangle}({\bm a}|\bm x) 
&= \BF_5^{\langle 124\rangle}((\cycle_5)^0 \circ {\bm a}|(\cycle_5)^0 \circ \bm x)\,, \\
\BF_5^{\langle 235\rangle}({\bm a}|\bm x) 
&= \BF_5^{\langle 124\rangle}((\cycle_5)^1 \circ {\bm a}|(\cycle_5)^1 \circ \bm x)\,, \\
\BF_5^{\langle 134\rangle}({\bm a}|\bm x) 
&= \BF_5^{\langle 124\rangle}((\cycle_5)^2 \circ {\bm a}|(\cycle_5)^2 \circ \bm x)\,, \\
\BF_5^{\langle 245\rangle}({\bm a}|\bm x) 
&= \BF_5^{\langle 124\rangle}((\cycle_5)^3 \circ {\bm a}|(\cycle_5)^3 \circ \bm x)\,, \\
\BF_5^{\langle 135\rangle}({\bm a}|\bm x) 
&= \BF_5^{\langle 124\rangle}((\cycle_5)^4 \circ {\bm a}|(\cycle_5)^4 \circ \bm x)\,.
\end{split}
\ee
\end{itemize}

\noindent Each basis function in \eqref{basis_5_orbit_123}--\eqref{basis_5_orbit_124} is a triple product \eqref{rep2}. The triangle-factors are directly read off from the general formula \eqref{star_n}, while explicit expressions for the leg-factors and the polygonal functions are  given in Appendix \bref{app:pentagon}.

The convergence of the master  polygonal functions \eqref{H5_123} and \eqref{H5_122} can be studied  along the lines of section \bref{sec:domain}. In particular, all convergence indices for these functions are zero. Notably,  \eqref{H5_123} is the Srivastava-Daoust hypergeometric function \cite{SrivastavaDaoust, Srivastava1985MultipleGH}. The convergence of this function was considered  in \cite{Srivastava1972}, but no explicit expressions for the convergence domains were found. The coordinate domain of the pentagon reconstruction formula \eqref{pentagon_reconstruction} is given by intersecting   convergence domains of 10 basis functions since all of them are power series in different sets of variables, see Appendix \bref{app:pentagon}. This drastically complicates the entire analysis compared to the box conformal integral for which all the basis functions depend on the same variables, cf. section \bref{sec:box}.

The reconstruction formula \eqref{pentagon_reconstruction} was originally  derived  using the bipartite Mellin-Barnes representation in \cite{Alkalaev:2025fgn}. As discussed in section \bref{sec:rec}, the formula passes two non-trivial checks. Firstly, setting one of  propagator powers to zero, the general formulas \eqref{reduction_prefactors1}--\eqref{reduction} demonstrate that 6 of 10 basis functions vanish, while a linear combination of the remaining  4 functions gives the box conformal integral \eqref{box_reconstruction}. Secondly, when $a_i=1$, $\forall i = 1,...,5$, the reconstruction formula \eqref{pentagon_reconstruction} reproduces the non-parametric conformal integral evaluated as the $\ZZ_5$-invariant sum of 10 logarithms in Ref. \cite{Nandan:2013ip}.

\subsection{Hexagon conformal integral}
\label{sec:hex}

In the six-point case, the  set of all ordered index triples $\rR_6$ consists of  twenty  elements:
\be
\label{R_6}
\begin{split}
\rR_6 = \Big\{ 
&\triple{123}, \triple{234}, \triple{345}, \triple{456}, \triple{156}, \triple{126}, \\
&\triple{124}, \triple{235}, \triple{346}, \triple{145}, \triple{256}, \triple{136}, \\ 
&\triple{125}, \triple{236}, \triple{134}, \triple{245}, \triple{356}, \triple{146}, \\ 
&\triple{135}, \triple{246}
\Big\}\,.
\end{split}
\ee
The four lines here  are four $\ZZ_6$-orbits. Note that there is the orbit shortening as described below eq. \eqref{cord}.  By choosing  representatives we compose the set 
\be
\rT_6 = \Big\{\triple{123}, \triple{124}, \triple{125}, \triple{135} \Big\}\,.
\ee
Thus, there are  four  master functions:
\be
\label{master_6}
\begin{split}
\BF_6^{\langle 123\rangle}({\bm a}|\bm x)
&=  
\rS_6^{\triple{123}}(\bm a)\, 
\rV_6^{\triple{123}}(\bm a| \bm x)\,
\HG_6^{\triple{123}}\left( {\bm a}\big|\rY^{\langle 123 \rangle }_1\,, ...\,, \rY^{\langle 123 \rangle }_9 \right), \\
\BF_6^{\langle 124\rangle}({\bm a}|\bm x) 
&=  
\rS_6^{\triple{124}}(\bm a)\, 
\rV_6^{\triple{124}}(\bm a| \bm x)\,
\HG_6^{\triple{124}}\left( {\bm a}\big|\rY^{\langle 124 \rangle }_1\,, ...\,, \rY^{\langle 124 \rangle }_9 \right), \\
\BF_6^{\langle 125\rangle}({\bm a}|\bm x) 
&=  
\rS_6^{\triple{125}}(\bm a)\, 
\rV_6^{\triple{125}}(\bm a| \bm x)\,
\HG_6^{\triple{125}}\left( {\bm a}\big|\rY^{\langle 125 \rangle }_1\,, ...\,, \rY^{\langle 125 \rangle }_9 \right), \\ 
\BF_6^{\langle 135\rangle}({\bm a}|\bm x) 
&=  
\rS_6^{\triple{135}}(\bm a)\, 
\rV_6^{\triple{135}}(\bm a| \bm x)\,
\HG_6^{\triple{135}}\left( {\bm a}\big|\rY^{\langle 135 \rangle }_1\,, ...\,, \rY^{\langle 135 \rangle }_9 \right).
\end{split}
\ee
The triangle-factors read off from the general formula \eqref{star_n} are given by 
\be
\label{S_6}
\begin{split}
\rS_6^{\triple{123}} &= \Gamma
\left[
\begin{array}{l l}
-|\bm a_{1,2}|', -|\bm a_{2,3}|', -|\bm a_{3,1}|' \\
\qquad a_1, \qquad a_2, \qquad a_3
\end{array}
\right],
\quad
\rS_6^{\triple{124}} = \Gamma
\left[
\begin{array}{l l}
-|\bm a_{1,2}|', -|\bm a_{2,4}|', -|\bm a_{4,1}|' \\
\qquad a_1, \qquad a_2, \qquad a_4
\end{array}
\right], \\
\rS_6^{\triple{125}} &= \Gamma
\left[
\begin{array}{l l}
-|\bm a_{1,2}|', -|\bm a_{2,5}|', -|\bm a_{5,1}|' \\
\qquad a_1,\qquad  a_2, \qquad a_5
\end{array}
\right],
\quad
\rS_6^{\triple{135}} = \Gamma
\left[
\begin{array}{l l}
-|\bm a_{1,3}|', -|\bm a_{3,5}|', -|\bm a_{5,1}|' \\
\qquad a_1, \qquad  a_3, \qquad  a_5
\end{array}
\right].
\end{split}
\ee
Note that despite the same arguments, the six-point triangle-factors in \eqref{S_6} are different from the four-point and five-point triangle-factors \eqref{S_4_123} and \eqref{S_5}, because the sums $|\bm a_{p,q}|'$ generally depend on additional parameter  $a_6$, cf. \eqref{notations_params}.

\begin{figure}
\centering
\includegraphics[scale=0.15]{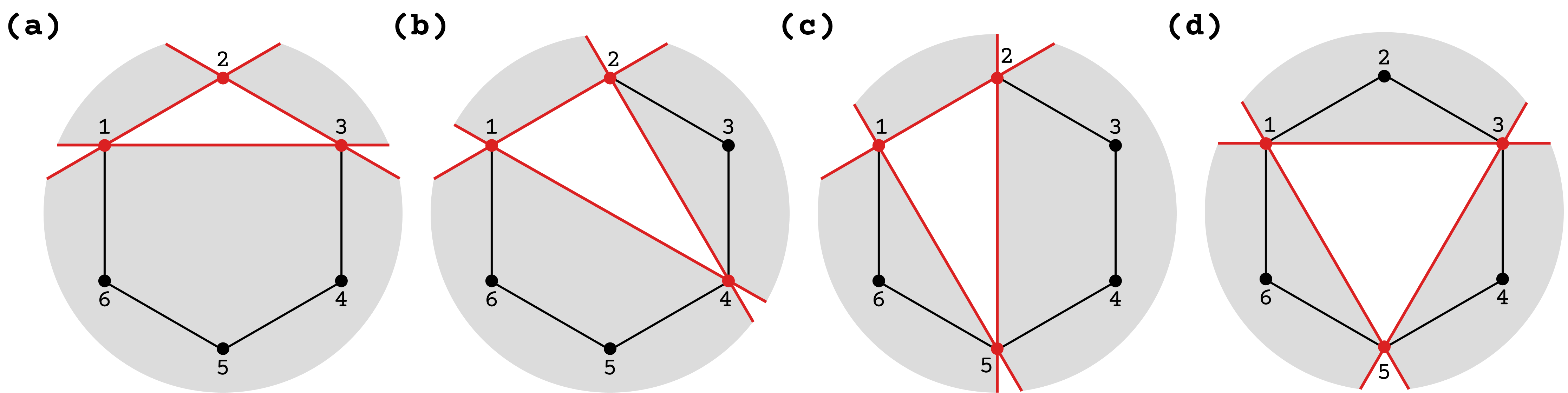}
\caption{Four basis triangles inscribed in the  conformal polygon $P_6$. Cubic cross-ratios are possible in the case of the basis triangles shown in {\bf (a)}, {\bf (b)}, {\bf (c)}. The basis triangle  {\bf (d)} has more rotational symmetry than other triangles that results in the orbit shortening.  In particular, all chambers in {\bf (d)} contain equal number of vertices, $L_6^{(1)} = L_6^{(3)}= L_6^{(5)}=1$.} 
\label{fig:triangles_hexagon}
\end{figure}

Now, one constructs arguments of the polygonal functions  in \eqref{master_6}.

\begin{itemize}

\item The basis triangle $\triangle_6^{\triple{123}} \subset P_6$, see  fig. \bref{fig:triangles_hexagon} {\bf (a)}. The only non-empty open chamber here is $\cham_6^{(2)}$, which contains vertices $4,5,6$. Thus, from the general formula \eqref{V_n} one derives   the corresponding leg-factor
\be
\label{V_6_123}
\rV_6^{\triple{123}}(\bm a|\bm x) 
=
X_{12}^{|\bm a_{1,2}|'} X_{23}^{|\bm a_{2,3}|'}  X_{31}^{|\bm a_{3,1}|'}\, X_{24}^{-a_4} X_{25}^{-a_5} X_{26}^{-a_6}\,.
\ee
The cross-ratio set $\bm \rY_6^{\triple{123}}$:
\be
\label{cross_6_123}
\begin{split}
\rY^{\langle 123 \rangle }_1 &= \frac{X_{12} X_{34} }{X_{13} X_{24}}\,, \quad 
\rY^{\langle 123 \rangle }_2 =  \frac{X_{23} X_{14} }{X_{13} X_{24}}\,, \qquad
\, \, \,
\rY^{\langle 123 \rangle }_3 =  \frac{X_{12} X_{35} }{X_{13} X_{25}}\,, \\
\rY^{\langle 123 \rangle }_4 &= \frac{X_{23} X_{15} }{X_{13} X_{25}}\,,  \quad
\rY^{\langle 123 \rangle }_5 =  \frac{X_{12} X_{23} X_{45} }{X_{13} X_{24} X_{25}}\,, \quad
\rY^{\langle 123 \rangle }_6 =  \frac{X_{12} X_{36} }{X_{13} X_{26} }\,, \\
\rY^{\langle 123 \rangle }_7 &= \frac{X_{23} X_{16} }{X_{13} X_{26}}\,,  \quad
\rY^{\langle 123 \rangle }_8 =  \frac{X_{12} X_{23} X_{46} }{X_{13} X_{24} X_{26}} \,, \quad
\rY^{\langle 123 \rangle }_9 =  \frac{X_{12} X_{23} X_{56} }{X_{13} X_{25} X_{26}} \,.
\end{split}
\ee
The respective cross-ratio diagrams are shown in fig. \bref{fig:cross_hexagon}. Note that  $\bm \rY_5^{\triple{123}} \subset \bm \rY_6^{\triple{123}}$, cf. \eqref{cross_5_123}. 
\begin{figure}
\centering
\includegraphics[scale=0.15]{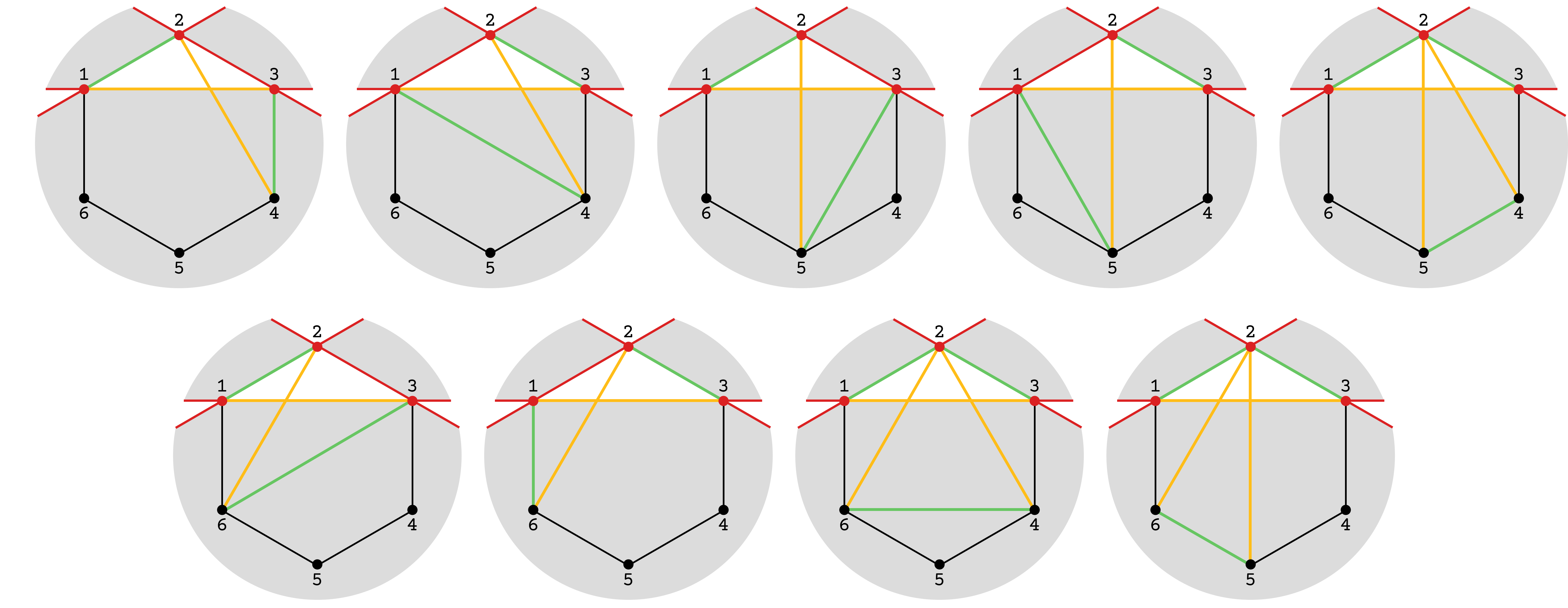}
\caption{Cross-ratios for the polygonal  function labelled by $\triple{123} \in \rT_6$.} 
\label{fig:cross_hexagon}
\end{figure}
The transianic matrix  $\cB\big(\bm \rY_6^{\triple{123}}\big)  =$
\be  
\begin{tabular}{|c||c|c|c|c|c|c|c|c|c|}
\hline
 & $\rY^{\triple{123} }_1$ & $\rY^{\triple{123} }_2$ & $\rY^{\triple{123} }_3$ & $\rY^{\triple{123} }_4$ & $\rY^{\triple{123} }_5$ & $\rY^{\triple{123} }_6$ & $\rY^{\triple{123} }_7$ & $\rY^{\triple{123} }_8$ & $\rY^{\triple{123} }_9$  \\
\hline
\hline
$b_4$ & $1$ & $1$ & $0$ & $0$ & $1$ & $0$ & $0$ & $1$ & $0$ \\
\hline
\hline
$b_5$ & $0$ & $0$ & $1$ & $1$ & $1$ & $0$ & $0$ & $0$ & $1$ \\
\hline
\hline
$b_6$ & $0$ & $0$ & $0$ & $0$ & $0$ & $1$ & $1$ & $1$ & $1$ \\
\hline
\hline
$b_{12}$ & $1$ & $0$ & $1$ & $0$ & $1$ & $1$ & $0$ & $1$ & $1$ \\
\hline
\hline
$b_{23}$ & $0$ & $1$ & $0$ & $1$ & $1$ & $0$ & $1$ & $1$ & $1$  \\
\hline
\hline
$b_{31}$ & $-1$ & $-1$ & $-1$ & $-1$ & $-1$ & $-1$ & $-1$ & $-1$ & $-1$ \\
\hline
\end{tabular}
\ee

\item The basis triangle $\triangle_6^{\triple{124}}$, see fig. \bref{fig:triangles_hexagon} {\bf (b)}. The open chamber $\cham_6^{(4)}$ is empty, while vertices  $3$ and $5,6$ are contained in $\cham_6^{(1)}$ and $\cham_6^{(2)}$, respectively. From the general formula \eqref{V_n} one derives   the corresponding leg-factor
\be
\label{V_6_124}
\rV_6^{\triple{124}}(\bm a|\bm x) 
=
X_{12}^{|\bm a_{1,2}|'} X_{24}^{|\bm a_{2,4}|'} X_{41}^{|\bm a_{4,1}|'}\, X_{13}^{-a_3} X_{25}^{-a_5} X_{26}^{-a_6} \,.
\ee
The cross-ratio set  $\bm \rY_6^{\triple{124}}$:
\be
\label{cross_6_124}
\begin{split}
\rY^{\langle 124 \rangle}_1 &= \frac{X_{23} X_{14}}{X_{13} X_{24}} \,, \quad
\rY^{\langle 124 \rangle}_2 = \frac{X_{12} X_{34}}{X_{13} X_{24}} \,, \quad
\rY^{\langle 124 \rangle}_3 =  \frac{X_{24} X_{15}}{X_{14} X_{25}} \,, \\
\rY^{\langle 124 \rangle}_4 &= \frac{X_{12} X_{45}}{X_{14} X_{25}} \,, \quad
\rY^{\langle 124 \rangle}_5 =  \frac{X_{12} X_{35}}{X_{13} X_{25}} \,, \quad
\rY^{\langle 124 \rangle}_6 =  \frac{X_{12} X_{46}}{X_{14} X_{26}} \,, \\
\rY^{\langle 124 \rangle}_7 &= \frac{X_{24} X_{16}}{X_{14} X_{26}} \,, \quad
\rY^{\langle 124 \rangle}_8 =  \frac{X_{12} X_{36}}{X_{13} X_{26}} \,, \quad
\rY^{\langle 124 \rangle}_9 =  \frac{X_{12} X_{24} X_{56}}{X_{14} X_{25} X_{26}} \,.
\end{split}
\ee
Note that  $\bm \rY_5^{\triple{124}} \subset \bm \rY_6^{\triple{124}}$, cf. \eqref{cross_5_124}.  The transianic matrix $\cB\big(\bm \rY_6^{\triple{124}}\big)  =$
\be
\begin{tabular}{|c||c|c|c|c|c|c|c|c|c|}
\hline
 & $\rY^{\triple{124} }_1$ & $\rY^{\triple{124} }_2$ & $\rY^{\triple{124} }_3$ & $\rY^{\triple{124} }_4$ & $\rY^{\triple{124} }_5$ & $\rY^{\triple{124} }_6$ & $\rY^{\triple{124} }_7$ & $\rY^{\triple{124} }_8$ & $\rY^{\triple{124} }_9$  \\
\hline
\hline
$b_3$ & $1$ & $1$ & $0$ & $0$ & $1$ & $0$ & $0$ & $1$ & $0$ \\
\hline
\hline
$b_5$ & $0$ & $0$ & $1$ & $1$ & $1$ & $0$ & $0$ & $0$ & $1$ \\
\hline
\hline
$b_6$ & $0$ & $0$ & $0$ & $0$ & $0$ & $1$ & $1$ & $1$ & $1$ \\
\hline
\hline
$b_{12}$ & $0$ & $1$ & $0$ & $1$ & $1$ & $1$ & $0$ & $1$ & $1$ \\
\hline
\hline
$b_{24}$ & $-1$ & $-1$ & $1$ & $0$ & $0$ & $0$ & $1$ & $0$ & $1$  \\
\hline
\hline
$b_{41}$ & $1$ & $0$ & $-1$ & $-1$ & $0$ & $-1$ & $-1$ & $0$ & $-1$ \\
\hline
\end{tabular}
\ee

\item The basis triangle $\triangle_6^{\triple{125}}$, see fig. \bref{fig:triangles_hexagon} {\bf (c)}. There are two non-empty open  chambers: $\cham_6^{(1)}$ contains vertices $3,4$, and $\cham_6^{(2)}$ contains vertex $6$. From the general formula \eqref{V_n} one derives   the corresponding leg-factor 
\be
\label{V_6_125}
\rV_6^{\triple{125}}(\bm a|\bm x) 
=
X_{12}^{|\bm a_{1,2}|'} X_{25}^{|\bm a_{2,5}|'} X_{51}^{|\bm a_{5,1}|'}\, X_{13}^{-a_3} X_{14}^{-a_4} X_{26}^{-a_6} \,.
\ee
The cross-ratio set  $\bm \rY_6^{\triple{125}}$:
\be
\label{cross_6_125}
\ba{lll}
\dps
\rY^{\langle 125 \rangle}_1 = \frac{X_{23} X_{15}}{X_{13} X_{25}} \,, 
&\dps\qquad\rY^{\langle 125 \rangle}_2 = \frac{X_{12} X_{35}}{X_{13} X_{25}} \,, 
&\dps\qquad \rY^{\langle 125 \rangle}_3 =  \frac{X_{24} X_{15}}{X_{14} X_{25}} \,, 
\vspace{2mm}
\\
\dps
\rY^{\langle 125 \rangle}_4 = \frac{X_{12} X_{45}}{X_{14} X_{25}} \,, 
&\dps\qquad\rY^{\langle 125 \rangle}_5 =  \frac{X_{12} X_{15} X_{34}}{X_{13} X_{14} X_{25}} \,, \qquad
&\dps\qquad\rY^{\langle 125 \rangle}_6 =  \frac{X_{12} X_{56}}{X_{26} X_{15}} \,, 
\vspace{2mm}
\\
\dps
\rY^{\langle 125 \rangle}_7 =  \frac{X_{25} X_{16}}{X_{26} X_{15}} \,, 
&\dps\qquad\rY^{\langle 125 \rangle}_8 =   \frac{X_{12} X_{36}}{X_{13} X_{26}} \,, 
&\dps\qquad\rY^{\langle 125 \rangle}_9 =   \frac{X_{12} X_{46}}{X_{14} X_{26}} \,.
\ea
\ee
Note that  $\bm \rY_5^{\triple{125}} \subset \bm \rY_6^{\triple{125}}$. The transianic matrix $\cB\big(\bm \rY_6^{\triple{125}}\big)  =$
\be
\begin{tabular}{|c||c|c|c|c|c|c|c|c|c|}
\hline
 & $\rY^{\triple{125} }_1$ & $\rY^{\triple{125} }_2$ & $\rY^{\triple{125} }_3$ & $\rY^{\triple{125} }_4$ & $\rY^{\triple{125} }_5$ & $\rY^{\triple{125} }_6$ & $\rY^{\triple{125} }_7$ & $\rY^{\triple{125} }_8$ & $\rY^{\triple{125} }_9$  \\
\hline
\hline
$b_3$ & $1$ & $1$ & $0$ & $0$ & $1$ & $0$ & $0$ & $1$ & $0$ \\
\hline
\hline
$b_4$ & $0$ & $0$ & $1$ & $1$ & $1$ & $0$ & $0$ & $0$ & $1$ \\
\hline
\hline
$b_6$ & $0$ & $0$ & $0$ & $0$ & $0$ & $1$ & $1$ & $1$ & $1$ \\
\hline
\hline
$b_{12}$ & $0$ & $1$ & $0$ & $1$ & $1$ & $1$ & $0$ & $1$ & $1$ \\
\hline
\hline
$b_{25}$ & $-1$ & $-1$ & $-1$ & $-1$ & $-1$ & $0$ & $1$ & $0$ & $0$  \\
\hline
\hline
$b_{51}$ & $1$ & $0$ & $1$ & $0$ & $1$ & $-1$ & $-1$ & $0$ & $0$ \\
\hline
\end{tabular}
\ee

\item The basis triangle $\triangle_6^{\triple{135}}$, see fig. \bref{fig:triangles_hexagon} {\bf (d)}. This case is exceptional because all open chambers contain the same number of vertices (one vertex per chamber). It follows that the leg-factor read off from the general formula \eqref{V_n} is more symmetric compared to the previous cases,  
\be
\label{V_6_135}
\rV_6^{\triple{135}}(\bm a|\bm x) 
=
X_{13}^{|\bm a_{1,3}|'} X_{35}^{|\bm a_{3,5}|'} X_{51}^{|\bm a_{5,1}|'}\, X_{14}^{-a_4} X_{36}^{-a_6} X_{25}^{-a_2} \,.
\ee
Since each open chamber contains only one vertex, then there are no cubic cross-ratios and the cross-ratio set  $\bm \rY_6^{\triple{135}}$ is given by 
\be
\label{cross_6_135}
\begin{split}
\rY^{\langle 135 \rangle}_1 &=  \frac{X_{35} X_{12}}{X_{25} X_{13}} \,, \quad
\rY^{\langle 135 \rangle}_2 = \frac{X_{15} X_{23}}{X_{13} X_{25}} \,, \quad
\rY^{\langle 135 \rangle}_3 =  \frac{ X_{13} X_{45}}{ X_{14} X_{35} } \,, \\
\rY^{\langle 135 \rangle}_4 &= \frac{X_{15} X_{34}}{X_{14} X_{35}} \,, \quad
\rY^{\langle 135 \rangle}_5 =  \frac{X_{15} X_{24}}{X_{14} X_{25}} \,, \quad
\rY^{\langle 135 \rangle}_6 = \frac{X_{13} X_{56}}{X_{15} X_{36}} \,,   \\
\rY^{\langle 135 \rangle}_7 &=  \frac{X_{35} X_{16}}{X_{15} X_{36}} \,, \quad
\rY^{\langle 135 \rangle}_8 = \frac{X_{35} X_{26}}{X_{25} X_{36}} \,, \quad
\rY^{\langle 135 \rangle}_9 = \frac{X_{13} X_{46}}{X_{14} X_{36}} \,,
\end{split}
\ee
Note that  $\bm \rY_5^{\triple{135}} \subset \bm \rY_6^{\triple{135}}$. The transianic matrix $\cB\big(\bm \rY_6^{\triple{135}}\big)  =$
\be
\begin{tabular}{|c||c|c|c|c|c|c|c|c|c|}
\hline
 & $\rY^{\triple{135} }_1$ & $\rY^{\triple{135} }_2$ & $\rY^{\triple{135} }_3$ & $\rY^{\triple{135} }_4$ & $\rY^{\triple{135} }_5$ & $\rY^{\triple{135} }_6$ & $\rY^{\triple{135} }_7$ & $\rY^{\triple{135} }_8$ & $\rY^{\triple{135} }_9$  \\
\hline
\hline
$b_2$ & $1$ & $1$ & $0$ & $0$ & $1$ & $0$ & $0$ & $1$ & $0$ \\
\hline
\hline
$b_4$ & $0$ & $0$ & $1$ & $1$ & $1$ & $0$ & $0$ & $0$ & $1$ \\
\hline
\hline
$b_6$ & $0$ & $0$ & $0$ & $0$ & $0$ & $1$ & $1$ & $1$ & $1$ \\
\hline
\hline
$b_{13}$ & $-1$ & $-1$ & $1$ & $0$ & $0$ & $1$ & $0$ & $0$ & $1$ \\
\hline
\hline
$b_{35}$ & $1$ & $0$ & $-1$ & $-1$ & $0$ & $0$ & $1$ & $1$ & $0$  \\
\hline
\hline
$b_{51}$ & $0$ & $1$ & $0$ & $1$ & $1$ & $-1$ & $-1$ & $0$ & $0$ \\
\hline
\end{tabular}
\ee

\end{itemize}

\noindent Using the general formulas \eqref{bare_n_1}--\eqref{bare_n_3} we build  four  polygonal functions:\footnote{The power series \eqref{H6_123} is of the Srivastava-Daoust type \cite{SrivastavaDaoust, Srivastava1985MultipleGH}.}
\begin{multline}
\label{H6_123}
\HG_6^{\triple{123}}\left( {\bm a}\big|\rY^{\langle 123 \rangle }_1\,, ...\,, \rY^{\langle 123 \rangle }_9 \right)
=
\sum_{m_1,..., m_9=0}^\infty \;\; \prod_{l=1}^9 \frac{\left( \rY_l^{\langle 123 \rangle }\right)^{m_l}}{m_l!}
\\
\times
\frac{
(a_4)_{m_1+m_2+m_5+m_8} 
(a_5)_{m_3+m_4+m_5+m_9} 
(a_6)_{m_6+m_7+m_8+m_9}  \,  
(-)^{m_1+m_2+m_3+m_4+m_6+m_7} 
}{
(1+|\bm a_{1,2}|')_{m_1+m_3+m_5+m_6+m_8+m_9}
(1+|\bm a_{2,3}|')_{m_2+m_4+m_5+m_7+m_8+m_9} 
(1+|\bm a_{3,1}|')_{-m_1-...-m_9} 
} \,.
\end{multline}
\begin{multline}
\label{H6_124}
\HG_6^{\triple{124}}\left( {\bm a}\big|\rY^{\langle 124 \rangle }_1\,, ...\,, \rY^{\langle 124 \rangle }_9 \right)
=
\sum_{m_1,..., m_9=0}^\infty \;\; \prod_{l=1}^9 \frac{\left( \rY_l^{\langle 124 \rangle }\right)^{m_l}}{m_l!}
\\
\times
\frac{
(a_3)_{m_1+m_2+m_5+m_8} 
(a_5)_{m_3+m_4+m_5+m_9} 
(a_6)_{m_6+m_7+m_8+m_9}  \,  
(-)^{m_1+m_2+m_3+m_4+m_6+m_7} 
}{
(1+|\bm a_{1,2}|')_{m_2+m_4+m_5+m_6+m_8+m_9}
(1+|\bm a_{2,4}|')_{-m_1-m_2+m_3+m_7+m_9} 
(1+|\bm a_{4,1}|')_{m_1-m_3-m_4-m_6-m_7-m_9} 
} \,.
\end{multline}
\begin{multline}
\label{H6_125}
\HG_6^{\triple{125}}\left( {\bm a}\big|\rY^{\triple{125} }_1\,, ...\,, \rY^{\triple{125} }_9 \right)
=
\sum_{m_1,..., m_9=0}^\infty \;\; \prod_{l=1}^9 \frac{\left( \rY_l^{\triple{125} }\right)^{m_l}}{m_l!}
\\
\times
\frac{
(a_3)_{m_1+m_2+m_5+m_8} 
(a_4)_{m_3+m_4+m_5+m_9} 
(a_6)_{m_6+m_7+m_8+m_9}  \,  
(-)^{m_1+m_2+m_3+m_4+m_6+m_7} 
}{
(1+|\bm a_{1,2}|')_{m_2+m_4+m_5+m_6+m_8+m_9}
(1+|\bm a_{2,5}|')_{-m_1-m_2-m_3-m_4-m_5+m_7} 
(1+|\bm a_{5,1}|')_{m_1+m_3+m_5-m_6-m_7} 
} \,.
\end{multline}
\begin{multline}
\label{H6_135}
\HG_6^{\triple{135}}\left( {\bm a}\big|\rY^{\triple{135} }_1\,, ...\,, \rY^{\triple{135} }_9 \right)
=
\sum_{m_1,..., m_9=0}^\infty  \;\;\prod_{l=1}^9 \frac{\left( \rY_l^{\triple{135} }\right)^{m_l}}{m_l!}
\\
\times
\frac{
(a_2)_{m_1+m_2+m_5+m_8} 
(a_4)_{m_3+m_4+m_5+m_9} 
(a_6)_{m_6+m_7+m_8+m_9}  \,  
(-)^{m_1+m_2+m_3+m_4+m_6+m_7} 
}{
(1+|\bm a_{1,3}|')_{-m_1-m_2+m_3+m_6+m_9}
(1+|\bm a_{3,5}|')_{m_1-m_3-m_4+m_7+m_8} 
(1+|\bm a_{5,1}|')_{m_2+m_4+m_5-m_6-m_7} 
} \,.
\end{multline}
This step completes the construction of master functions \eqref{master_6}. The reconstruction formula \eqref{rep2} now represents the hexagon conformal integral as 
\be
\label{hexagon_reconstruction}
\ba{l}
\dps
I_6^{\bm a}(\bm x) 
=  \sum_{l=0}^{5} (\cycle_6)^l \circ \left( \BF_6^{\triple{123}}({\bm a}|\bm x)  + \BF_6^{\triple{124}}({\bm a}|\bm x) + \BF_6^{\triple{125}}({\bm a}|\bm x) + \BF_6^{\triple{135}}({\bm a}|\bm x) \right) \\
=
\BF_6^{\triple{123}}({\bm a}|\bm x) + \BF_6^{\triple{234}}({\bm a}|\bm x) + \BF_6^{\triple{345}}({\bm a}|\bm x) + \BF_6^{\triple{456}}({\bm a}|\bm x) + \BF_6^{\triple{156}}({\bm a}|\bm x)+\BF_6^{\triple{126}}({\bm a}|\bm x) \\
+  
\BF_6^{\triple{124}}({\bm a}|\bm x) + \BF_6^{\triple{235}}({\bm a}|\bm x) + \BF_6^{\triple{346}}({\bm a}|\bm x) + \BF_6^{\triple{145}}({\bm a}|\bm x) + \BF_6^{\triple{256}}({\bm a}|\bm x)+\BF_6^{\triple{136}}({\bm a}|\bm x) \\
+  
\BF_6^{\triple{125}}({\bm a}|\bm x) + \BF_6^{\triple{236}}({\bm a}|\bm x) + \BF_6^{\triple{134}}({\bm a}|\bm x) + \BF_6^{\triple{245}}({\bm a}|\bm x) + \BF_6^{\triple{356}}({\bm a}|\bm x)+\BF_6^{\triple{146}}({\bm a}|\bm x) \\
+  
\BF_6^{\triple{135}}({\bm a}|\bm x) + \BF_6^{\triple{246}}({\bm a}|\bm x)\,.
\ea
\ee
The orbit structure of the index set $\rR_6$ \eqref{R_6}  is inherited here. The $\ZZ_6$-orbits of basis functions are listed below.  

\begin{itemize}
\item Basis functions labelled by elements of the orbit $\{\ZZ_6 \circ \triple{123} \}$:
\be
\label{basis_6_orbit_123}
\begin{split}
\BF_6^{\triple{123}}({\bm a}|\bm x) &= \BF_6^{\triple{123}}((\cycle_6)^0 \circ {\bm a}|(\cycle_6)^0 \circ \bm x)\,, \\
\BF_6^{\triple{234}}({\bm a}|\bm x) &= \BF_6^{\triple{123}}((\cycle_6)^1 \circ {\bm a}|(\cycle_6)^1 \circ \bm x)\,, \\
\BF_6^{\triple{345}}({\bm a}|\bm x) &= \BF_6^{\triple{123}}((\cycle_6)^2 \circ {\bm a}|(\cycle_6)^2 \circ \bm x)\,, \\
\BF_6^{\triple{456}}({\bm a}|\bm x) &= \BF_6^{\triple{123}}((\cycle_6)^3 \circ {\bm a}|(\cycle_6)^3 \circ \bm x)\,, \\
\BF_6^{\triple{156}}({\bm a}|\bm x) &= \BF_6^{\triple{123}}((\cycle_6)^4 \circ {\bm a}|(\cycle_6)^4 \circ \bm x)\,, \\
\BF_6^{\triple{126}}({\bm a}|\bm x) &= \BF_6^{\triple{123}}((\cycle_6)^5 \circ {\bm a}|(\cycle_6)^5 \circ \bm x)\,.
\end{split}
\ee

\item Basis functions labelled by elements of the orbit $\{\ZZ_6 \circ \triple{124} \}$:
\be
\label{basis_6_orbit_124}
\begin{split}
\BF_6^{\triple{124}}({\bm a}|\bm x) &= \BF_6^{\triple{124}}((\cycle_6)^0 \circ {\bm a}|(\cycle_6)^0 \circ \bm x)\,, \\
\BF_6^{\triple{235}}({\bm a}|\bm x) &= \BF_6^{\triple{124}}((\cycle_6)^1 \circ {\bm a}|(\cycle_6)^1 \circ \bm x)\,, \\
\BF_6^{\triple{346}}({\bm a}|\bm x) &= \BF_6^{\triple{124}}((\cycle_6)^2 \circ {\bm a}|(\cycle_6)^2 \circ \bm x)\,, \\
\BF_6^{\triple{145}}({\bm a}|\bm x) &= \BF_6^{\triple{124}}((\cycle_6)^3 \circ {\bm a}|(\cycle_6)^3 \circ \bm x)\,, \\
\BF_6^{\triple{256}}({\bm a}|\bm x) &= \BF_6^{\triple{124}}((\cycle_6)^4 \circ {\bm a}|(\cycle_6)^4 \circ \bm x)\,, \\
\BF_6^{\triple{136}}({\bm a}|\bm x) &= \BF_6^{\triple{124}}((\cycle_6)^5 \circ {\bm a}|(\cycle_6)^5 \circ \bm x)\,.
\end{split}
\ee

\item Basis functions labelled by elements of the orbit $\{\ZZ_6 \circ \triple{125} \}$: 
\be
\label{basis_6_orbit_125}
\begin{split}
\BF_6^{\triple{125}}({\bm a}|\bm x) &= \BF_6^{\triple{125}}((\cycle_6)^0 \circ {\bm a}|(\cycle_6)^0 \circ \bm x)\,, \\
\BF_6^{\triple{236}}({\bm a}|\bm x) &= \BF_6^{\triple{125}}((\cycle_6)^1 \circ {\bm a}|(\cycle_6)^1 \circ \bm x)\,, \\
\BF_6^{\triple{134}}({\bm a}|\bm x) &= \BF_6^{\triple{125}}((\cycle_6)^2 \circ {\bm a}|(\cycle_6)^2 \circ \bm x)\,, \\
\BF_6^{\triple{245}}({\bm a}|\bm x) &= \BF_6^{\triple{125}}((\cycle_6)^3 \circ {\bm a}|(\cycle_6)^3 \circ \bm x)\,, \\
\BF_6^{\triple{356}}({\bm a}|\bm x) &= \BF_6^{\triple{125}}((\cycle_6)^4 \circ {\bm a}|(\cycle_6)^4 \circ \bm x)\,, \\
\BF_6^{\triple{146}}({\bm a}|\bm x) &= \BF_6^{\triple{125}}((\cycle_6)^5 \circ {\bm a}|(\cycle_6)^5 \circ \bm x)\,.
\end{split}
\ee

\item Basis functions labelled by elements of the orbit $\{\ZZ_6 \circ \triple{135} \}$: 
\be
\label{basis_6_orbit_135}
\begin{split}
\BF_6^{\triple{135}}({\bm a}|\bm x) = \BF_6^{\triple{135}}((\cycle_6)^0 \circ {\bm a}|(\cycle_6)^0 \circ \bm x)\,, \\
\BF_6^{\triple{246}}({\bm a}|\bm x) = \BF_6^{\triple{135}}((\cycle_6)^1 \circ {\bm a}|(\cycle_6)^1 \circ \bm x)\,.
\end{split}
\ee
\end{itemize}
\noindent The basis functions in \eqref{basis_6_orbit_123}--\eqref{basis_6_orbit_135} are triple products \eqref{rep2}. The triangle-factors are directly read off from \eqref{star_n}, while the leg-factors and the polygonal functions are explicitly given in Appendix \bref{app:hexagon}. 

The convergence domains can be examined along the lines of  section \bref{sec:domain}. In particular, all convergence indices for the master functions are zero. However, similar to the pentagon case, further analysis is quite difficult  due to the large set of variables and basis functions.

The 18 basis functions \eqref{basis_6_orbit_123}, \eqref{basis_6_orbit_124},  \eqref{basis_6_orbit_125} were obtained  in \cite{Alkalaev:2025fgn} using the bipartite representation, where it was also argued that their linear combination violates the reduction condition \eqref{in_reduction}. In this paper, following the diagrammatic algorithm, we  found 2 more basis functions \eqref{basis_6_orbit_135}. Thus, the reconstruction formula \eqref{hexagon_reconstruction} represents the hexagon conformal integral as the sum of 20 basis functions. Given the general reduction properties \eqref{reduction_prefactors1}--\eqref{reduction} it is very straightforward to check that when any one of propagator powers is set to zero the complete formula \eqref{hexagon_reconstruction} reproduces the 10-term expression for the pentagon conformal integral \eqref{pentagon_reconstruction}.

Note that previously the hexagon conformal integral was evaluated  as the sum of 26  and 25  multivariate hypergeometric series in \cite{Ananthanarayan:2020ncn} and \cite{Banik:2023rrz},  respectively. It would be important to find  suitable  transformations of  the corresponding hypergeometric series  that  would reconcile these three representations of the hexagon conformal integral.\footnote{This can be quite a difficult task. However,  significant progress has been made recently in the study of transformation formulas  and analytic  continuation formulas for various hypergeometric functions, see e.g. \cite{Bezrodnykh:2018review, Ananthanarayan:2020xpd, Ananthanarayan:2021yar}} On the other hand, it is natural to assume that among all possible representations of conformal integrals by linear combinations of generalized hypergeometric functions there is a sort of {\it canonical} representation, which is expected to contain a particular set of basis functions in a given convergence domain such that the required permutation symmetries and the reduction conditions are {\it manifest}. We believe that our way to enumerate basis functions by index triples is a reasonable candidate.

\section{Conclusions and outlooks}
\label{sec:concl}

The diagrammatic algorithm proposed in this paper introduces a new parametric class of polygonal hypergeometric functions. They are completely defined in terms of plane geometry by considering  conformal polygons, inscribed triangles, cross-ratio colored diagrams and calculating transianic indices, or, equivalently, transianic matrices.\footnote{In this respect, the transianic matrix is similar to the toric matrix used in the GKZ hypergeometric systems.} All these ingredients are associated to some Baxter lattice.

The  explicit geometric origin of the diagrammatic approach  suggests that polygonal functions are deeply related to a number of issues arising in the study of conformal integrals.  E.g., in this paper we argue    that the polygonal functions calculate the one-loop multipoint conformal integrals by means of the conjectured reconstruction formula. Furthermore, the diagrammatic algorithm should manifest itself within the Yangian bootstrap \cite{Chicherin:2017frs, Chicherin:2017cns, Loebbert:2019vcj} since the basis functions being part of the conformal integrals should solve the Yangian constraints. On the other hand, we believe that our geometric constructions involving elements of the Baxter integrable lattices \cite{Baxter:1978xr, Zamolodchikov:1980mb} and, more generally, of the loom construction \cite{Kazakov:2022dbd,Kazakov:2023nyu,Alfimov:2023vev,Levkovich-Maslyuk:2024zdy} should be  related to representing conformal integrals as volumes of simplices in   spaces of constant curvature  \cite{Davydychev:1997wa,Mason:2010pg,Schnetz:2010pd,Nandan:2013ip, Bourjaily:2019exo, Ren:2023tuj}, because the conformal polygons  can be viewed as sections of polytopes.

The present construction of polygonal functions is formal in the sense that in order to define a power series completely one also has to specify its convergence region. For the conformal integrals this would mean that one can describe fractions of the kinematical space covered by the reconstruction formula \eqref{conj_fin}. Indeed, the convergence domain of the conformal integral is determined by intersecting convergence domains of basis functions involved in the reconstruction formula. Beyond the box conformal integrals given by hypergeometric double  series this problem is poorly understood. There are general theorems in the theory of multivariate hypergeometric functions (of Horn's type) which  make it possible (in principle) to establish domains of convergence, see e.g. \cite{exton1976multiple,Gelfand1990,Gelfand1992,tsikh}. In practice, finding convergence domains for a given power series explicitly is a highly non-trivial problem due to intricate parameterizations of the convergence radii in each variable. Partial results for particular  multivariate hypergeometric  functions can be found e.g. in \cite{Srivastava1972,Friot:2011ic,Bezrodnykh:2018review} . In the present context, already the pentagon conformal integral is given by ten parametric power series of five variables which obviously requires a separate study. In section \bref{sec:domain} we took the first steps in this direction and introduced the convergence indices for the polygonal functions. Based on the lower-point examples in section \bref{sec:examples} we expect that the diagrammatic algorithm necessarily leads to zero convergence indices, which implies that the basis functions have some non-trivial convergence domains in the coordinate space. We plan to consider these issues in more detail elsewhere.


There are a few possible future directions for this work. E.g. one can extend our analysis to conformal integrals in  $D$-dimensional spaces of  Lorentz signature to see what modifications are possible, see e.g. \cite{Corcoran:2020akn,Corcoran:2020epz,Corcoran:2023ljn}. Also, the multi-loop conformal integrals \cite{Paulos:2012nu,Basso:2017jwq,Derkachov:2018rot,Duhr:2023bku,Loebbert:2024fsj, Olivucci:2021cfy, Derkachov:2021ufp, Olivucci:2023tnw, Aprile:2023gnh, Loebbert:2025abz} is another  interesting point of application. Here, the diagrammatic algorithm is expected to give new classes of hypergeometric functions as well as the reconstruction formulas. Also, one may consider non-conformal integrals as some of them are related to conformal integrals with one point sent to infinity, see e.g. \cite{Drummond:2006rz, Usyukina:1992jd}. It should be stressed, however, that the starting point of our diagrammatic algorithm is the  Baxter lattice which encodes the conformality constraint through the total angle condition  \eqref{angle}. Nonetheless, it would be interesting to consider possible geometric implementations of non-conformality. Finally, one may wonder how the polygonal functions are defined within the GKZ hypergeometric systems. Among other things, this may help to find their domains of convergence  discussed above, see e.g. \cite{tsikh,delaCruz:2019skx,Klausen:2019hrg,Weinzierl:2022eaz}.

\vspace{4mm} 

\noindent \textbf{Acknowledgements.} We are grateful to Mikhail Alfimov, Sergey Derkachev, Alexey Isaev, Wladyslaw Wachowski for useful discussions and to Ekaterina Mandrygina for her help in drawing diagrams. S.M. also thanks his mother Galina for pointing out the possibility to come up with new terminology. Our work was supported by the Foundation for the Advancement of Theoretical Physics and Mathematics “BASIS”.

\appendix

\section{Cross-ratio orbits}
\label{app:orbit}

Let $X = \{x_g: g\in G\}$ be a set of elements parametrized by the symmetric group $G = \cS_p$. By construction,  the set  $X$ forms (the basis of) the regular representation of the symmetric group and, therefore,  $|X| = p!$.  To describe the action of $G$ on $X$ one uses the orbit-stabilizer theorem which says that the set of cosets $G/G_x$ for the stabilizer subgroup $G_x\subset G$ is in one-to-one correspondence with the $G$-orbit $G\cdot x$. The orbit length is $|G\cdot x| = p!/|G_x|$. 

Now we apply this standard construction to the cross-ratio functions \eqref{U_W}, where a set $X$ will be identified with functions $U[i_1,..., i_4]$ or $W[j_1,..., j_5]$, while $G$ will be  $\cS_4$ or $\cS_5$. We will see that  cross-ratios for a given set of indices are enumerated by the respective cosets. This approach describes all possible cross-ratios which can be built from four or five points in $\RR^D$.  

Let $X_4$ and $X_5$ be two sets of cross-ratios which form the regular representations of $\cS_4$ and $\cS_5$. The stabilizers are defined as follows 
\be
\ba{l}
\label{sym_UW}
\tilde\pi \in G_x \;: \qquad  \tilde\pi \circ U[i_1, i_2, i_3, i_4]  = U[i_1, i_2, i_3, i_4]\,,
\vspace{2mm}
\\
\tilde\pi \in G_x\; : \qquad  \tilde\pi \circ W[j_1, j_2, j_3, j_4, j_5]  = W[j_1, j_2, j_3, j_4, j_5]\,, 
\ea
\ee
where both stabilizers turn out to be the Klein four-group $G_x \cong V \cong  \ZZ_2 \times \ZZ_2$ which has two different permutation realizations
\be
V  = \{e, \,\sigma_{12}\sigma_{34},\, \sigma_{13}\sigma_{42},\, \sigma_{14}\sigma_{23}\}\subset \cS_4
\quad \text{and}\quad
V =  \{e,\, \sigma_{13}\,,\sigma_{45}\,,\sigma_{13}\sigma_{45}\} \subset \cS_5\,.
\ee
Since $|V| = 4$ one finds out that the orbit lengths are equal to $4!/4 = 6$ and $5!/4 = 30$.

\begin{figure}
\centering
\includegraphics[scale=0.15]{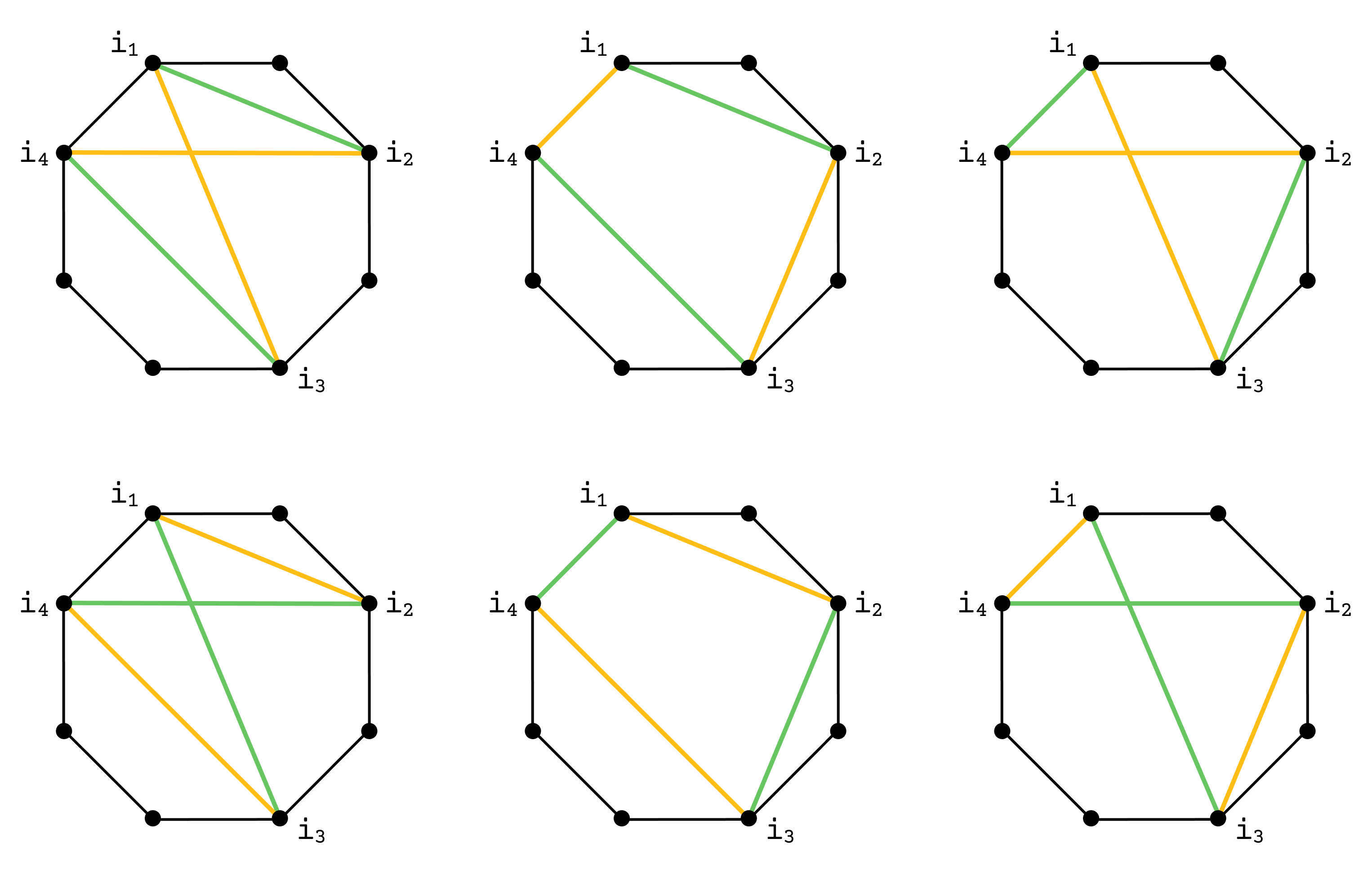}
\caption{The orbit of the symmetric group $\cS_4$ acting  on the set of quadratic cross-ratios associated to four vertices labelled by  $i_1, ..., i_4$. The orbit length equals $6$.  The second raw is obtained from the first one by inverting the colors.} 
\label{fig:quadra}
\end{figure}
\begin{figure}
\centering
\includegraphics[scale=0.15]{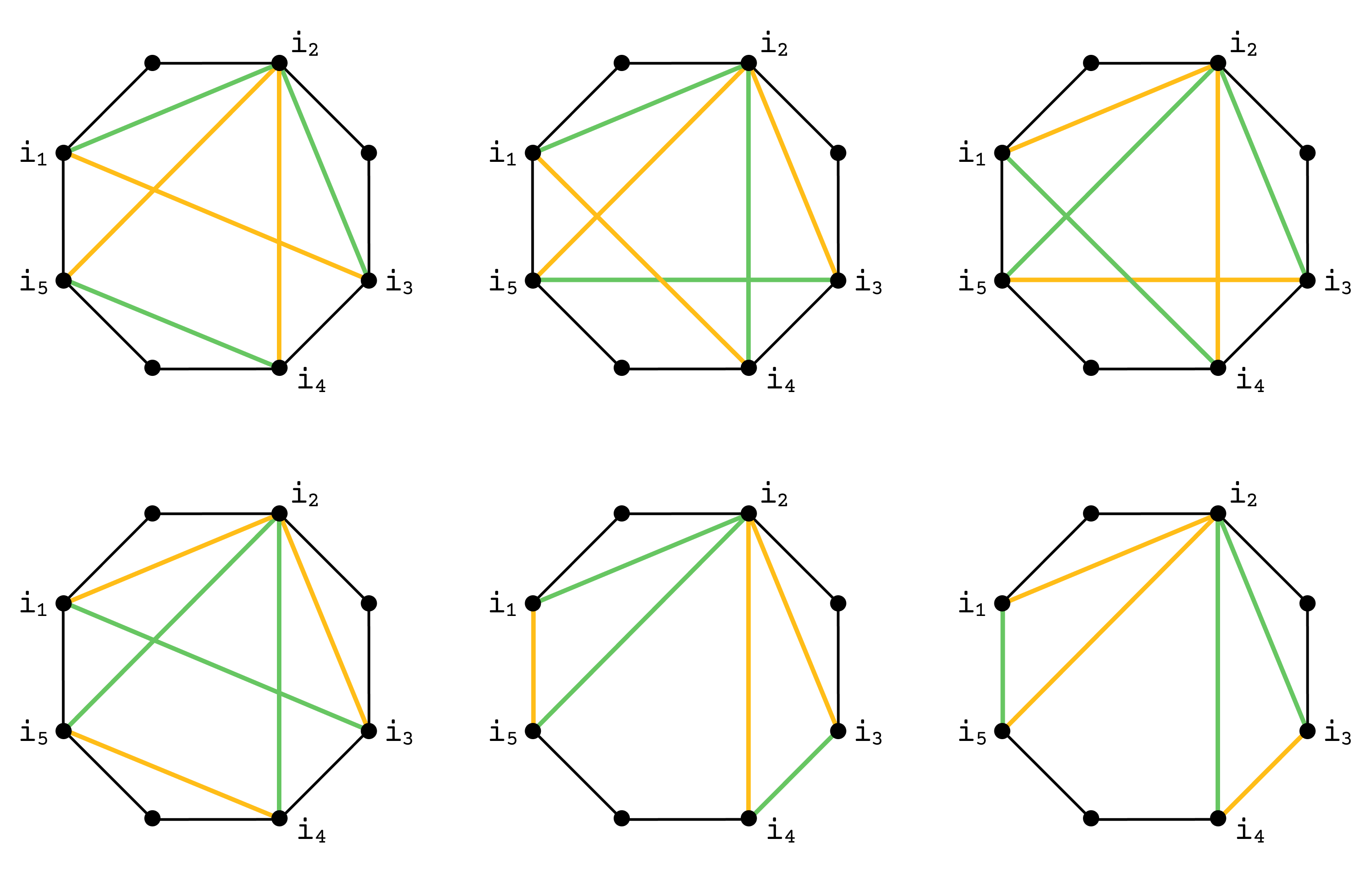}
\caption{The first six cubic cross-ratios in the $\cS_5$-orbit. All other elements  are obtained by simultaneous rotations of the pentagon vertices.} 
\label{fig:pentas}
\end{figure}

The graphical realization of  the orbits is now manifest. E.g. the $\cS_4$-orbit for a quadratic cross-ratio $U[i_1, i_2, i_3, i_4] \in X_4$ is built by acting with permutations from the coset set $\cS_4/V$: 
\be
\label{six_U}
\ba{c}
U[i_1, i_2, i_3, i_4]  = e \circ  U[i_1, i_2, i_3, i_4]\;, 
\quad 
U[i_2, i_1, i_3, i_4]  = \sigma_{12} \circ  U[i_1, i_2, i_3, i_4]\;,
\\
U[i_3, i_2, i_1, i_4]  = \sigma_{13} \circ  U[i_1, i_2, i_3, i_4]\;,
\quad
U[i_1, i_3, i_2, i_4]  = \sigma_{23} \circ  U[i_1, i_2, i_3, i_4]\;,
\\
U[i_1, i_4, i_2, i_3]  = \sigma_{14} \circ  U[i_1, i_2, i_3, i_4]\;,
\quad 
U[i_1, i_4, i_3, i_2]  = \sigma_{24} \circ  U[i_1, i_2, i_3, i_4]\;.

\ea
\ee
The corresponding cross-ratio diagrams are listed  in fig. \bref{fig:quadra} from left to right.  

Similarly, one builds the $\cS_5$-orbit of  $W[i_1, i_2, i_3, i_4, i_5] \in X_5$ which is generated by the cosets $\cS_5/V$. The orbit length equals 30 and the first 6 elements shown in fig. \bref{fig:pentas} (from left to right) are obtained  as follows
\be
\label{six_W}
\ba{l}
W[i_1, i_2, i_3, i_4, i_5]  = e \circ  W[i_1, i_2, i_3, i_4, i_5]\;, 
\quad \;\;\;
W[i_1, i_2, i_4, i_3, i_5]  = \sigma_{34} \circ  W[i_1, i_2, i_3, i_4, i_5]\;,
\\
W[i_5, i_2, i_3, i_4, i_1]  = \sigma_{15} \circ  W[i_1, i_2, i_3, i_4, i_5]\;,
\quad
W[i_4, i_2, i_5, i_1, i_3]  = \sigma_{14}\sigma_{35}\circ  W[i_1, i_2, i_3, i_4, i_5]\;,
\\
W[i_1, i_2, i_5, i_4, i_3]  = \sigma_{35} \circ  W[i_1, i_2, i_3, i_4, i_5]\;,
\quad 
W[i_4, i_2, i_3, i_1, i_5]  = \sigma_{14} \circ  W[i_1, i_2, i_3, i_4, i_5]\;.
\ea
\ee
Other 4 series of 6 cross-ratios are obtained by cyclic permutations $C_5 \in \cS_5$ which sequentially rotate  vertices  as $j_2 \to j_3 \to j_4 \to j_5 \to j_1$.

\section{Explicit expressions for the lower-point polygonal functions}
\label{app:expl}

\subsection{Pentagon}
\label{app:pentagon}

\subsubsection*{Orbit $\{\ZZ_5 \circ \triple{123} \}$:}

\paragraph{$\triple{234}$.} The leg-factor is 
\be
\label{V5_234}
\rV_5^{\triple{234}}(\bm a|\bm x) =
X_{23}^{|\bm a_{2,3}|'} X_{34}^{|\bm a_{3,4}|'}  X_{42}^{|\bm a_{4,2}|'}\, X_{35}^{-a_5} X_{13}^{-a_1}\,.
\ee
The cross-ratio set is produced from \eqref{cross_5_123}: $\bm \rY_5^{\triple{234}} = \{ \cycle_5 \circ \rY^{\triple{123}}_l, l=1,...,5\}$. The polygonal function is
\begin{multline}
\label{H5_234}
\HG_5^{\triple{234}}\left(\bm a| \bm \rY_5^{\triple{234}} \right)  
= \sum_{m_1,...,m_5=0}^{\infty} \;\;
\prod_{l=1}^{5} \frac{\left(\cycle_5 \circ \rY^{\triple{123}}_l \right)^{m_l}}{m_l!}\\
\times \frac{(-)^{m_1+m_2+m_3+m_4} (a_5)_{m_1+m_2+m_5} (a_1)_{m_3+m_4+m_5} }{(1+|\bm a_{2,3}|')_{m_1+m_3+m_5} (1+|\bm a_{3,4}|')_{m_2+m_4+m_5} (1+|\bm a_{4,2}|')_{-m_1-m_2-m_3-m_4-m_5} } \,,
\end{multline}
where 
\be
\label{cross_5_234}
\begin{split}
\cycle_5 \circ \rY^{\langle 123 \rangle }_1 &=  \frac{X_{23} X_{45} }{X_{24} X_{35}}\,, 
\qquad 
\cycle_5 \circ \rY^{\langle 123 \rangle }_2  =  \frac{X_{34} X_{25} }{X_{24} X_{35}}\,, \\
\cycle_5 \circ \rY^{\langle 123 \rangle }_3 &=  \frac{X_{23} X_{14} }{X_{24} X_{13}}\,, 
\qquad
\cycle_5 \circ \rY^{\langle 123 \rangle }_4  = \frac{X_{34} X_{12} }{X_{24} X_{13}}\,,  \\
\cycle_5 \circ \rY^{\langle 123 \rangle }_5 &=  \frac{X_{23} X_{34} X_{15} }{X_{24} X_{35} X_{13}}\,.
\end{split}
\ee

\paragraph{$\triple{345}$.}The leg-factor is 
\be
\label{V5_345}
\rV_5^{\triple{345}}(\bm a|\bm x) =
X_{34}^{|\bm a_{3,4}|'} X_{45}^{|\bm a_{4,5}|'}  X_{53}^{|\bm a_{5,3}|'}\, X_{14}^{-a_1} X_{24}^{-a_2}\,.
\ee
The cross-ratio set is produced from \eqref{cross_5_123}: $\bm \rY_5^{\triple{345}} = \{ (\cycle_5)^2 \circ \rY^{\triple{123}}_l, l=1,...,5\}$. The polygonal  function is
\begin{multline}
\label{H5_345}
\HG_5^{\triple{345}}\left(\bm a| \bm \rY_5^{\triple{345}} \right)  
= \sum_{m_1,...,m_5=0}^{\infty} \;\;
\prod_{l=1}^{5} \frac{\left((\cycle_5)^2 \circ \rY^{\triple{123}}_l \right)^{m_l}}{m_l!}\\
\times \frac{(-)^{m_1+m_2+m_3+m_4} (a_1)_{m_1+m_2+m_5} (a_2)_{m_3+m_4+m_5} }{(1+|\bm a_{3,4}|')_{m_1+m_3+m_5} (1+|\bm a_{4,5}|')_{m_2+m_4+m_5} (1+|\bm a_{5,3}|')_{-m_1-m_2-m_3-m_4-m_5} } \,,
\end{multline}
where 
\be
\label{cross_5_345}
\begin{split}
(\cycle_5)^2 \circ \rY^{\langle 123 \rangle }_1 &=  \frac{X_{34} X_{15} }{X_{35} X_{14}}\,, 
\qquad 
(\cycle_5)^2 \circ \rY^{\langle 123 \rangle }_2  =  \frac{X_{45} X_{13} }{X_{35} X_{14}}\,, \\
(\cycle_5)^2 \circ \rY^{\langle 123 \rangle }_3 &=  \frac{X_{34} X_{25} }{X_{35} X_{24}}\,, 
\qquad
(\cycle_5)^2 \circ \rY^{\langle 123 \rangle }_4  = \frac{X_{45} X_{23} }{X_{35} X_{24}}\,,  \\
(\cycle_5)^2 \circ \rY^{\langle 123 \rangle }_5 &=  \frac{X_{34} X_{45} X_{12} }{X_{35} X_{14} X_{24}}\,.
\end{split}
\ee

\paragraph{$\triple{145}$.}The leg-factor is 
\be
\label{V5_145}
\rV_5^{\triple{145}}(\bm a|\bm x) =
X_{45}^{|\bm a_{4,5}|'} X_{51}^{|\bm a_{5,1}|'}  X_{14}^{|\bm a_{1,4}|'}\, X_{25}^{-a_2} X_{35}^{-a_3}\,.
\ee
The cross-ratio set is produced from \eqref{cross_5_123}: $\bm \rY_5^{\triple{145}} = \{ (\cycle_5)^3 \circ \rY^{\triple{123}}_l, l=1,...,5\}$. The polygonal  function is
\begin{multline}
\label{H5_145}
\HG_5^{\triple{145}}\left(\bm a| \bm \rY_5^{\triple{145}} \right)  
= \sum_{m_1,...,m_5=0}^{\infty} \;\;
\prod_{l=1}^{5} \frac{\left((\cycle_5)^3 \circ \rY^{\triple{123}}_l \right)^{m_l}}{m_l!}\\
\times \frac{(-)^{m_1+m_2+m_3+m_4} (a_2)_{m_1+m_2+m_5} (a_3)_{m_3+m_4+m_5} }{(1+|\bm a_{4,5}|')_{m_1+m_3+m_5} (1+|\bm a_{5,1}|')_{m_2+m_4+m_5} (1+|\bm a_{1,4}|')_{-m_1-m_2-m_3-m_4-m_5} } \,,
\end{multline}
where
\be
\label{cross_5_145}
\begin{split}
(\cycle_5)^3 \circ \rY^{\langle 123 \rangle }_1 &=  \frac{X_{45} X_{12} }{X_{14} X_{25}}\,, 
\qquad
(\cycle_5)^3 \circ \rY^{\langle 123 \rangle }_2  =  \frac{X_{15} X_{24} }{X_{14} X_{25}}\,, \\
(\cycle_5)^3 \circ \rY^{\langle 123 \rangle }_3 &=  \frac{X_{45} X_{13} }{X_{14} X_{35}}\,, 
\qquad
(\cycle_5)^3 \circ \rY^{\langle 123 \rangle }_4  = \frac{X_{15} X_{34} }{X_{14} X_{35}}\,,  \\
(\cycle_5)^3 \circ \rY^{\langle 123 \rangle }_5 &=  \frac{X_{45} X_{15} X_{23} }{X_{14} X_{25} X_{35}}\,.
\end{split}
\ee

\paragraph{$\triple{125}$.}The leg-factor is 
\be
\label{V5_125}
\rV_5^{\triple{125}}(\bm a|\bm x) =
X_{51}^{|\bm a_{5,1}|'} X_{12}^{|\bm a_{1,2}|'}  X_{25}^{|\bm a_{2,5}|'}\, X_{13}^{-a_3} X_{14}^{-a_4}\,.
\ee
The cross-ratio set is produced from \eqref{cross_5_123}: $\bm \rY_5^{\triple{125}} = \{ (\cycle_5)^4 \circ \rY^{\triple{123}}_l, l=1,...,5\}$. The polygonal  function is
\begin{multline}
\label{H5_125}
\HG_5^{\triple{125}}\left(\bm a| \bm \rY_5^{\triple{125}} \right)  
= \sum_{m_1,...,m_5=0}^{\infty} \;\;
\prod_{l=1}^{5} \frac{\left((\cycle_5)^4 \circ \rY^{\triple{123}}_l \right)^{m_l}}{m_l!}\\
\times \frac{(-)^{m_1+m_2+m_3+m_4} (a_3)_{m_1+m_2+m_5} (a_4)_{m_3+m_4+m_5} }{(1+|\bm a_{5,1}|')_{m_1+m_3+m_5} (1+|\bm a_{1,2}|')_{m_2+m_4+m_5} (1+|\bm a_{2,5}|')_{-m_1-m_2-m_3-m_4-m_5} } \,,
\end{multline}
where
\be
\label{cross_5_125}
\begin{split}
(\cycle_5)^4 \circ \rY^{\langle 123 \rangle }_1 &=  \frac{X_{15} X_{23} }{X_{25} X_{13}}\,, 
\qquad
(\cycle_5)^4 \circ \rY^{\langle 123 \rangle }_2  =  \frac{X_{12} X_{35} }{X_{25} X_{13}}\,, \\
(\cycle_5)^4 \circ \rY^{\langle 123 \rangle }_3 &=  \frac{X_{15} X_{24} }{X_{25} X_{14}}\,, 
\qquad
(\cycle_5)^4 \circ \rY^{\langle 123 \rangle }_4  = \frac{X_{12} X_{45} }{X_{25} X_{14}}\,,  \\
(\cycle_5)^4 \circ \rY^{\langle 123 \rangle }_5 &=  \frac{X_{15} X_{12} X_{34} }{X_{25} X_{13} X_{14}}\,.
\end{split}
\ee

\subsubsection*{Orbit $\{\ZZ_5 \circ \triple{124} \}$:}

\paragraph{$\triple{235}$.}The leg-factor is
\be
\label{V5_235}
\rV_5^{\triple{235}}(\bm a|\bm x) 
=
X_{23}^{|\bm a_{2,3}|'} X_{35}^{|\bm a_{3,5}|'} X_{52}^{|\bm a_{5,2}|'}\, X_{24}^{-a_4} X_{13}^{-a_1}\,.
\ee
The cross-ratio set is produced from \eqref{cross_5_124}: $\bm \rY_5^{\triple{235}} = \{ \cycle_5 \circ \rY^{\triple{124}}_l, l=1,...,5\}$. The polygonal  function is
\begin{multline}
\label{H5_235}
\HG_5^{\triple{235}}\left(\bm a| \bm \rY_5^{\triple{235}} \right)  
= \sum_{m_1,...,m_5=0}^{\infty} \;\;
\prod_{l=1}^{5} \frac{\left(\cycle_5 \circ \rY^{\triple{124}}_l \right)^{m_l}}{m_l!}
\\ \times 
\frac{(-)^{m_1+m_2+m_3+m_4} (a_4)_{m_1+m_2+m_5} (a_1)_{m_3+m_4+m_5}  }{(1+|\bm a_{2,3}|')_{m_2+m_4+m_5} (1+|\bm a_{3,5}|')_{m_3-m_1-m_2} (1+|\bm a_{5,2}|')_{m_1-m_3-m_4} }  \,,
\end{multline}
where
\be
\label{cross_5_235}
\begin{split}
\cycle_5 \circ \rY^{\langle 124 \rangle }_1 &= \frac{X_{34} X_{25} }{X_{24} X_{35}} \,,
\qquad
\cycle_5 \circ \rY^{\langle 124 \rangle }_2  =  \frac{X_{23} X_{45} }{X_{24} X_{35}} \,,  \\
\cycle_5 \circ \rY^{\langle 124 \rangle }_3 &= \frac{X_{35} X_{12} }{X_{25} X_{13}} \,, 
\qquad
\cycle_5 \circ \rY^{\langle 124 \rangle }_4  =  \frac{X_{23} X_{15} }{X_{25} X_{13}} \,,   \\
\cycle_5 \circ \rY^{\langle 124 \rangle }_5 &=  \frac{X_{23} X_{14} }{X_{24} X_{13}} \,. 
\end{split}
\ee

\paragraph{$\triple{134}$.}The leg-factor is
\be
\label{V5_134}
\rV_5^{\triple{134}}(\bm a|\bm x) 
=
X_{34}^{|\bm a_{3,4}|'} X_{41}^{|\bm a_{4,1}|'} X_{13}^{|\bm a_{1,3}|'}\, X_{35}^{-a_5} X_{24}^{-a_2}\,.
\ee
The cross-ratio set is produced from \eqref{cross_5_124}: $\bm \rY_5^{\triple{134}} = \{ (\cycle_5)^2 \circ \rY^{\triple{124}}_l, l=1,...,5\}$. The polygonal  function is
\begin{multline}
\label{H5_134}
\HG_5^{\triple{134}}\left(\bm a| \bm \rY_5^{\triple{134}} \right)  
= \sum_{m_1,...,m_5=0}^{\infty} \;\;
\prod_{l=1}^{5} \frac{\left((\cycle_5)^2 \circ \rY^{\triple{124}}_l \right)^{m_l}}{m_l!}
\\ \times 
\frac{(-)^{m_1+m_2+m_3+m_4} (a_5)_{m_1+m_2+m_5} (a_2)_{m_3+m_4+m_5}  }{(1+|\bm a_{3,4}|')_{m_2+m_4+m_5} (1+|\bm a_{4,1}|')_{m_3-m_1-m_2} (1+|\bm a_{1,3}|')_{m_1-m_3-m_4} }  \,,
\end{multline}
where
\be
\label{cross_5_134}
\begin{split}
(\cycle_5)^2 \circ \rY^{\langle 124 \rangle }_1 &= \frac{X_{45} X_{13} }{X_{35} X_{14}} \,,
\qquad
(\cycle_5)^2 \circ \rY^{\langle 124 \rangle }_2  =  \frac{X_{34} X_{15} }{X_{35} X_{14}} \,,  \\
(\cycle_5)^2 \circ \rY^{\langle 124 \rangle }_3 &= \frac{X_{14} X_{23} }{X_{13} X_{24}} \,, 
\qquad
(\cycle_5)^2 \circ \rY^{\langle 124 \rangle }_4  =  \frac{X_{34} X_{12} }{X_{13} X_{24}} \,,   \\
(\cycle_5)^2 \circ \rY^{\langle 124 \rangle }_5 &=  \frac{X_{34} X_{25} }{X_{35} X_{24}} \,. 
\end{split}
\ee

\paragraph{$\triple{245}$.}The leg-factor is
\be
\label{V5_245}
\rV_5^{\triple{245}}(\bm a|\bm x) 
=
X_{45}^{|\bm a_{4,5}|'} X_{52}^{|\bm a_{5,2}|'} X_{24}^{|\bm a_{2,4}|'}\, X_{14}^{-a_1} X_{35}^{-a_3}\,.
\ee
The cross-ratio set is produced from \eqref{cross_5_124}: $\bm \rY_5^{\triple{245}} = \{ (\cycle_5)^3 \circ \rY^{\triple{124}}_l, l=1,...,5\}$. The polygonal  function is
\begin{multline}
\label{H5_245}
\HG_5^{\triple{245}}\left(\bm a| \bm \rY_5^{\triple{245}} \right)  
= \sum_{m_1,...,m_5=0}^{\infty} \;\;
\prod_{l=1}^{5} \frac{\left((\cycle_5)^3 \circ \rY^{\triple{124}}_l \right)^{m_l}}{m_l!}
\\ \times 
\frac{(-)^{m_1+m_2+m_3+m_4} (a_1)_{m_1+m_2+m_5} (a_3)_{m_3+m_4+m_5}  }{(1+|\bm a_{4,5}|')_{m_2+m_4+m_5} (1+|\bm a_{5,2}|')_{m_3-m_1-m_2} (1+|\bm a_{2,4}|')_{m_1-m_3-m_4} }  \,,
\end{multline}
where
\be
\label{cross_5_245}
\begin{split}
(\cycle_5)^3 \circ \rY^{\langle 124 \rangle }_1 &= \frac{X_{15} X_{24} }{X_{14} X_{25}} \,,
\qquad
(\cycle_5)^3 \circ \rY^{\langle 124 \rangle }_2  =  \frac{X_{45} X_{12} }{X_{14} X_{25}} \,,  \\
(\cycle_5)^3 \circ \rY^{\langle 124 \rangle }_3 &= \frac{X_{25} X_{34} }{X_{24} X_{35}} \,, 
\qquad
(\cycle_5)^3 \circ \rY^{\langle 124 \rangle }_4  =  \frac{X_{45} X_{23} }{X_{24} X_{35}} \,,   \\
(\cycle_5)^3 \circ \rY^{\langle 124 \rangle }_5 &=  \frac{X_{45} X_{13} }{X_{14} X_{35}} \,. 
\end{split}
\ee

\paragraph{$\triple{135}$.}The leg-factor is
\be
\label{V5_135}
\rV_5^{\triple{135}}(\bm a|\bm x) 
=
X_{51}^{|\bm a_{5,1}|'} X_{13}^{|\bm a_{1,3}|'} X_{35}^{|\bm a_{3,5}|'}\, X_{25}^{-a_2} X_{14}^{-a_4}\,.
\ee
The cross-ratio set is produced from \eqref{cross_5_124}: $\bm \rY_5^{\triple{135}} = \{ (\cycle_5)^4 \circ \rY^{\triple{124}}_l, l=1,...,5\}$. The polygonal  function is
\begin{multline}
\label{H5_135}
\HG_5^{\triple{135}}\left(\bm a| \bm \rY_5^{\triple{135}} \right)  
= \sum_{m_1,...,m_5=0}^{\infty} \;\;
\prod_{l=1}^{5} \frac{\left((\cycle_5)^4 \circ \rY^{\triple{124}}_l \right)^{m_l}}{m_l!}
\\ \times 
\frac{(-)^{m_1+m_2+m_3+m_4} (a_2)_{m_1+m_2+m_5} (a_4)_{m_3+m_4+m_5}  }{(1+|\bm a_{5,1}|')_{m_2+m_4+m_5} (1+|\bm a_{1,3}|')_{m_3-m_1-m_2} (1+|\bm a_{3,5}|')_{m_1-m_3-m_4} }  \,,
\end{multline}
where 
\be
\label{cross_5_135}
\begin{split}
(\cycle_5)^4 \circ \rY^{\langle 124 \rangle }_1 &= \frac{X_{12} X_{35} }{X_{25} X_{13}} \,,
\qquad
(\cycle_5)^4 \circ \rY^{\langle 124 \rangle }_2  =  \frac{X_{15} X_{23} }{X_{25} X_{13}} \,,  \\
(\cycle_5)^4 \circ \rY^{\langle 124 \rangle }_3 &= \frac{X_{13} X_{45} }{X_{35} X_{14}} \,, 
\qquad
(\cycle_5)^4 \circ \rY^{\langle 124 \rangle }_4  =  \frac{X_{15} X_{34} }{X_{35} X_{14}} \,,   \\
(\cycle_5)^4 \circ \rY^{\langle 124 \rangle }_5 &=  \frac{X_{15} X_{24} }{X_{25} X_{14}} \,. 
\end{split}
\ee

\subsection{Hexagon}
\label{app:hexagon}

\subsubsection*{Orbit $\{\ZZ_6 \circ \triple{123} \}$:}

\paragraph{$\triple{234}$.}The leg-factor is
\be
\label{V_6_234}
\rV_6^{\triple{234}}(\bm a|\bm x) 
=
X_{23}^{|\bm a_{2,3}|'} X_{34}^{|\bm a_{3,4}|'}  X_{42}^{|\bm a_{4,2}|'}\, X_{35}^{-a_5} X_{36}^{-a_6} X_{13}^{-a_1}\,.
\ee
The cross-ratio set is produced from \eqref{cross_6_123}: $\bm \rY_6^{\triple{234}} = \{ \cycle_6 \circ \rY^{\triple{123}}_l, l=1,...,9\}$. The polygonal  function is
\begin{multline}
\label{H6_234}
\HG_6^{\triple{234}}\left( {\bm a}\big|\bm \rY^{\triple{234} }  \right)
=
\sum_{m_1,..., m_9=0}^\infty \;\; \prod_{l=1}^9 \frac{\left(\cycle_6 \circ \rY_l^{\triple{123} }\right)^{m_l}}{m_l!}
\\
\times
\frac{
(a_5)_{m_1+m_2+m_5+m_8} 
(a_6)_{m_3+m_4+m_5+m_9} 
(a_1)_{m_6+m_7+m_8+m_9}  \,  
(-)^{m_1+m_2+m_3+m_4+m_6+m_7} 
}{
(1+|\bm a_{2,3}|')_{m_1+m_3+m_5+m_6+m_8+m_9}
(1+|\bm a_{3,4}|')_{m_2+m_4+m_5+m_7+m_8+m_9} 
(1+|\bm a_{4,2}|')_{-m_1-...-m_9} 
} \,,
\end{multline}
where 
\be
\label{cross_6_234}
\begin{split}
\cycle_6 \circ \rY^{\langle 123 \rangle }_1 &= \frac{X_{23} X_{45} }{X_{24} X_{35}}\,, \quad 
\cycle_6 \circ \rY^{\langle 123 \rangle }_2 =  \frac{X_{34} X_{25} }{X_{24} X_{35}}\,, \qquad\;\;
\cycle_6 \circ \rY^{\langle 123 \rangle }_3 =  \frac{X_{23} X_{46} }{X_{24} X_{36}}\,, \\
\cycle_6 \circ \rY^{\langle 123 \rangle }_4 &= \frac{X_{34} X_{26} }{X_{24} X_{36}}\,,  \quad
\cycle_6 \circ \rY^{\langle 123 \rangle }_5 =  \frac{X_{23} X_{34} X_{56} }{X_{24} X_{35} X_{36}}\,, \quad
\cycle_6 \circ \rY^{\langle 123 \rangle }_6 =  \frac{X_{23} X_{14} }{X_{24} X_{13} }\,, \\
\cycle_6 \circ \rY^{\langle 123 \rangle }_7 &= \frac{X_{34} X_{12} }{X_{24} X_{13}}\,,  \quad
\cycle_6 \circ \rY^{\langle 123 \rangle }_8 =  \frac{X_{23} X_{34} X_{15} }{X_{24} X_{35} X_{13}} \,, \quad
\cycle_6 \circ \rY^{\langle 123 \rangle }_9 =  \frac{X_{23} X_{34} X_{16} }{X_{13} X_{24} X_{36}} \,.
\end{split}
\ee

\paragraph{$\triple{345}$.}The leg-factor is
\be
\label{V_6_345}
\rV_6^{\triple{345}}(\bm a|\bm x) 
=
X_{34}^{|\bm a_{3,4}|'} X_{45}^{|\bm a_{4,5}|'}  X_{53}^{|\bm a_{5,3}|'}\, X_{46}^{-a_6} X_{14}^{-a_1} X_{24}^{-a_2}\,.
\ee
The cross-ratio set is produced from \eqref{cross_6_123}: $\bm \rY_6^{\triple{345}} = \{ (\cycle_6)^2 \circ \rY^{\triple{123}}_l, l=1,...,9\}$. The polygonal  function is
\begin{multline}
\label{H6_345}
\HG_6^{\triple{345}}\left( {\bm a}\big|\bm \rY^{\triple{345} }  \right)
=
\sum_{m_1,..., m_9=0}^\infty \;\; \prod_{l=1}^9 \frac{\left((\cycle_6)^2 \circ \rY_l^{\triple{123} }\right)^{m_l}}{m_l!}
\\
\times
\frac{
(a_6)_{m_1+m_2+m_5+m_8} 
(a_1)_{m_3+m_4+m_5+m_9} 
(a_2)_{m_6+m_7+m_8+m_9}  \,  
(-)^{m_1+m_2+m_3+m_4+m_6+m_7} 
}{
(1+|\bm a_{3,4}|')_{m_1+m_3+m_5+m_6+m_8+m_9}
(1+|\bm a_{4,5}|')_{m_2+m_4+m_5+m_7+m_8+m_9} 
(1+|\bm a_{5,3}|')_{-m_1-...-m_9} 
} \,,
\end{multline}
where 
\be
\label{cross_6_345}
\begin{split}
(\cycle_6)^2 \circ \rY^{\langle 123 \rangle }_1 &= \frac{X_{34} X_{56} }{X_{46} X_{35}}\,, \quad 
(\cycle_6)^2 \circ \rY^{\langle 123 \rangle }_2 =  \frac{X_{45} X_{36} }{X_{46} X_{35}}\,, \qquad\;\;
(\cycle_6)^2 \circ \rY^{\langle 123 \rangle }_3 =  \frac{X_{34} X_{15} }{X_{14} X_{35}}\,, \\
(\cycle_6)^2 \circ \rY^{\langle 123 \rangle }_4 &= \frac{X_{45} X_{13} }{X_{14} X_{35}}\,,  \quad
(\cycle_6)^2 \circ \rY^{\langle 123 \rangle }_5 =  \frac{X_{34} X_{45} X_{16} }{X_{14} X_{35} X_{46}}\,, \quad
(\cycle_6)^2 \circ \rY^{\langle 123 \rangle }_6 =  \frac{X_{34} X_{25} }{X_{24} X_{35} }\,, \\
(\cycle_6)^2 \circ \rY^{\langle 123 \rangle }_7 &= \frac{X_{45} X_{23} }{X_{24} X_{35}}\,,  \quad
(\cycle_6)^2 \circ \rY^{\langle 123 \rangle }_8 =  \frac{X_{34} X_{45} X_{26} }{X_{24} X_{35} X_{46}} \,, \quad
(\cycle_6)^2 \circ \rY^{\langle 123 \rangle }_9 =  \frac{X_{12} X_{34} X_{45} }{X_{14} X_{24} X_{35}} \,.
\end{split}
\ee

\paragraph{$\triple{456}$.}The leg-factor is
\be
\label{V_6_456}
\rV_6^{\triple{456}}(\bm a|\bm x) 
=
X_{45}^{|\bm a_{4,5}|'} X_{56}^{|\bm a_{5,6}|'}  X_{64}^{|\bm a_{6,4}|'}\, X_{15}^{-a_1} X_{25}^{-a_2} X_{35}^{-a_3}\,.
\ee
The cross-ratio set is produced from \eqref{cross_6_123}: $\bm \rY_6^{\triple{456}} = \{ (\cycle_6)^3 \circ \rY^{\triple{123}}_l, l=1,...,9\}$. The polygonal  function is
\begin{multline}
\label{H6_456}
\HG_6^{\triple{456}}\left( {\bm a}\big|\bm \rY^{\triple{456} }  \right)
=
\sum_{m_1,..., m_9=0}^\infty \;\; \prod_{l=1}^9 \frac{\left((\cycle_6)^3 \circ \rY_l^{\triple{123} }\right)^{m_l}}{m_l!}
\\
\times
\frac{
(a_1)_{m_1+m_2+m_5+m_8} 
(a_2)_{m_3+m_4+m_5+m_9} 
(a_3)_{m_6+m_7+m_8+m_9}  \,  
(-)^{m_1+m_2+m_3+m_4+m_6+m_7} 
}{
(1+|\bm a_{4,5}|')_{m_1+m_3+m_5+m_6+m_8+m_9}
(1+|\bm a_{5,6}|')_{m_2+m_4+m_5+m_7+m_8+m_9} 
(1+|\bm a_{6,4}|')_{-m_1-...-m_9} 
} \,,
\end{multline}
where 
\be
\label{cross_6_456}
\begin{split}
(\cycle_6)^3 \circ \rY^{\langle 123 \rangle }_1 &= \frac{X_{45} X_{16} }{X_{46} X_{15}}\,, \quad 
(\cycle_6)^3 \circ \rY^{\langle 123 \rangle }_2 =  \frac{X_{56} X_{14} }{X_{46} X_{15}}\,, \qquad\;\;
(\cycle_6)^3 \circ \rY^{\langle 123 \rangle }_3 =  \frac{X_{45} X_{26} }{X_{46} X_{25}}\,, \\
(\cycle_6)^3 \circ \rY^{\langle 123 \rangle }_4 &= \frac{X_{56} X_{24} }{X_{46} X_{25}}\,,  \quad
(\cycle_6)^3 \circ \rY^{\langle 123 \rangle }_5 =  \frac{X_{45} X_{56} X_{12} }{X_{15} X_{25} X_{46}}\,, \quad
(\cycle_6)^3 \circ \rY^{\langle 123 \rangle }_6 =  \frac{X_{45} X_{36} }{X_{46} X_{35} }\,, \\
(\cycle_6)^3 \circ \rY^{\langle 123 \rangle }_7 &= \frac{X_{56} X_{34} }{X_{46} X_{35}}\,,  \quad
(\cycle_6)^3 \circ \rY^{\langle 123 \rangle }_8 =  \frac{X_{13} X_{45} X_{56} }{X_{15} X_{35} X_{46}} \,, \quad
(\cycle_6)^3 \circ \rY^{\langle 123 \rangle }_9 =  \frac{X_{23} X_{45} X_{56} }{X_{46} X_{25} X_{35}} \,.
\end{split}
\ee

\paragraph{$\triple{156}$.}The leg-factor is
\be
\label{V_6_156}
\rV_6^{\triple{156}}(\bm a|\bm x) 
=
X_{56}^{|\bm a_{5,6}|'} X_{61}^{|\bm a_{6,1}|'}  X_{15}^{|\bm a_{1,5}|'}\, X_{26}^{-a_2} X_{36}^{-a_3} X_{46}^{-a_4}\,.
\ee
The cross-ratio set is produced from \eqref{cross_6_123}: $\bm \rY_6^{\triple{156}} = \{ (\cycle_6)^4 \circ \rY^{\triple{123}}_l, l=1,...,9\}$. The polygonal  function is
\begin{multline}
\label{H6_156}
\HG_6^{\triple{156}}\left( {\bm a}\big|\bm \rY^{\triple{156} }  \right)
=
\sum_{m_1,..., m_9=0}^\infty \;\; \prod_{l=1}^9 \frac{\left((\cycle_6)^4 \circ \rY_l^{\triple{123} }\right)^{m_l}}{m_l!}
\\
\times
\frac{
(a_2)_{m_1+m_2+m_5+m_8} 
(a_3)_{m_3+m_4+m_5+m_9} 
(a_4)_{m_6+m_7+m_8+m_9}  \,  
(-)^{m_1+m_2+m_3+m_4+m_6+m_7} 
}{
(1+|\bm a_{5,6}|')_{m_1+m_3+m_5+m_6+m_8+m_9}
(1+|\bm a_{6,1}|')_{m_2+m_4+m_5+m_7+m_8+m_9} 
(1+|\bm a_{1,5}|')_{-m_1-...-m_9} 
} \,,
\end{multline}
where 
\be
\label{cross_6_156}
\begin{split}
(\cycle_6)^4 \circ \rY^{\langle 123 \rangle }_1 &= \frac{X_{56} X_{12} }{X_{26} X_{15}}\,, \quad 
(\cycle_6)^4 \circ \rY^{\langle 123 \rangle }_2 =  \frac{X_{16} X_{25} }{X_{26} X_{15}}\,, \qquad\;\;
(\cycle_6)^4 \circ \rY^{\langle 123 \rangle }_3 =  \frac{X_{56} X_{13} }{X_{36} X_{15}}\,, \\
(\cycle_6)^4 \circ \rY^{\langle 123 \rangle }_4 &= \frac{X_{16} X_{35} }{X_{36} X_{15}}\,,  \quad
(\cycle_6)^4 \circ \rY^{\langle 123 \rangle }_5 =  \frac{X_{56} X_{16} X_{23} }{X_{15} X_{26} X_{36}}\,, \quad
(\cycle_6)^4 \circ \rY^{\langle 123 \rangle }_6 =  \frac{X_{56} X_{14} }{X_{46} X_{15} }\,, \\
(\cycle_6)^4 \circ \rY^{\langle 123 \rangle }_7 &= \frac{X_{16} X_{45} }{X_{46} X_{15}}\,,  \quad
(\cycle_6)^4 \circ \rY^{\langle 123 \rangle }_8 =  \frac{X_{24} X_{56} X_{16} }{X_{15} X_{26} X_{46}} \,, \quad
(\cycle_6)^4 \circ \rY^{\langle 123 \rangle }_9 =  \frac{X_{34} X_{56} X_{16} }{X_{46} X_{15} X_{36}} \,.
\end{split}
\ee

\paragraph{$\triple{126}$.}The leg-factor is
\be
\label{V_6_126}
\rV_6^{\triple{126}}(\bm a|\bm x) 
=
X_{61}^{|\bm a_{6,1}|'} X_{12}^{|\bm a_{1,2}|'}  X_{26}^{|\bm a_{2,6}|'}\, X_{13}^{-a_3} X_{14}^{-a_4} X_{15}^{-a_5}\,.
\ee
The cross-ratio set is produced from \eqref{cross_6_123}: $\bm \rY_6^{\triple{126}} = \{ (\cycle_6)^5 \circ \rY^{\triple{123}}_l, l=1,...,9\}$. The polygonal  function is
\begin{multline}
\label{H6_126}
\HG_6^{\triple{126}}\left( {\bm a}\big|\bm \rY^{\triple{126} }  \right)
=
\sum_{m_1,..., m_9=0}^\infty \;\; \prod_{l=1}^9 \frac{\left((\cycle_6)^5 \circ \rY_l^{\triple{123} }\right)^{m_l}}{m_l!}
\\
\times
\frac{
(a_3)_{m_1+m_2+m_5+m_8} 
(a_4)_{m_3+m_4+m_5+m_9} 
(a_5)_{m_6+m_7+m_8+m_9}  \,  
(-)^{m_1+m_2+m_3+m_4+m_6+m_7} 
}{
(1+|\bm a_{6,1}|')_{m_1+m_3+m_5+m_6+m_8+m_9}
(1+|\bm a_{1,2}|')_{m_2+m_4+m_5+m_7+m_8+m_9} 
(1+|\bm a_{2,6}|')_{-m_1-...-m_9} 
} \,,
\end{multline}
where 
\be
\label{cross_6_126}
\begin{split}
(\cycle_6)^5 \circ \rY^{\langle 123 \rangle }_1 &= \frac{X_{16} X_{23} }{X_{26} X_{13}}\,, \quad 
(\cycle_6)^5 \circ \rY^{\langle 123 \rangle }_2 =  \frac{X_{36} X_{12} }{X_{26} X_{13}}\,, \qquad\;\;
(\cycle_6)^5 \circ \rY^{\langle 123 \rangle }_3 =  \frac{X_{16} X_{24} }{X_{26} X_{14}}\,, \\
(\cycle_6)^5 \circ \rY^{\langle 123 \rangle }_4 &= \frac{X_{46} X_{12} }{X_{26} X_{14}}\,,  \quad
(\cycle_6)^5 \circ \rY^{\langle 123 \rangle }_5 =  \frac{X_{12} X_{34} X_{16} }{X_{13} X_{26} X_{14}}\,, \quad
(\cycle_6)^5 \circ \rY^{\langle 123 \rangle }_6 =  \frac{X_{16} X_{25} }{X_{26} X_{15} }\,, \\
(\cycle_6)^5 \circ \rY^{\langle 123 \rangle }_7 &= \frac{X_{12} X_{56} }{X_{26} X_{15}}\,,  \quad
(\cycle_6)^5 \circ \rY^{\langle 123 \rangle }_8 =  \frac{X_{12} X_{35} X_{16} }{X_{15} X_{26} X_{13}} \,, \quad
(\cycle_6)^5 \circ \rY^{\langle 123 \rangle }_9 =  \frac{X_{45} X_{12} X_{16} }{X_{14} X_{15} X_{26}} \,.
\end{split}
\ee

\subsubsection*{Orbit $\{\ZZ_6 \circ \triple{124} \}$:}

\paragraph{$\triple{235}$.}The leg-factor is
\be
\label{V_6_235}
\rV_6^{\triple{235}}(\bm a|\bm x) 
=
X_{23}^{|\bm a_{2,3}|'} X_{35}^{|\bm a_{3,5}|'} X_{52}^{|\bm a_{5,2}|'}\, X_{24}^{-a_4} X_{36}^{-a_6} X_{13}^{-a_1} \,.
\ee
The cross-ratio set is produced from \eqref{cross_6_124}: $\bm \rY_6^{\triple{235}} = \{ \cycle_6 \circ \rY^{\triple{124}}_l, l=1,...,9\}$. The polygonal  function is
\begin{multline}
\label{H6_235}
\HG_6^{\triple{235}}\left( {\bm a}\big|\bm \rY^{\triple{235}}\right)
=
\sum_{m_1,..., m_9=0}^\infty \;\; \prod_{l=1}^9 \frac{\left(\cycle_6 \circ \rY_l^{\triple{124}}\right)^{m_l}}{m_l!}
\\
\times
\frac{
(a_4)_{m_1+m_2+m_5+m_8} 
(a_6)_{m_3+m_4+m_5+m_9} 
(a_1)_{m_6+m_7+m_8+m_9}  \,  
(-)^{m_1+m_2+m_3+m_4+m_6+m_7} 
}{
(1+|\bm a_{2,3}|')_{m_2+m_4+m_5+m_6+m_8+m_9}
(1+|\bm a_{3,5}|')_{-m_1-m_2+m_3+m_7+m_9} 
(1+|\bm a_{5,2}|')_{m_1-m_3-m_4-m_6-m_7-m_9} 
} \,,
\end{multline}
where 
\be
\label{cross_6_235}
\begin{split}
\cycle_6 \circ \rY^{\langle 124 \rangle}_1 &= \frac{X_{34} X_{25}}{X_{35} X_{24}} \,, \quad
\cycle_6 \circ \rY^{\langle 124 \rangle}_2 = \frac{X_{23} X_{45}}{X_{35} X_{24}} \,, \quad
\cycle_6 \circ \rY^{\langle 124 \rangle}_3 =  \frac{X_{35} X_{26}}{X_{36} X_{25}} \,, \\
\cycle_6 \circ \rY^{\langle 124 \rangle}_4 &= \frac{X_{23} X_{56}}{X_{36} X_{25}} \,, \quad
\cycle_6 \circ \rY^{\langle 124 \rangle}_5 =  \frac{X_{23} X_{46}}{X_{24} X_{36}} \,, \quad
\cycle_6 \circ \rY^{\langle 124 \rangle}_6 =  \frac{X_{23} X_{15}}{X_{25} X_{13}} \,, \\
\cycle_6 \circ \rY^{\langle 124 \rangle}_7 &= \frac{X_{35} X_{12}}{X_{25} X_{13}} \,, \quad
\cycle_6 \circ \rY^{\langle 124 \rangle}_8 =  \frac{X_{23} X_{14}}{X_{24} X_{13}} \,, \quad
\cycle_6 \circ \rY^{\langle 124 \rangle}_9 =  \frac{X_{23} X_{35} X_{16}}{X_{25} X_{36} X_{13}} \,.
\end{split}
\ee

\paragraph{$\triple{346}$.}The leg-factor is
\be
\label{V_6_346}
\rV_6^{\triple{346}}(\bm a|\bm x) 
=
X_{34}^{|\bm a_{3,4}|'} X_{46}^{|\bm a_{4,6}|'} X_{63}^{|\bm a_{6,3}|'}\, X_{35}^{-a_5} X_{14}^{-a_1} X_{24}^{-a_2} \,.
\ee
The cross-ratio set is produced from \eqref{cross_6_124}: $\bm \rY_6^{\triple{346}} = \{ (\cycle_6)^2 \circ \rY^{\triple{124}}_l, l=1,...,9\}$. The polygonal  function is
\begin{multline}
\label{H6_346}
\HG_6^{\triple{346}}\left( {\bm a}\big|\bm \rY^{\triple{346}}\right)
=
\sum_{m_1,..., m_9=0}^\infty \;\; \prod_{l=1}^9 \frac{\left((\cycle_6)^2 \circ \rY_l^{\triple{124}}\right)^{m_l}}{m_l!}
\\
\times
\frac{
(a_5)_{m_1+m_2+m_5+m_8} 
(a_1)_{m_3+m_4+m_5+m_9} 
(a_2)_{m_6+m_7+m_8+m_9}  \,  
(-)^{m_1+m_2+m_3+m_4+m_6+m_7} 
}{
(1+|\bm a_{3,4}|')_{m_2+m_4+m_5+m_6+m_8+m_9}
(1+|\bm a_{4,6}|')_{-m_1-m_2+m_3+m_7+m_9} 
(1+|\bm a_{6,3}|')_{m_1-m_3-m_4-m_6-m_7-m_9} 
} \,,
\end{multline}
where 
\be
\label{cross_6_346}
\begin{split}
(\cycle_6)^2 \circ \rY^{\langle 124 \rangle}_1 &= \frac{X_{45} X_{36}}{X_{46} X_{35}} \,, \quad
(\cycle_6)^2 \circ \rY^{\langle 124 \rangle}_2 = \frac{X_{34} X_{56}}{X_{46} X_{35}} \,, \quad
(\cycle_6)^2 \circ \rY^{\langle 124 \rangle}_3 =  \frac{X_{46} X_{13}}{X_{14} X_{36}} \,, \\
(\cycle_6)^2 \circ \rY^{\langle 124 \rangle}_4 &= \frac{X_{34} X_{16}}{X_{14} X_{36}} \,, \quad
(\cycle_6)^2 \circ \rY^{\langle 124 \rangle}_5 =  \frac{X_{34} X_{15}}{X_{35} X_{14}} \,, \quad
(\cycle_6)^2 \circ \rY^{\langle 124 \rangle}_6 =  \frac{X_{34} X_{26}}{X_{36} X_{24}} \,, \\
(\cycle_6)^2 \circ \rY^{\langle 124 \rangle}_7 &= \frac{X_{46} X_{23}}{X_{36} X_{24}} \,, \quad
(\cycle_6)^2 \circ \rY^{\langle 124 \rangle}_8 =  \frac{X_{34} X_{25}}{X_{35} X_{24}} \,, \quad
(\cycle_6)^2 \circ \rY^{\langle 124 \rangle}_9 =  \frac{X_{34} X_{46} X_{12}}{X_{36} X_{14} X_{24}} \,.
\end{split}
\ee

\paragraph{$\triple{145}$.}The leg-factor is
\be
\label{V_6_145}
\rV_6^{\triple{145}}(\bm a|\bm x) 
=
X_{45}^{|\bm a_{4,5}|'} X_{51}^{|\bm a_{5,1}|'} X_{14}^{|\bm a_{1,4}|'}\, X_{46}^{-a_6} X_{25}^{-a_2} X_{35}^{-a_3} \,.
\ee
The cross-ratio set is produced from \eqref{cross_6_124}: $\bm \rY_6^{\triple{145}} = \{ (\cycle_6)^3 \circ \rY^{\triple{124}}_l, l=1,...,9\}$. The polygonal  function is
\begin{multline}
\label{H6_145}
\HG_6^{\triple{145}}\left( {\bm a}\big|\bm \rY^{\triple{145}}\right)
=
\sum_{m_1,..., m_9=0}^\infty \;\; \prod_{l=1}^9 \frac{\left((\cycle_6)^3 \circ \rY_l^{\triple{124}}\right)^{m_l}}{m_l!}
\\
\times
\frac{
(a_6)_{m_1+m_2+m_5+m_8} 
(a_2)_{m_3+m_4+m_5+m_9} 
(a_3)_{m_6+m_7+m_8+m_9}  \,  
(-)^{m_1+m_2+m_3+m_4+m_6+m_7} 
}{
(1+|\bm a_{4,5}|')_{m_2+m_4+m_5+m_6+m_8+m_9}
(1+|\bm a_{5,1}|')_{-m_1-m_2+m_3+m_7+m_9} 
(1+|\bm a_{1,4}|')_{m_1-m_3-m_4-m_6-m_7-m_9} 
} \,,
\end{multline}
where 
\be
\label{cross_6_145}
\begin{split}
(\cycle_6)^3 \circ \rY^{\langle 124 \rangle}_1 &= \frac{X_{56} X_{14}}{X_{15} X_{46}} \,, \quad
(\cycle_6)^3 \circ \rY^{\langle 124 \rangle}_2 = \frac{X_{45} X_{16}}{X_{15} X_{46}} \,, \quad
(\cycle_6)^3 \circ \rY^{\langle 124 \rangle}_3 =  \frac{X_{15} X_{24}}{X_{25} X_{14}} \,, \\
(\cycle_6)^3 \circ \rY^{\langle 124 \rangle}_4 &= \frac{X_{45} X_{12}}{X_{25} X_{14}} \,, \quad
(\cycle_6)^3 \circ \rY^{\langle 124 \rangle}_5 =  \frac{X_{45} X_{26}}{X_{46} X_{25}} \,, \quad
(\cycle_6)^3 \circ \rY^{\langle 124 \rangle}_6 =  \frac{X_{45} X_{13}}{X_{14} X_{35}} \,, \\
(\cycle_6)^3 \circ \rY^{\langle 124 \rangle}_7 &= \frac{X_{15} X_{34}}{X_{14} X_{35}} \,, \quad
(\cycle_6)^3 \circ \rY^{\langle 124 \rangle}_8 =  \frac{X_{45} X_{36}}{X_{46} X_{35}} \,, \quad
(\cycle_6)^3 \circ \rY^{\langle 124 \rangle}_9 =  \frac{X_{45} X_{15} X_{23}}{X_{14} X_{25} X_{35}} \,.
\end{split}
\ee

\paragraph{$\triple{256}$.}The leg-factor is
\be
\label{V_6_256}
\rV_6^{\triple{256}}(\bm a|\bm x) 
=
X_{56}^{|\bm a_{5,6}|'} X_{62}^{|\bm a_{6,2}|'} X_{25}^{|\bm a_{2,5}|'}\, X_{15}^{-a_1} X_{36}^{-a_3} X_{46}^{-a_4} \,.
\ee
The cross-ratio set is produced from \eqref{cross_6_124}: $\bm \rY_6^{\triple{256}} = \{ (\cycle_6)^4 \circ \rY^{\triple{124}}_l, l=1,...,9\}$. The polygonal  function is
\begin{multline}
\label{H6_256}
\HG_6^{\triple{256}}\left( {\bm a}\big|\bm \rY^{\triple{256}}\right)
=
\sum_{m_1,..., m_9=0}^\infty  \;\;\prod_{l=1}^9 \frac{\left((\cycle_6)^4 \circ \rY_l^{\triple{124}}\right)^{m_l}}{m_l!}
\\
\times
\frac{
(a_1)_{m_1+m_2+m_5+m_8} 
(a_3)_{m_3+m_4+m_5+m_9} 
(a_4)_{m_6+m_7+m_8+m_9}  \,  
(-)^{m_1+m_2+m_3+m_4+m_6+m_7} 
}{
(1+|\bm a_{5,6}|')_{m_2+m_4+m_5+m_6+m_8+m_9}
(1+|\bm a_{6,2}|')_{-m_1-m_2+m_3+m_7+m_9} 
(1+|\bm a_{2,5}|')_{m_1-m_3-m_4-m_6-m_7-m_9} 
} \,,
\end{multline}
where 
\be
\label{cross_6_256}
\begin{split}
(\cycle_6)^4 \circ \rY^{\langle 124 \rangle}_1 &= \frac{X_{16} X_{25}}{X_{26} X_{15}} \,, \quad
(\cycle_6)^4 \circ \rY^{\langle 124 \rangle}_2 = \frac{X_{56} X_{12}}{X_{26} X_{15}} \,, \quad
(\cycle_6)^4 \circ \rY^{\langle 124 \rangle}_3 =  \frac{X_{26} X_{35}}{X_{36} X_{25}} \,, \\
(\cycle_6)^4 \circ \rY^{\langle 124 \rangle}_4 &= \frac{X_{56} X_{23}}{X_{36} X_{25}} \,, \quad
(\cycle_6)^4 \circ \rY^{\langle 124 \rangle}_5 =  \frac{X_{56} X_{13}}{X_{15} X_{36}} \,, \quad
(\cycle_6)^4 \circ \rY^{\langle 124 \rangle}_6 =  \frac{X_{56} X_{24}}{X_{25} X_{46}} \,, \\
(\cycle_6)^4 \circ \rY^{\langle 124 \rangle}_7 &= \frac{X_{26} X_{45}}{X_{25} X_{46}} \,, \quad
(\cycle_6)^4 \circ \rY^{\langle 124 \rangle}_8 =  \frac{X_{56} X_{14}}{X_{15} X_{46}} \,, \quad
(\cycle_6)^4 \circ \rY^{\langle 124 \rangle}_9 =  \frac{X_{56} X_{26} X_{34}}{X_{25} X_{36} X_{46}} \,.
\end{split}
\ee

\paragraph{$\triple{136}$.}The leg-factor is
\be
\label{V_6_136}
\rV_6^{\triple{136}}(\bm a|\bm x) 
=
X_{61}^{|\bm a_{6,1}|'} X_{13}^{|\bm a_{1,3}|'} X_{36}^{|\bm a_{3,6}|'}\, X_{26}^{-a_2} X_{14}^{-a_4} X_{15}^{-a_5} \,.
\ee
The cross-ratio set is produced from \eqref{cross_6_124}: $\bm \rY_6^{\triple{136}} = \{ (\cycle_6)^5 \circ \rY^{\triple{124}}_l, l=1,...,9\}$. The polygonal  function is
\begin{multline}
\label{H6_136}
\HG_6^{\triple{136}}\left( {\bm a}\big|\bm \rY^{\triple{136}}\right)
=
\sum_{m_1,..., m_9=0}^\infty \;\; \prod_{l=1}^9 \frac{\left((\cycle_6)^5 \circ \rY_l^{\triple{124}}\right)^{m_l}}{m_l!}
\\
\times
\frac{
(a_2)_{m_1+m_2+m_5+m_8} 
(a_4)_{m_3+m_4+m_5+m_9} 
(a_5)_{m_6+m_7+m_8+m_9}  \,  
(-)^{m_1+m_2+m_3+m_4+m_6+m_7} 
}{
(1+|\bm a_{6,1}|')_{m_2+m_4+m_5+m_6+m_8+m_9}
(1+|\bm a_{1,3}|')_{-m_1-m_2+m_3+m_7+m_9} 
(1+|\bm a_{3,6}|')_{m_1-m_3-m_4-m_6-m_7-m_9} 
} \,,
\end{multline}
where 
\be
\label{cross_6_136}
\begin{split}
(\cycle_6)^5 \circ \rY^{\langle 124 \rangle}_1 &= \frac{X_{12} X_{36}}{X_{13} X_{26}} \,, \quad
(\cycle_6)^5 \circ \rY^{\langle 124 \rangle}_2 = \frac{X_{16} X_{23}}{X_{13} X_{26}} \,, \quad
(\cycle_6)^5 \circ \rY^{\langle 124 \rangle}_3 =  \frac{X_{13} X_{46}}{X_{14} X_{36}} \,, \\
(\cycle_6)^5 \circ \rY^{\langle 124 \rangle}_4 &= \frac{X_{16} X_{34}}{X_{14} X_{36}} \,, \quad
(\cycle_6)^5 \circ \rY^{\langle 124 \rangle}_5 =  \frac{X_{16} X_{24}}{X_{26} X_{14}} \,, \quad
(\cycle_6)^5 \circ \rY^{\langle 124 \rangle}_6 =  \frac{X_{16} X_{35}}{X_{36} X_{15}} \,, \\
(\cycle_6)^5 \circ \rY^{\langle 124 \rangle}_7 &= \frac{X_{13} X_{56}}{X_{36} X_{15}} \,, \quad
(\cycle_6)^5 \circ \rY^{\langle 124 \rangle}_8 =  \frac{X_{16} X_{25}}{X_{26} X_{15}} \,, \quad
(\cycle_6)^5 \circ \rY^{\langle 124 \rangle}_9 =  \frac{X_{16} X_{13} X_{45}}{X_{36} X_{14} X_{15}} \,.
\end{split}
\ee

\subsubsection*{Orbit $\{\ZZ_6 \circ \triple{125} \}$:}

\paragraph{$\triple{236}$.}The leg-factor is
\be
\label{V_6_236}
\rV_6^{\triple{236}}(\bm a|\bm x) 
=
X_{23}^{|\bm a_{2,3}|'} X_{36}^{|\bm a_{3,6}|'} X_{62}^{|\bm a_{6,2}|'}\, X_{24}^{-a_4} X_{25}^{-a_5} X_{13}^{-a_1} \,.
\ee
The cross-ratio set is produced from \eqref{cross_6_125}: $\bm \rY_6^{\triple{236}} = \{ \cycle_6\circ \rY^{\triple{125}}_l, l=1,...,9\}$. The polygonal  function is
\begin{multline}
\label{H6_236}
\HG_6^{\triple{236}}\left( {\bm a}\big|\bm \rY^{\triple{236} }\right)
=
\sum_{m_1,..., m_9=0}^\infty \;\; \prod_{l=1}^9 \frac{\left(\cycle_6 \circ \rY_l^{\triple{125} }\right)^{m_l}}{m_l!}
\\
\times
\frac{
(a_4)_{m_1+m_2+m_5+m_8} 
(a_5)_{m_3+m_4+m_5+m_9} 
(a_1)_{m_6+m_7+m_8+m_9}  \,  
(-)^{m_1+m_2+m_3+m_4+m_6+m_7} 
}{
(1+|\bm a_{2,3}|')_{m_2+m_4+m_5+m_6+m_8+m_9}
(1+|\bm a_{3,6}|')_{-m_1-m_2-m_3-m_4-m_5+m_7} 
(1+|\bm a_{6,2}|')_{m_1+m_3+m_5-m_6-m_7} 
} \,,
\end{multline}
where 
\be
\label{cross_6_236}
\begin{split}
\cycle_6 \circ \rY^{\langle 125 \rangle}_1 &= \frac{X_{34} X_{26}}{X_{24} X_{36}} \,, \quad
\cycle_6 \circ \rY^{\langle 125 \rangle}_2 = \frac{X_{23} X_{46}}{X_{24} X_{36}} \,, \qquad\;\;
\cycle_6 \circ \rY^{\langle 125 \rangle}_3 =  \frac{X_{35} X_{26}}{X_{25} X_{36}} \,, \\
\cycle_6 \circ \rY^{\langle 125 \rangle}_4 &= \frac{X_{23} X_{56}}{X_{25} X_{36}} \,, \quad
\cycle_6 \circ \rY^{\langle 125 \rangle}_5 =  \frac{X_{23} X_{26} X_{45}}{X_{24} X_{25} X_{36}} \,, \quad
\cycle_6 \circ \rY^{\langle 125 \rangle}_6 =  \frac{X_{23} X_{16}}{X_{13} X_{26}} \,, \\
\cycle_6 \circ \rY^{\langle 125 \rangle}_7 &=  \frac{X_{36} X_{12}}{X_{13} X_{26}} \,, \quad
\cycle_6 \circ \rY^{\langle 125 \rangle}_8 =   \frac{X_{23} X_{14}}{X_{24} X_{13}} \,, \qquad\;\;\;
\cycle_6 \circ \rY^{\langle 125 \rangle}_9 =   \frac{X_{23} X_{15}}{X_{25} X_{13}} \,.
\end{split}
\ee

\paragraph{$\triple{134}$.}The leg-factor is
\be
\label{V_6_134}
\rV_6^{\triple{134}}(\bm a|\bm x) 
=
X_{34}^{|\bm a_{3,4}|'} X_{41}^{|\bm a_{4,1}|'} X_{13}^{|\bm a_{1,3}|'}\, X_{35}^{-a_5} X_{36}^{-a_6} X_{24}^{-a_2} \,.
\ee
The cross-ratio set is produced from \eqref{cross_6_125}: $\bm \rY_6^{\triple{134}} = \{ (\cycle_6)^2 \circ \rY^{\triple{125}}_l, l=1,...,9\}$. The polygonal  function is
\begin{multline}
\label{H6_134}
\HG_6^{\triple{134}}\left( {\bm a}\big|\bm \rY^{\triple{134} }\right)
=
\sum_{m_1,..., m_9=0}^\infty \;\; \prod_{l=1}^9 \frac{\left((\cycle_6)^2 \circ \rY_l^{\triple{125} }\right)^{m_l}}{m_l!}
\\
\times
\frac{
(a_5)_{m_1+m_2+m_5+m_8} 
(a_6)_{m_3+m_4+m_5+m_9} 
(a_2)_{m_6+m_7+m_8+m_9}  \,  
(-)^{m_1+m_2+m_3+m_4+m_6+m_7} 
}{
(1+|\bm a_{3,4}|')_{m_2+m_4+m_5+m_6+m_8+m_9}
(1+|\bm a_{4,1}|')_{-m_1-m_2-m_3-m_4-m_5+m_7} 
(1+|\bm a_{1,3}|')_{m_1+m_3+m_5-m_6-m_7} 
} \,,
\end{multline}
where 
\be
\label{cross_6_134}
\begin{split}
(\cycle_6)^2 \circ \rY^{\langle 125 \rangle}_1 &= \frac{X_{45} X_{13}}{X_{35} X_{14}} \,, \quad
(\cycle_6)^2 \circ \rY^{\langle 125 \rangle}_2 = \frac{X_{34} X_{15}}{X_{35} X_{14}} \,, \qquad\;\;
(\cycle_6)^2 \circ \rY^{\langle 125 \rangle}_3 =  \frac{X_{46} X_{13}}{X_{36} X_{14}} \,, \\
(\cycle_6)^2 \circ \rY^{\langle 125 \rangle}_4 &= \frac{X_{34} X_{16}}{X_{36} X_{14}} \,, \quad
(\cycle_6)^2 \circ \rY^{\langle 125 \rangle}_5 =  \frac{X_{34} X_{13} X_{56}}{X_{35} X_{36} X_{14}} \,, \quad
(\cycle_6)^2 \circ \rY^{\langle 125 \rangle}_6 =  \frac{X_{34} X_{12}}{X_{24} X_{13}} \,, \\
(\cycle_6)^2 \circ \rY^{\langle 125 \rangle}_7 &=  \frac{X_{14} X_{23}}{X_{24} X_{13}} \,, \quad
(\cycle_6)^2 \circ \rY^{\langle 125 \rangle}_8 =   \frac{X_{34} X_{25}}{X_{35} X_{24}} \,, \qquad\;\;\;
(\cycle_6)^2 \circ \rY^{\langle 125 \rangle}_9 =   \frac{X_{34} X_{26}}{X_{36} X_{24}} \,.
\end{split}
\ee

\paragraph{$\triple{245}$.}The leg-factor is
\be
\label{V_6_245}
\rV_6^{\triple{245}}(\bm a|\bm x) 
=
X_{45}^{|\bm a_{4,5}|'} X_{52}^{|\bm a_{5,2}|'} X_{24}^{|\bm a_{2,4}|'}\, X_{46}^{-a_6} X_{14}^{-a_1} X_{35}^{-a_3} \,.
\ee
The cross-ratio set is produced from \eqref{cross_6_125}: $\bm \rY_6^{\triple{245}} = \{ (\cycle_6)^3 \circ \rY^{\triple{125}}_l, l=1,...,9\}$. The polygonal  function is
\begin{multline}
\label{H6_245}
\HG_6^{\triple{245}}\left( {\bm a}\big|\bm \rY^{\triple{245} }\right)
=
\sum_{m_1,..., m_9=0}^\infty \;\; \prod_{l=1}^9 \frac{\left((\cycle_6)^3 \circ \rY_l^{\triple{125} }\right)^{m_l}}{m_l!}
\\
\times
\frac{
(a_6)_{m_1+m_2+m_5+m_8} 
(a_1)_{m_3+m_4+m_5+m_9} 
(a_3)_{m_6+m_7+m_8+m_9}  \,  
(-)^{m_1+m_2+m_3+m_4+m_6+m_7} 
}{
(1+|\bm a_{4,5}|')_{m_2+m_4+m_5+m_6+m_8+m_9}
(1+|\bm a_{5,2}|')_{-m_1-m_2-m_3-m_4-m_5+m_7} 
(1+|\bm a_{2,4}|')_{m_1+m_3+m_5-m_6-m_7} 
} \,,
\end{multline}
where 
\be
\label{cross_6_245}
\begin{split}
(\cycle_6)^3 \circ \rY^{\langle 125 \rangle}_1 &= \frac{X_{56} X_{24}}{X_{46} X_{25}} \,, \quad
(\cycle_6)^3 \circ \rY^{\langle 125 \rangle}_2 = \frac{X_{45} X_{26}}{X_{46} X_{25}} \,, \qquad\;\;
(\cycle_6)^3 \circ \rY^{\langle 125 \rangle}_3 =  \frac{X_{15} X_{24}}{X_{14} X_{25}} \,, \\
(\cycle_6)^3 \circ \rY^{\langle 125 \rangle}_4 &= \frac{X_{45} X_{12}}{X_{14} X_{25}} \,, \quad
(\cycle_6)^3 \circ \rY^{\langle 125 \rangle}_5 =  \frac{X_{45} X_{24} X_{16}}{X_{46} X_{14} X_{25}} \,, \quad
(\cycle_6)^3 \circ \rY^{\langle 125 \rangle}_6 =  \frac{X_{45} X_{23}}{X_{35} X_{24}} \,, \\
(\cycle_6)^3 \circ \rY^{\langle 125 \rangle}_7 &=  \frac{X_{25} X_{34}}{X_{35} X_{24}} \,, \quad
(\cycle_6)^3 \circ \rY^{\langle 125 \rangle}_8 =   \frac{X_{45} X_{36}}{X_{46} X_{35}} \,, \qquad\;\;\;
(\cycle_6)^3 \circ \rY^{\langle 125 \rangle}_9 =   \frac{X_{45} X_{13}}{X_{14} X_{35}} \,.
\end{split}
\ee

\paragraph{$\triple{356}$.}The leg-factor is
\be
\label{V_6_356}
\rV_6^{\triple{356}}(\bm a|\bm x) 
=
X_{56}^{|\bm a_{5,6}|'} X_{63}^{|\bm a_{6,3}|'} X_{35}^{|\bm a_{3,5}|'}\, X_{15}^{-a_1} X_{25}^{-a_2} X_{46}^{-a_4} \,.
\ee
The cross-ratio set is produced from \eqref{cross_6_125}: $\bm \rY_6^{\triple{356}} = \{ (\cycle_6)^4 \circ \rY^{\triple{125}}_l, l=1,...,9\}$. The polygonal  function is
\begin{multline}
\label{H6_356}
\HG_6^{\triple{356}}\left( {\bm a}\big|\bm \rY^{\triple{356} }\right)
=
\sum_{m_1,..., m_9=0}^\infty \;\; \prod_{l=1}^9 \frac{\left((\cycle_6)^4 \circ \rY_l^{\triple{125} }\right)^{m_l}}{m_l!}
\\
\times
\frac{
(a_1)_{m_1+m_2+m_5+m_8} 
(a_2)_{m_3+m_4+m_5+m_9} 
(a_4)_{m_6+m_7+m_8+m_9}  \,  
(-)^{m_1+m_2+m_3+m_4+m_6+m_7} 
}{
(1+|\bm a_{5,6}|')_{m_2+m_4+m_5+m_6+m_8+m_9}
(1+|\bm a_{6,3}|')_{-m_1-m_2-m_3-m_4-m_5+m_7} 
(1+|\bm a_{3,5}|')_{m_1+m_3+m_5-m_6-m_7} 
} \,,
\end{multline}
where 
\be
\label{cross_6_356}
\begin{split}
(\cycle_6)^4 \circ \rY^{\langle 125 \rangle}_1 &= \frac{X_{16} X_{35}}{X_{15} X_{36}} \,, \quad
(\cycle_6)^4 \circ \rY^{\langle 125 \rangle}_2 = \frac{X_{56} X_{13}}{X_{15} X_{36}} \,, \qquad\;\;
(\cycle_6)^4 \circ \rY^{\langle 125 \rangle}_3 =  \frac{X_{26} X_{35}}{X_{25} X_{36}} \,, \\
(\cycle_6)^4 \circ \rY^{\langle 125 \rangle}_4 &= \frac{X_{56} X_{23}}{X_{25} X_{36}} \,, \quad
(\cycle_6)^4 \circ \rY^{\langle 125 \rangle}_5 =  \frac{X_{56} X_{35} X_{12}}{X_{15} X_{25} X_{36}} \,, \quad
(\cycle_6)^4 \circ \rY^{\langle 125 \rangle}_6 =  \frac{X_{56} X_{34}}{X_{46} X_{35}} \,, \\
(\cycle_6)^4 \circ \rY^{\langle 125 \rangle}_7 &=  \frac{X_{36} X_{45}}{X_{46} X_{35}} \,, \quad
(\cycle_6)^4 \circ \rY^{\langle 125 \rangle}_8 =   \frac{X_{56} X_{14}}{X_{15} X_{46}} \,, \qquad\;\;\;
(\cycle_6)^4 \circ \rY^{\langle 125 \rangle}_9 =   \frac{X_{56} X_{24}}{X_{25} X_{46}} \,.
\end{split}
\ee

\paragraph{$\triple{146}$.}The leg-factor is
\be
\label{V_6_146}
\rV_6^{\triple{146}}(\bm a|\bm x) 
=
X_{61}^{|\bm a_{6,1}|'} X_{14}^{|\bm a_{1,4}|'} X_{46}^{|\bm a_{4,6}|'}\, X_{26}^{-a_2} X_{36}^{-a_3} X_{15}^{-a_5} \,.
\ee
The cross-ratio set is produced from \eqref{cross_6_125}: $\bm \rY_6^{\triple{146}} = \{ (\cycle_6)^5 \circ \rY^{\triple{125}}_l, l=1,...,9\}$. The polygonal  function is
\begin{multline}
\label{H6_146}
\HG_6^{\triple{146}}\left( {\bm a}\big|\bm \rY^{\triple{146} }\right)
=
\sum_{m_1,..., m_9=0}^\infty \;\; \prod_{l=1}^9 \frac{\left((\cycle_6)^5 \circ \rY_l^{\triple{125} }\right)^{m_l}}{m_l!}
\\
\times
\frac{
(a_2)_{m_1+m_2+m_5+m_8} 
(a_3)_{m_3+m_4+m_5+m_9} 
(a_5)_{m_6+m_7+m_8+m_9}  \,  
(-)^{m_1+m_2+m_3+m_4+m_6+m_7} 
}{
(1+|\bm a_{6,1}|')_{m_2+m_4+m_5+m_6+m_8+m_9}
(1+|\bm a_{1,4}|')_{-m_1-m_2-m_3-m_4-m_5+m_7} 
(1+|\bm a_{4,6}|')_{m_1+m_3+m_5-m_6-m_7} 
} \,,
\end{multline}
where 
\be
\label{cross_6_146}
\begin{split}
(\cycle_6)^5 \circ \rY^{\langle 125 \rangle}_1 &= \frac{X_{12} X_{46}}{X_{26} X_{14}} \,, \quad
(\cycle_6)^5 \circ \rY^{\langle 125 \rangle}_2 = \frac{X_{16} X_{24}}{X_{26} X_{14}} \,, \qquad\;\;
(\cycle_6)^5 \circ \rY^{\langle 125 \rangle}_3 =  \frac{X_{13} X_{46}}{X_{36} X_{14}} \,, \\
(\cycle_6)^5 \circ \rY^{\langle 125 \rangle}_4 &= \frac{X_{16} X_{34}}{X_{36} X_{14}} \,, \quad
(\cycle_6)^5 \circ \rY^{\langle 125 \rangle}_5 =  \frac{X_{16} X_{46} X_{23}}{X_{26} X_{36} X_{14}} \,, \quad
(\cycle_6)^5 \circ \rY^{\langle 125 \rangle}_6 =  \frac{X_{16} X_{45}}{X_{15} X_{46}} \,, \\
(\cycle_6)^5 \circ \rY^{\langle 125 \rangle}_7 &=  \frac{X_{14} X_{56}}{X_{15} X_{46}} \,, \quad
(\cycle_6)^5 \circ \rY^{\langle 125 \rangle}_8 =   \frac{X_{16} X_{25}}{X_{26} X_{15}} \,, \qquad\;\;\;
(\cycle_6)^5 \circ \rY^{\langle 125 \rangle}_9 =   \frac{X_{16} X_{35}}{X_{36} X_{15}} \,.
\end{split}
\ee

\subsubsection*{Orbit $\{\ZZ_6 \circ \triple{135} \}$:}

\paragraph{$\triple{246}$.}The leg-factor is
\be
\label{V_6_246}
\rV_6^{\triple{246}}(\bm a|\bm x) 
=
X_{24}^{|\bm a_{2,4}|'} X_{46}^{|\bm a_{4,6}|'} X_{62}^{|\bm a_{6,2}|'}\, X_{25}^{-a_5} X_{14}^{-a_1} X_{36}^{-a_3} \,.
\ee
The cross-ratio set is produced from \eqref{cross_6_135}: $\bm \rY_6^{\triple{246}} = \{ \cycle_6 \circ \rY^{\triple{135}}_l, l=1,...,9\}$. The polygonal  function is
\begin{multline}
\label{H6_246}
\HG_6^{\triple{246}}\left( {\bm a}\big|\bm \rY^{\triple{246}}\right)
=
\sum_{m_1,..., m_9=0}^\infty \;\; \prod_{l=1}^9 \frac{\left(\cycle_6 \circ \rY_l^{\triple{135} }\right)^{m_l}}{m_l!}
\\
\times
\frac{
(a_3)_{m_1+m_2+m_5+m_8} 
(a_5)_{m_3+m_4+m_5+m_9} 
(a_1)_{m_6+m_7+m_8+m_9}  \,  
(-)^{m_1+m_2+m_3+m_4+m_6+m_7} 
}{
(1+|\bm a_{2,4}|')_{-m_1-m_2+m_3+m_6+m_9}
(1+|\bm a_{4,6}|')_{m_1-m_3-m_4+m_7+m_8} 
(1+|\bm a_{6,2}|')_{m_2+m_4+m_5-m_6-m_7} 
} \,,
\end{multline}
where 
\be
\label{cross_6_246}
\begin{split}
\cycle_6 \circ \rY^{\langle 135 \rangle}_1 &=  \frac{X_{46} X_{23}}{X_{36} X_{24}} \,, \quad
\cycle_6 \circ \rY^{\langle 135 \rangle}_2 = \frac{X_{26} X_{34}}{X_{24} X_{36}} \,, \quad
\cycle_6 \circ \rY^{\langle 135 \rangle}_3 =  \frac{ X_{24} X_{56}}{ X_{25} X_{46} } \,, \\
\cycle_6 \circ \rY^{\langle 135 \rangle}_4 &= \frac{X_{26} X_{45}}{X_{25} X_{46}} \,, \quad
\cycle_6 \circ \rY^{\langle 135 \rangle}_5 =  \frac{X_{26} X_{35}}{X_{25} X_{36}} \,, \quad
\cycle_6 \circ \rY^{\langle 135 \rangle}_6 = \frac{X_{24} X_{16}}{X_{26} X_{14}} \,,   \\
\cycle_6 \circ \rY^{\langle 135 \rangle}_7 &=  \frac{X_{46} X_{12}}{X_{26} X_{14}} \,, \quad
\cycle_6 \circ \rY^{\langle 135 \rangle}_8 = \frac{X_{46} X_{13}}{X_{36} X_{14}} \,, \quad
\cycle_6 \circ \rY^{\langle 135 \rangle}_9 = \frac{X_{24} X_{15}}{X_{25} X_{14}} \,.
\end{split}
\ee


\begin{thebibliography}{10}

\bibitem{Alkalaev:2025fgn}
K.~Alkalaev and S.~Mandrygin, \emph{{Multipoint conformal integrals in D
  dimensions. Part I. Bipartite Mellin-Barnes representation and
  reconstruction}},
  \href{http://dx.doi.org/10.1007/JHEP09(2025)118}{\emph{JHEP} {\bf 09} (2025)
  118}, [\href{http://arxiv.org/abs/2502.12127}{{\tt 2502.12127}}].

\bibitem{Smirnov:2004ym}
V.~A. Smirnov, \emph{{Evaluating Feynman integrals}}, {\emph{Springer Tracts
  Mod. Phys.} {\bf 211} (2004) 1--244}.

\bibitem{Weinzierl:2022eaz}
S.~Weinzierl, \emph{{Feynman Integrals. A Comprehensive Treatment for Students
  and Researchers}}.
\newblock UNITEXT for Physics. Springer, 2022,
  \href{http://dx.doi.org/10.1007/978-3-030-99558-4}{10.1007/978-3-030-99558-4}.

\bibitem{Kalmykov:2008gq}
M.~Y. Kalmykov, B.~A. Kniehl, B.~F.~L. Ward and S.~A. Yost,
  \emph{{Hypergeometric functions, their epsilon expansions and Feynman
  diagrams}},  in \emph{{15th International Seminar on High Energy Physics}},
  10, 2008.
\newblock \href{http://arxiv.org/abs/0810.3238}{{\tt 0810.3238}}.

\bibitem{Bourjaily:2022bwx}
J.~L. Bourjaily et~al., \emph{{Functions Beyond Multiple Polylogarithms for
  Precision Collider Physics}},  in \emph{{Snowmass 2021}}, 3, 2022.
\newblock \href{http://arxiv.org/abs/2203.07088}{{\tt 2203.07088}}.

\bibitem{Usyukina:1992jd}
N.~I. Usyukina and A.~I. Davydychev, \emph{{An Approach to the evaluation of
  three and four point ladder diagrams}},
  \href{http://dx.doi.org/10.1016/0370-2693(93)91834-A}{\emph{Phys. Lett. B}
  {\bf 298} (1993) 363--370}.

\bibitem{Symanzik:1972wj}
K.~Symanzik, \emph{{On Calculations in conformal invariant field theories}},
  \href{http://dx.doi.org/10.1007/BF02824349}{\emph{Lett. Nuovo Cim.} {\bf 3}
  (1972) 734--738}.

\bibitem{Ferrara:1972uq}
S.~Ferrara, A.~F. Grillo, G.~Parisi and R.~Gatto, \emph{{The shadow operator
  formalism for conformal algebra. Vacuum expectation values and operator
  products}}, \href{http://dx.doi.org/10.1007/BF02907130}{\emph{Lett. Nuovo
  Cim.} {\bf 4S2} (1972) 115--120}.

\bibitem{Ferrara:1972kab}
S.~Ferrara, A.~F. Grillo, G.~Parisi and R.~Gatto, \emph{{Covariant expansion of
  the conformal four-point function}},
  \href{http://dx.doi.org/10.1016/0550-3213(73)90467-7}{\emph{Nucl. Phys. B}
  {\bf 49} (1972) 77--98}.

\bibitem{Dolan:2000uw}
F.~A. Dolan and H.~Osborn, \emph{{Implications of N=1 superconformal symmetry
  for chiral fields}},
  \href{http://dx.doi.org/10.1016/S0550-3213(00)00553-8}{\emph{Nucl. Phys. B}
  {\bf 593} (2001) 599--633}, [\href{http://arxiv.org/abs/hep-th/0006098}{{\tt
  hep-th/0006098}}].

\bibitem{Dolan:2000ut}
F.~A. Dolan and H.~Osborn, \emph{{Conformal four point functions and the
  operator product expansion}},
  \href{http://dx.doi.org/10.1016/S0550-3213(01)00013-X}{\emph{Nucl. Phys. B}
  {\bf 599} (2001) 459--496}, [\href{http://arxiv.org/abs/hep-th/0011040}{{\tt
  hep-th/0011040}}].

\bibitem{Dolan:2011dv}
F.~Dolan and H.~Osborn, \emph{{Conformal Partial Waves: Further Mathematical
  Results}},  \href{http://arxiv.org/abs/1108.6194}{{\tt 1108.6194}}.

\bibitem{Fateev:2011qa}
V.~Fateev and S.~Ribault, \emph{{The Large central charge limit of conformal
  blocks}}, \href{http://dx.doi.org/10.1007/JHEP02(2012)001}{\emph{JHEP} {\bf
  02} (2012) 001}, [\href{http://arxiv.org/abs/1109.6764}{{\tt 1109.6764}}].

\bibitem{SimmonsDuffin:2012uy}
D.~Simmons-Duffin, \emph{{Projectors, Shadows, and Conformal Blocks}},
  \href{http://dx.doi.org/10.1007/JHEP04(2014)146}{\emph{JHEP} {\bf 04} (2014)
  146}, [\href{http://arxiv.org/abs/1204.3894}{{\tt 1204.3894}}].

\bibitem{Rosenhaus:2018zqn}
V.~Rosenhaus, \emph{{Multipoint Conformal Blocks in the Comb Channel}},
  \href{http://dx.doi.org/10.1007/JHEP02(2019)142}{\emph{JHEP} {\bf 02} (2019)
  142}, [\href{http://arxiv.org/abs/1810.03244}{{\tt 1810.03244}}].

\bibitem{Petkou:2021zhg}
A.~C. Petkou, \emph{{Thermal one-point functions and single-valued
  polylogarithms}},
  \href{http://dx.doi.org/10.1016/j.physletb.2021.136467}{\emph{Phys. Lett. B}
  {\bf 820} (2021) 136467}, [\href{http://arxiv.org/abs/2105.03530}{{\tt
  2105.03530}}].

\bibitem{Karydas:2023ufs}
M.~Karydas, S.~Li, A.~C. Petkou and M.~Vilatte, \emph{{Conformal Graphs as
  Twisted Partition Functions}},
  \href{http://dx.doi.org/10.1103/PhysRevLett.132.231601}{\emph{Phys. Rev.
  Lett.} {\bf 132} (2024) 231601}, [\href{http://arxiv.org/abs/2312.00135}{{\tt
  2312.00135}}].

\bibitem{Alkalaev:2023evp}
K.~Alkalaev and S.~Mandrygin, \emph{{Torus shadow formalism and exact global
  conformal blocks}},
  \href{http://dx.doi.org/10.1007/JHEP11(2023)157}{\emph{JHEP} {\bf 11} (2023)
  157}, [\href{http://arxiv.org/abs/2307.12061}{{\tt 2307.12061}}].

\bibitem{Alkalaev:2024jxh}
K.~Alkalaev and S.~Mandrygin, \emph{{One-point thermal conformal blocks from
  four-point conformal integrals}},
  \href{http://dx.doi.org/10.1007/JHEP10(2024)241}{\emph{JHEP} {\bf 10} (2024)
  241}, [\href{http://arxiv.org/abs/2407.01741}{{\tt 2407.01741}}].

\bibitem{Belavin:2024mzd}
V.~Belavin, J.~Ramos~Cabezas and B.~Runov, \emph{{Shadow formalism for
  supersymmetric conformal blocks}},
  \href{http://dx.doi.org/10.1007/JHEP11(2024)048}{\emph{JHEP} {\bf 11} (2024)
  048}, [\href{http://arxiv.org/abs/2408.07684}{{\tt 2408.07684}}].

\bibitem{Belavin:2024nnw}
V.~Belavin and J.~Ramos~Cabezas, \emph{{Global conformal blocks via shadow
  formalism}}, \href{http://dx.doi.org/10.1007/JHEP02(2024)167}{\emph{JHEP}
  {\bf 02} (2024) 167}, [\href{http://arxiv.org/abs/2401.02580}{{\tt
  2401.02580}}].

\bibitem{Drummond:2006rz}
J.~M. Drummond, J.~Henn, V.~A. Smirnov and E.~Sokatchev, \emph{{Magic
  identities for conformal four-point integrals}},
  \href{http://dx.doi.org/10.1088/1126-6708/2007/01/064}{\emph{JHEP} {\bf 01}
  (2007) 064}, [\href{http://arxiv.org/abs/hep-th/0607160}{{\tt
  hep-th/0607160}}].

\bibitem{Drummond:2008vq}
J.~M. Drummond, J.~Henn, G.~P. Korchemsky and E.~Sokatchev, \emph{{Dual
  superconformal symmetry of scattering amplitudes in N=4 super-Yang-Mills
  theory}},
  \href{http://dx.doi.org/10.1016/j.nuclphysb.2009.11.022}{\emph{Nucl. Phys. B}
  {\bf 828} (2010) 317--374}, [\href{http://arxiv.org/abs/0807.1095}{{\tt
  0807.1095}}].

\bibitem{Zamolodchikov:1980mb}
A.~B. Zamolodchikov, \emph{{'Fishnet' diagrams as a completely integrable
  system}}, \href{http://dx.doi.org/10.1016/0370-2693(80)90547-X}{\emph{Phys.
  Lett. B} {\bf 97} (1980) 63--66}.

\bibitem{Gurdogan:2015csr}
O.~G\"urdo\u{g}an and V.~Kazakov, \emph{{New Integrable 4D Quantum Field
  Theories from Strongly Deformed Planar $\mathcal N = $ 4 Supersymmetric
  Yang-Mills Theory}},
  \href{http://dx.doi.org/10.1103/PhysRevLett.117.201602}{\emph{Phys. Rev.
  Lett.} {\bf 117} (2016) 201602}, [\href{http://arxiv.org/abs/1512.06704}{{\tt
  1512.06704}}].

\bibitem{Loebbert:2022nfu}
F.~Loebbert, \emph{{Integrability for Feynman integrals}},
  \href{http://dx.doi.org/10.21468/SciPostPhysProc.14.008}{\emph{SciPost Phys.
  Proc.} {\bf 14} (2023) 008}, [\href{http://arxiv.org/abs/2212.09636}{{\tt
  2212.09636}}].

\bibitem{Osborn}
H.~Osborn, ``Lectures on conformal field theories in more than two
  dimensions.'' \url{https://www.damtp.cam.ac.uk/user/ho/CFTNotes.pdf}.

\bibitem{Buric:2021kgy}
I.~Buric, S.~Lacroix, J.~A. Mann, L.~Quintavalle and V.~Schomerus,
  \emph{{Gaudin models and multipoint conformal blocks III: comb channel
  coordinates and OPE factorisation}},
  \href{http://dx.doi.org/10.1007/JHEP06(2022)144}{\emph{JHEP} {\bf 06} (2022)
  144}, [\href{http://arxiv.org/abs/2112.10827}{{\tt 2112.10827}}].

\bibitem{DelDuca:2011wh}
V.~Del~Duca, L.~J. Dixon, J.~M. Drummond, C.~Duhr, J.~M. Henn and V.~A.
  Smirnov, \emph{{The one-loop six-dimensional hexagon integral with three
  massive corners}},
  \href{http://dx.doi.org/10.1103/PhysRevD.84.045017}{\emph{Phys. Rev. D} {\bf
  84} (2011) 045017}, [\href{http://arxiv.org/abs/1105.2011}{{\tt 1105.2011}}].

\bibitem{Nandan:2013ip}
D.~Nandan, M.~F. Paulos, M.~Spradlin and A.~Volovich, \emph{{Star Integrals,
  Convolutions and Simplices}},
  \href{http://dx.doi.org/10.1007/JHEP05(2013)105}{\emph{JHEP} {\bf 05} (2013)
  105}, [\href{http://arxiv.org/abs/1301.2500}{{\tt 1301.2500}}].

\bibitem{Loebbert:2019vcj}
F.~Loebbert, D.~M\"uller and H.~M\"unkler, \emph{{Yangian Bootstrap for
  Conformal Feynman Integrals}},
  \href{http://dx.doi.org/10.1103/PhysRevD.101.066006}{\emph{Phys. Rev. D} {\bf
  101} (2020) 066006}, [\href{http://arxiv.org/abs/1912.05561}{{\tt
  1912.05561}}].

\bibitem{Davydychev:1997wa}
A.~I. Davydychev and R.~Delbourgo, \emph{{A Geometrical angle on Feynman
  integrals}}, \href{http://dx.doi.org/10.1063/1.532513}{\emph{J. Math. Phys.}
  {\bf 39} (1998) 4299--4334}, [\href{http://arxiv.org/abs/hep-th/9709216}{{\tt
  hep-th/9709216}}].

\bibitem{Mason:2010pg}
L.~Mason and D.~Skinner, \emph{{Amplitudes at Weak Coupling as Polytopes in
  AdS$_{5}$}}, \href{http://dx.doi.org/10.1088/1751-8113/44/13/135401}{\emph{J.
  Phys. A} {\bf 44} (2011) 135401}, [\href{http://arxiv.org/abs/1004.3498}{{\tt
  1004.3498}}].

\bibitem{Schnetz:2010pd}
O.~Schnetz, \emph{{The geometry of one-loop amplitudes}},
  \href{http://arxiv.org/abs/1010.5334}{{\tt 1010.5334}}.

\bibitem{Bourjaily:2019exo}
J.~L. Bourjaily, E.~Gardi, A.~J. McLeod and C.~Vergu, \emph{{All-mass $n$-gon
  integrals in $n$ dimensions}},
  \href{http://dx.doi.org/10.1007/JHEP08(2020)029}{\emph{JHEP} {\bf 08} (2020)
  029}, [\href{http://arxiv.org/abs/1912.11067}{{\tt 1912.11067}}].

\bibitem{Ren:2023tuj}
L.~Ren, M.~Spradlin, C.~Vergu and A.~Volovich, \emph{{One-loop integrals from
  volumes of orthoschemes}},
  \href{http://dx.doi.org/10.1007/JHEP05(2024)104}{\emph{JHEP} {\bf 05} (2024)
  104}, [\href{http://arxiv.org/abs/2306.04630}{{\tt 2306.04630}}].

\bibitem{Ananthanarayan:2020ncn}
B.~Ananthanarayan, S.~Banik, S.~Friot and S.~Ghosh, \emph{{Double box and
  hexagon conformal Feynman integrals}},
  \href{http://dx.doi.org/10.1103/PhysRevD.102.091901}{\emph{Phys. Rev. D} {\bf
  102} (2020) 091901}, [\href{http://arxiv.org/abs/2007.08360}{{\tt
  2007.08360}}].

\bibitem{Ananthanarayan:2020fhl}
B.~Ananthanarayan, S.~Banik, S.~Friot and S.~Ghosh, \emph{{Multiple Series
  Representations of N-fold Mellin-Barnes Integrals}},
  \href{http://dx.doi.org/10.1103/PhysRevLett.127.151601}{\emph{Phys. Rev.
  Lett.} {\bf 127} (2021) 151601}, [\href{http://arxiv.org/abs/2012.15108}{{\tt
  2012.15108}}].

\bibitem{Banik:2023rrz}
S.~Banik and S.~Friot, \emph{{Multiple Mellin-Barnes integrals and
  triangulations of point configurations}},
  \href{http://dx.doi.org/10.1103/PhysRevD.110.036002}{\emph{Phys. Rev. D} {\bf
  110} (2024) 036002}, [\href{http://arxiv.org/abs/2309.00409}{{\tt
  2309.00409}}].

\bibitem{Banik:2024ann}
S.~Banik and S.~Friot, \emph{{Analytic Evaluation of Multiple Mellin-Barnes
  Integrals}},  in \emph{{Loops and Legs in Quantum Field Theory}}, 7, 2024.
\newblock \href{http://arxiv.org/abs/2407.20120}{{\tt 2407.20120}}.

\bibitem{Pal:2021llg}
A.~Pal and K.~Ray, \emph{{Conformal integrals in four dimensions}},
  \href{http://dx.doi.org/10.1007/JHEP10(2022)087}{\emph{JHEP} {\bf 10} (2022)
  087}, [\href{http://arxiv.org/abs/2109.09379}{{\tt 2109.09379}}].

\bibitem{Pal:2023kgu}
A.~Pal and K.~Ray, \emph{{Conformal integrals in all dimensions as generalized
  hypergeometric functions and Clifford groups}},
  \href{http://dx.doi.org/10.1063/5.0222880}{\emph{J. Math. Phys.} {\bf 66}
  (2025) 043504}, [\href{http://arxiv.org/abs/2303.17326}{{\tt 2303.17326}}].

\bibitem{Levkovich-Maslyuk:2024zdy}
F.~Levkovich-Maslyuk and V.~Mishnyakov, \emph{{Yangian symmetry, GKZ equations
  and integrable Feynman graphs in conformal variables}},
  \href{http://arxiv.org/abs/2412.19296}{{\tt 2412.19296}}.

\bibitem{Chicherin:2017cns}
D.~Chicherin, V.~Kazakov, F.~Loebbert, D.~M\"uller and D.-l. Zhong,
  \emph{{Yangian Symmetry for Bi-Scalar Loop Amplitudes}},
  \href{http://dx.doi.org/10.1007/JHEP05(2018)003}{\emph{JHEP} {\bf 05} (2018)
  003}, [\href{http://arxiv.org/abs/1704.01967}{{\tt 1704.01967}}].

\bibitem{Chicherin:2017frs}
D.~Chicherin, V.~Kazakov, F.~Loebbert, D.~M\"uller and D.-l. Zhong,
  \emph{{Yangian Symmetry for Fishnet Feynman Graphs}},
  \href{http://dx.doi.org/10.1103/PhysRevD.96.121901}{\emph{Phys. Rev. D} {\bf
  96} (2017) 121901}, [\href{http://arxiv.org/abs/1708.00007}{{\tt
  1708.00007}}].

\bibitem{Isaev_Rubakov_1}
A.~P. Isaev and V.~A. Rubakov, \emph{Theory of Groups and Symmetries}.
\newblock WORLD SCIENTIFIC, 2018,
  \href{http://dx.doi.org/10.1142/10898}{10.1142/10898}.

\bibitem{Dubovyk:2022obc}
I.~Dubovyk, J.~Gluza and G.~Somogyi, \emph{{Mellin-Barnes Integrals: A Primer
  on Particle Physics Applications}},
  \href{http://dx.doi.org/10.1007/978-3-031-14272-7}{\emph{Lect. Notes Phys.}
  {\bf 1008} (2022) pp.}, [\href{http://arxiv.org/abs/2211.13733}{{\tt
  2211.13733}}].

\bibitem{Zhdanov1998}
O.~N. Zhdanov and A.~K. Tsikh, \emph{Studying the multiple mellin-barnes
  integrals by means of multidimensional residues},
  \href{http://dx.doi.org/10.1007/bf02677509}{\emph{Siberian Mathematical
  Journal} {\bf 39} (Apr., 1998) 245–260}.

\bibitem{Friot:2011ic}
S.~Friot and D.~Greynat, \emph{{On convergent series representations of
  Mellin-Barnes integrals}},
  \href{http://dx.doi.org/10.1063/1.3679686}{\emph{J. Math. Phys.} {\bf 53}
  (2012) 023508}, [\href{http://arxiv.org/abs/1107.0328}{{\tt 1107.0328}}].

\bibitem{Baxter:1978xr}
R.~J. Baxter, \emph{{Solvable eight vertex model on an arbitrary planar
  lattice}}, \href{http://dx.doi.org/10.1098/rsta.1978.0062}{\emph{Phil. Trans.
  Roy. Soc. Lond. A} {\bf 289} (1978) 315--346}.

\bibitem{Kazakov:2022dbd}
V.~Kazakov and E.~Olivucci, \emph{{The loom for general fishnet CFTs}},
  \href{http://dx.doi.org/10.1007/JHEP06(2023)041}{\emph{JHEP} {\bf 06} (2023)
  041}, [\href{http://arxiv.org/abs/2212.09732}{{\tt 2212.09732}}].

\bibitem{Kazakov:2023nyu}
V.~Kazakov, F.~Levkovich-Maslyuk and V.~Mishnyakov, \emph{{Integrable Feynman
  graphs and Yangian symmetry on the loom}},
  \href{http://dx.doi.org/10.1007/JHEP06(2025)104}{\emph{JHEP} {\bf 06} (2025)
  104}, [\href{http://arxiv.org/abs/2304.04654}{{\tt 2304.04654}}].

\bibitem{Alfimov:2023vev}
M.~Alfimov, G.~Ferrando, V.~Kazakov and E.~Olivucci, \emph{{Checkerboard CFT}},
  \href{http://dx.doi.org/10.1007/JHEP01(2025)015}{\emph{JHEP} {\bf 01} (2025)
  015}, [\href{http://arxiv.org/abs/2311.01437}{{\tt 2311.01437}}].

\bibitem{Caratheodory:1919zz}
C.~Caratheodory, \emph{{Über den variabilitätsbereich der fourier’schen
  konstanten von positiven harmonischen funktionen}},
  \href{http://dx.doi.org/10.1007/BF03014795}{\emph{Rendiconti del Circolo
  Matematico di Palermo (1884-1940)} (1911) }.

\bibitem{Vasiliev:1981dg}
A.~N. Vasiliev, Y.~M. Pismak and Y.~R. Khonkonen, \emph{{1/$N$ Expansion:
  Calculation of the Exponents $\eta$ and Nu in the Order 1/$N^2$ for Arbitrary
  Number of Dimensions}},
  \href{http://dx.doi.org/10.1007/BF01019296}{\emph{Theor. Math. Phys.} {\bf
  47} (1981) 465--475}.

\bibitem{Kazakov:1983dyk}
D.~I. Kazakov, \emph{{The Method Of Uniqueness, A New Powerful Technique For
  Multiloop Calculations}},
  \href{http://dx.doi.org/10.1016/0370-2693(83)90816-X}{\emph{Phys. Lett. B}
  {\bf 133} (1983) 406--410}.

\bibitem{Gorishnii:1984te}
S.~G. Gorishnii and A.~P. Isaev, \emph{{On an Approach to the Calculation of
  Multiloop Massless Feynman Integrals}},
  \href{http://dx.doi.org/10.1007/BF01018263}{\emph{Theor. Math. Phys.} {\bf
  62} (1985) 232}.

\bibitem{Isaev:2003tk}
A.~P. Isaev, \emph{{Multiloop Feynman integrals and conformal quantum
  mechanics}},
  \href{http://dx.doi.org/10.1016/S0550-3213(03)00393-6}{\emph{Nucl. Phys. B}
  {\bf 662} (2003) 461--475}, [\href{http://arxiv.org/abs/hep-th/0303056}{{\tt
  hep-th/0303056}}].

\bibitem{Chicherin:2012yn}
D.~Chicherin, S.~Derkachov and A.~P. Isaev, \emph{{Conformal group: R-matrix
  and star-triangle relation}},
  \href{http://dx.doi.org/10.1007/JHEP04(2013)020}{\emph{JHEP} {\bf 04} (2013)
  020}, [\href{http://arxiv.org/abs/1206.4150}{{\tt 1206.4150}}].

\bibitem{Derkachov:2022ytx}
S.~Derkachov, A.~P. Isaev and L.~Shumilov, \emph{{Conformal triangles and
  zig-zag diagrams}},
  \href{http://dx.doi.org/10.1016/j.physletb.2022.137150}{\emph{Phys. Lett. B}
  {\bf 830} (2022) 137150}, [\href{http://arxiv.org/abs/2201.12232}{{\tt
  2201.12232}}].

\bibitem{Derkachov:2023xqq}
S.~E. Derkachov, A.~P. Isaev and L.~A. Shumilov, \emph{{Ladder and zig-zag
  Feynman diagrams, operator formalism and conformal triangles}},
  \href{http://dx.doi.org/10.1007/JHEP06(2023)059}{\emph{JHEP} {\bf 06} (2023)
  059}, [\href{http://arxiv.org/abs/2302.11238}{{\tt 2302.11238}}].

\bibitem{Horn1889}
J.~Horn, \emph{Ueber die convergenz der hypergeometrischen reihen zweier und
  dreier veranderlichen},
  \href{http://dx.doi.org/10.1007/bf01443681}{\emph{Mathematische Annalen} {\bf
  34} (Dec., 1889) 544–600}.

\bibitem{exton1976multiple}
H.~Exton, \emph{Multiple Hypergeometric Functions and Applications}.
\newblock Ellis Horwood series in mathematics and its applications. E. Horwood,
  1976.

\bibitem{Bateman:100233}
H.~Bateman and A.~Erdélyi, \emph{{Higher transcendental functions}}.
\newblock California Institute of technology. Bateman Manuscript project.
  McGraw-Hill, New York, NY, 1953.

\bibitem{Davydychev:1992eww}
A.~I. Davydychev, \emph{{Recursive algorithm for evaluating vertex-type Feynman
  integrals}}, \href{http://dx.doi.org/10.1088/0305-4470/25/21/017}{\emph{J.
  Phys. A} {\bf 25} (1992) 5587}.

\bibitem{SrivastavaDaoust}
H.~M. Srivastava and M.~C. Daoust, \emph{{Certain generalized Neumann
  expansions associated with the Kampe De Feriet function}}, {\emph{Nederl.
  Akad. Wetensch. Indag. Math.} {\bf 31} (1969) 449}.

\bibitem{Srivastava1985MultipleGH}
H.~M. Srivastava and P.~W. Karlsson, \emph{Multiple gaussian hypergeometric
  series},  1985.

\bibitem{Srivastava1972}
H.~M. Srivastava and M.~C. Daoust, \emph{A note on the convergence of
  {Kamp{\'e}} de {F{\'e}riet}'s double hypergeometrics series}, {\emph{Math.
  Nachr.} {\bf 53} (1972) 151--159}.

\bibitem{Bezrodnykh:2018review}
S.~I. Bezrodnykh, \emph{{The Lauricella hypergeometric function $F_D^{(N)}$,
  the Riemann--Hilbert problem, and some applications}},
  \href{http://dx.doi.org/10.1070/RM9841}{\emph{Uspekhi Mat. Nauk} {\bf 73}
  (2018) 3--94}.

\bibitem{Ananthanarayan:2020xpd}
B.~Ananthanarayan, S.~Banik, S.~Friot and S.~Ghosh, \emph{{Massive One-loop
  Conformal Feynman Integrals and Quadratic Transformations of Multiple
  Hypergeometric Series}},
  \href{http://dx.doi.org/10.1103/PhysRevD.103.096008}{\emph{Phys. Rev. D} {\bf
  103} (2021) 096008}, [\href{http://arxiv.org/abs/2012.15646}{{\tt
  2012.15646}}].

\bibitem{Ananthanarayan:2021yar}
B.~Ananthanarayan, S.~Bera, S.~Friot and T.~Pathak, \emph{{Olsson.wl \&
  ROC2.wl: Mathematica packages for transformations of multivariable
  hypergeometric functions \& regions of convergence for their series
  representations in the two variables case}},
  \href{http://dx.doi.org/10.1016/j.cpc.2024.109162}{\emph{Comput. Phys.
  Commun.} {\bf 300} (2024) 109162},
  [\href{http://arxiv.org/abs/2201.01189}{{\tt 2201.01189}}].

\bibitem{Gelfand1990}
I.~Gelfand, M.~Kapranov and A.~Zelevinsky, \emph{Generalized euler integrals
  and a-hypergeometric functions},
  \href{http://dx.doi.org/https://doi.org/10.1016/0001-8708(90)90048-R}{\emph{Advances
  in Mathematics} {\bf 84} (1990) 255--271}.

\bibitem{Gelfand1992}
I.~Gelfand, M.~Graev and V.~Retakh, \emph{General hypergeometric systems of
  equations and series of hypergeometric type},
  \href{http://dx.doi.org/10.1070/rm1992v047n04abeh000915}{\emph{Russian
  Mathematical Surveys} {\bf 47} (Aug., 1992) 1–88}.

\bibitem{tsikh}
L.~Nilsson, M.~Passare and A.~Tsikh, \emph{{Domains of convergence for
  $A$-hypergeometric series and integrals}}, {\emph{J. Sib. Fed. Univ. Math.
  Phys.} {\bf 12} (2019) 509}.

\bibitem{Corcoran:2020akn}
L.~Corcoran and M.~Staudacher, \emph{{The dual conformal box integral in
  Minkowski space}},
  \href{http://dx.doi.org/10.1016/j.nuclphysb.2021.115310}{\emph{Nucl. Phys. B}
  {\bf 964} (2021) 115310}, [\href{http://arxiv.org/abs/2006.11292}{{\tt
  2006.11292}}].

\bibitem{Corcoran:2020epz}
L.~Corcoran, F.~Loebbert, J.~Miczajka and M.~Staudacher, \emph{{Minkowski Box
  from Yangian Bootstrap}},
  \href{http://dx.doi.org/10.1007/JHEP04(2021)160}{\emph{JHEP} {\bf 04} (2021)
  160}, [\href{http://arxiv.org/abs/2012.07852}{{\tt 2012.07852}}].

\bibitem{Corcoran:2023ljn}
L.~Corcoran, \emph{{Conformal Feynman Integrals and Correlation Functions in
  Fishnet Theory}}.
\newblock PhD thesis, Humboldt U., Berlin, 2023.
\newblock 10.18452/25602.

\bibitem{Paulos:2012nu}
M.~F. Paulos, M.~Spradlin and A.~Volovich, \emph{{Mellin Amplitudes for Dual
  Conformal Integrals}},
  \href{http://dx.doi.org/10.1007/JHEP08(2012)072}{\emph{JHEP} {\bf 08} (2012)
  072}, [\href{http://arxiv.org/abs/1203.6362}{{\tt 1203.6362}}].

\bibitem{Basso:2017jwq}
B.~Basso and L.~J. Dixon, \emph{{Gluing Ladder Feynman Diagrams into
  Fishnets}},
  \href{http://dx.doi.org/10.1103/PhysRevLett.119.071601}{\emph{Phys. Rev.
  Lett.} {\bf 119} (2017) 071601}, [\href{http://arxiv.org/abs/1705.03545}{{\tt
  1705.03545}}].

\bibitem{Derkachov:2018rot}
S.~Derkachov, V.~Kazakov and E.~Olivucci, \emph{{Basso-Dixon Correlators in
  Two-Dimensional Fishnet CFT}},
  \href{http://dx.doi.org/10.1007/JHEP04(2019)032}{\emph{JHEP} {\bf 04} (2019)
  032}, [\href{http://arxiv.org/abs/1811.10623}{{\tt 1811.10623}}].

\bibitem{Duhr:2023bku}
C.~Duhr and F.~Porkert, \emph{{Feynman integrals in two dimensions and
  single-valued hypergeometric functions}},
  \href{http://dx.doi.org/10.1007/JHEP02(2024)179}{\emph{JHEP} {\bf 02} (2024)
  179}, [\href{http://arxiv.org/abs/2309.12772}{{\tt 2309.12772}}].

\bibitem{Loebbert:2024fsj}
F.~Loebbert and S.~F. Stawinski, \emph{{Conformal four-point integrals:
  recursive structure, Toda equations and double copy}},
  \href{http://dx.doi.org/10.1007/JHEP11(2024)092}{\emph{JHEP} {\bf 11} (2024)
  092}, [\href{http://arxiv.org/abs/2408.15331}{{\tt 2408.15331}}].

\bibitem{Olivucci:2021cfy}
E.~Olivucci, \emph{{Hexagonalization of Fishnet integrals. Part I. Mirror
  excitations}}, \href{http://dx.doi.org/10.1007/JHEP11(2021)204}{\emph{JHEP}
  {\bf 11} (2021) 204}, [\href{http://arxiv.org/abs/2107.13035}{{\tt
  2107.13035}}].

\bibitem{Derkachov:2021ufp}
S.~Derkachov, G.~Ferrando and E.~Olivucci, \emph{{Mirror channel eigenvectors
  of the d-dimensional fishnets}},
  \href{http://dx.doi.org/10.1007/JHEP12(2021)174}{\emph{JHEP} {\bf 12} (2021)
  174}, [\href{http://arxiv.org/abs/2108.12620}{{\tt 2108.12620}}].

\bibitem{Olivucci:2023tnw}
E.~Olivucci, \emph{{Hexagonalization of Fishnet integrals. Part II. Overlaps
  and multi-point correlators}},
  \href{http://dx.doi.org/10.1007/JHEP01(2024)081}{\emph{JHEP} {\bf 01} (2024)
  081}, [\href{http://arxiv.org/abs/2306.04503}{{\tt 2306.04503}}].

\bibitem{Aprile:2023gnh}
F.~Aprile and E.~Olivucci, \emph{{Multipoint fishnet Feynman diagrams:
  Sequential splitting}},
  \href{http://dx.doi.org/10.1103/PhysRevD.108.L121902}{\emph{Phys. Rev. D}
  {\bf 108} (2023) L121902}, [\href{http://arxiv.org/abs/2307.12984}{{\tt
  2307.12984}}].

\bibitem{Loebbert:2025abz}
F.~Loebbert, L.~R{\"u}enaufer and S.~F. Stawinski, \emph{{Non-Local Symmetries
  of Planar Feynman Integrals}},  \href{http://arxiv.org/abs/2505.05550}{{\tt
  2505.05550}}.

\bibitem{delaCruz:2019skx}
L.~de~la Cruz, \emph{{Feynman integrals as A-hypergeometric functions}},
  \href{http://dx.doi.org/10.1007/JHEP12(2019)123}{\emph{JHEP} {\bf 12} (2019)
  123}, [\href{http://arxiv.org/abs/1907.00507}{{\tt 1907.00507}}].

\bibitem{Klausen:2019hrg}
R.~P. Klausen, \emph{{Hypergeometric Series Representations of Feynman
  Integrals by GKZ Hypergeometric Systems}},
  \href{http://dx.doi.org/10.1007/JHEP04(2020)121}{\emph{JHEP} {\bf 04} (2020)
  121}, [\href{http://arxiv.org/abs/1910.08651}{{\tt 1910.08651}}].

\end{thebibliography}

\providecommand{\href}[2]{#2}\begingroup\raggedright\endgroup

\end{document}